\documentclass[sigconf,nonacm]{acmart}
\usepackage{hyperref}

\usepackage{amsfonts} 
\usepackage{mathtools}
\usepackage{graphicx}
\usepackage{algorithmic}
\usepackage{algorithm}

\usepackage{ulem}
\usepackage{nowidow}
\usepackage{wrapfig}

\usepackage[referable]{threeparttablex}
\usepackage{tabu}
\usepackage{multirow}
\usepackage{makecell}
\usepackage{longtable}
\usepackage{pifont}
\usepackage{tikz}
\usetikzlibrary{positioning,tikzmark}
\usetikzlibrary{arrows.meta}
\usepackage{xcolor}
\definecolor{green}{RGB}{    0,150,130} %
\definecolor{green70}{RGB}{ 76,181,167} %
\definecolor{blue}{RGB}{    70,100,170} %
\definecolor{blue70}{RGB}{ 125,146,195} %
\definecolor{blue50}{RGB}{ 162,177,212} %
\definecolor{blue30}{RGB}{ 199,208,229} %
\definecolor{blue15}{RGB}{ 227,232,242} %

\usepackage{colortbl}
\usepackage{booktabs}
\usepackage{blindtext}
\usepackage{caption}
\usepackage{calc}
\usepackage{pdflscape}
\usepackage{supertabular}
\definecolor{lightergray}{rgb}{0.86, 0.86, 0.86}
\usepackage[autolanguage]{numprint}
\npthousandsep{\,}

\AtBeginDocument{%
  \providecommand\BibTeX{{%
    \normalfont B\kern-0.5em{\scshape i\kern-0.25em b}\kern-0.8em\TeX}}}

\begin{document}

\title{Finding Near-Optimal Weight Independent Sets at Scale}

\author{Ernestine Großmann}%
\email{e.grossmann@informatik.uni-heidelberg.de}
\orcid{0000-0002-9678-0253}
\affiliation{%
	\institution{Heidelberg University}
	\city{Heidelberg}
	\country{Germany}
}
\author{Sebastian Lamm}%
\email{lamm@ira.uka.de}
\affiliation{%
	\institution{Karlsruhe Institute of Technology}
	\city{Karlsruhe}
	\country{Germany}
}
\orcid{0000-0001-7828-921X}

\author{Christian Schulz}%
\email{christian.schulz@informatik.uni-heidelberg.de}
\orcid{0000-0002-2823-3506}
\affiliation{%
	\institution{Heidelberg University}
	\city{Heidelberg}
	\country{Germany}
}
\author{Darren Strash}
\email{dstrash@hamilton.edu}
\affiliation{%
	\institution{Department of Computer Science, Hamilton College}
	\city{Clinton}
	\state{NY}
	\country{USA}
}
\orcid{0000-0001-7095-8749}

\renewcommand{\shortauthors}{Großmann, et al.}

\newtheorem{reduction}{Reduction}

\newcommand{\ie}{i.e.\ }
\newcommand{\etal}{et~al.}
\newcommand{\eg}{e.g.\ }
\newcommand{\niceremark}[3]{\textcolor{red}{\textsc{#1 #2: }}\textcolor{blue}{\textsf{#3}}}
\newcommand{\csch}[2][says]{\niceremark{Christian}{#1}{#2}}
\newcommand{\lamm}[2][says]{\niceremark{Sebastian}{#1}{#2}}
\newcommand{\darren}[2][says]{\niceremark{Darren}{#1}{#2}}
\newcommand{\ernestine}[2][says]{\niceremark{Ernestine}{#1}{#2}}
\newcommand{\red}[1]{{\color{red}{#1}}}

\newcommand{\I}{\mathcal{I}}
\newcommand{\hils}{\textsf{HILS}}
\newcommand{\htwis}{\textsf{HtWIS}}
\newcommand{\wmmis}{\textsf{m$^2$wis}}
\newcommand{\wmmiss}{\textsf{m$^2$wis + s}}

\newcommand{\mc}{\multicolumn}
\newcommand{\mr}{\multirow}

\begin{abstract}
Computing maximum weight independent sets in graphs is an important NP-hard optimization problem. 
The problem is particularly difficult to solve in large graphs for which data reduction techniques do not work well. 
To be more precise, state-of-the-art branch-and-reduce algorithms can solve many large-scale graphs if reductions are applicable.
Otherwise, their performance quickly degrades due to branching requiring exponential time. 
In this paper, we develop an advanced memetic algorithm to tackle the problem, which incorporates recent data reduction techniques to compute near-optimal weighted independent sets in huge sparse networks.  
More precisely, we use a memetic approach to recursively choose vertices that are likely to be in a large-weight independent set. We include these vertices into the solution, and further reduce the graph.
We show that identifying and removing vertices likely to be in large-weight independent sets opens up the reduction space and speeds up the computation of large-weight independent sets remarkably.
Our experimental evaluation indicates that we are able to outperform state-of-the-art algorithms. 
For example, our two algorithm configurations compute the best results among all competing algorithms for 205 out of 207 instances. Thus can be seen as a useful tool when large-weight independent sets need to be computed in~practice. 

\end{abstract}

        \maketitle

        \section{Introduction}

        \setcounter{page}{1}%
        \pagenumbering{arabic}

        For a given graph $G=(V,E)$ an \textit{independent set} (IS) is defined as a subset $I\subseteq V$ of all vertices such that each pair of vertices in $I$ are non adjacent. A \textit{maximum independent set} (MIS) describes an IS with highest possible cardinality. By transforming the graph~$G$ into the complement graph $\overline{G}$ the MIS problem results in the \textit{maximum clique} problem. However, for sparse graphs $G$, using a maximum clique solver is impractical as the complement $\overline{G}$ is very dense and therefore unlikely to fit in memory for all but the smallest instances. Another related problem is the \textit{minimum vertex cover} problem. Note that for a MIS $\I$ of $G$,  $V\setminus \I$ is a minimum vertex cover. For a weighted graph $G=(V,E,\omega)$ with non-negative vertex weights given by a function $\omega:V \rightarrow \mathbb{R}_{\geq 0}$, the \textit{maximum weight independent set} (MWIS) problem is to find an independent set $\I$ with maximum weight $\omega(\I) = \sum_{v\in \I} \omega(v)$. The applications of the MWIS problem, as well as the related problems addressed above, can be used for solving different application problems such as long-haul vehicle routing~\cite{DBLP:conf/esa/DongGNPRS22}, the winner determination problem~\cite{wu2015solving} or prediction of structural and functional sites in proteins~\cite{mascia2010predicting}. As a detailed example, consider an application of MWIS for map labeling, where displaying non-overlapping labels throughout dynamic map operations such as zooming and rotating~\cite{gemsa2016evaluation} or while tracking a physical movement of a user or set of moving entities~\cite{barth-2016} is of high interest in many applications. In the underlying map labeling problem, the labels are represented by vertices in a graph, weighted by importance. Each pair of vertices is connected by an edge if the two corresponding labels would overlap. %
        In this graph, a MWIS describes a high-quality set of labels, with regard to their importance level, that can be visualized without any overlap. 

        Since these problems are NP-hard~\cite{DBLP:books/fm/GareyJ79}, heuristic
        algorithms are used in practice to efficiently compute solutions of high quality on \emph{large} graphs~\cite{andrade-2012,grosso2008simple,DBLP:conf/www/XiaoHZD21}. 
        Depending on the definition of the neighborhood, local search algorithms are able to explore local solution spaces very effectively. 
        However, local search algorithms are also prone to get stuck in local optima. 
        As with many other heuristics, results can be improved if several repeated runs are made with some measures taken to diversify
        the search. 
        Still, even a large number of repeated executions can only scratch the surface of the huge space of possible independent sets for large-scale data sets. 

        Traditional branch-and-bound methods~\cite{segundo-recoloring,segundo-bitboard-2011,tomita-recoloring,DBLP:conf/alenex/GellnerLSSZ21,DBLP:conf/alenex/Lamm0SWZ19,DBLP:conf/siamcsc/HespeL0S20} may often solve
        small graphs with hundreds to thousands of vertices in practice, and medium-sized instances can be solved exactly in practice using reduction rules to reduce the graph. In particular, it has been observed that if data reductions work very well, then the instance is likely to be solved. If data reductions do not work very well, \ie the size of the reduced graph is large, then the instance can often not be solved.
        Even though new algorithms such as the struction algorithm~\cite{DBLP:conf/alenex/GellnerLSSZ21} already manage to solve a lot of large instances, some remain unsolved.

        In order to explore the global solution space extensively, more sophisticated metaheuristics, such as GRASP~\cite{DBLP:conf/esa/DongGNPRS22} or iterated local search~\cite{andrade-2012,hybrid-ils-2018}, have been used.
        In this work, we extend the set of metaheuristics used for the MWIS problem by introducing a novel memetic algorithm.
        Memetic algorithms (MAs) combine genetic algorithms with local search~\cite{KimHKM11} to effectively explore (global search) and exploit (local search) the solution space.
        The general idea behind genetic algorithms is to use mechanisms inspired by biological evolution such as selection, mutation, recombination, and survival of the fittest. 

        \paragraph*{Our Results.}
        Our contribution is two-fold: First, we develop a state-of-the-art memetic algorithm that is based on recombination operations employing graph partitioning techniques.
        It computes large-weight independent sets by incorporating a wide range of recently developed advanced reduction rules. 
        In particular, our algorithm uses a wide range of frequently used data reduction techniques from \cite{DBLP:conf/alenex/Lamm0SWZ19,DBLP:conf/alenex/GellnerLSSZ21} and also employs a number of recently proposed data reduction rules by Gu~\etal~\cite{gu2021towards}.

        The algorithm may be viewed as performing two functions simultaneously: (1) reduction rules for the weighted independent set problem are used to boost the performance of the memetic algorithm \emph{and} (2) the memetic algorithm opens up the opportunity for further reductions by selecting vertices that are likely to be in large-weight independent sets. In short, our method applies reduction rules to form a reduced graph, then computes vertices to insert into the final solution and removes their neighborhood (including the vertices themselves) from the graph. Thus further reductions can be applied. The process is then repeated recursively until the graph is empty. We show that this technique finds near-optimal weighted independent sets much faster than existing local search algorithms, is competitive with state-of-the-art exact algorithms for smaller graphs, and allows us to compute large-weight independent sets on huge sparse graphs. Overall, our algorithm configurations compute the best results among all competing algorithms for 205 out of 207 instances, and thus can be seen as the dominating tool when large weight independent sets need to be computed in practice. 

        Our second contribution in this work is the experimental evaluation of the orderings in which currently available data reductions are applied.  We examine the impact of different orderings on solution size and on running time. One outcome of this evaluation are robust orderings of reductions for exact reduction rules, as well as a specific ordering which can improve the solution quality further at an expense of computation time.

        \section{Preliminaries}
        In this work, a graph $G=(V,E)$ is an undirected graph with 
        ${n=|V|}$ and ${m = |E|}$, where $V =\{0,...,n-1\}$. The neighborhood $N(v)$ of a vertex $v \in V$ is defined as $N(v) = \{u \in V : (u,v) \in E\}$. 
        Additionally, $N[v]=N(v) \cup v$. The same sets are defined for the neighborhood $N(U)$ of a set of vertices $U \subset V$, \ie $N(U) = \cup_{v \in U} N(v)\setminus U$ 
        and $N[U] = N(U) \cup U$. The degree of a vertex $\mathrm{deg}(v)$ is defined as the number of its neighbors $\mathrm{deg}(v)=|N(v)|$. The complement graph is defined as $\overline{G}=(V,\overline{E})$, where $\overline{E}=\{(u,v): (u,v)\notin E\}$ is the set of edges not present in $G$.
        A set $I\subseteq V$ is called \textit{independent set} (IS) if for all vertices $v,u \in I$ there is no edge $(v,u)\in E$. For a given IS~$\I$ a vertex $v \notin \I$ is called free, if $\I\cup \{v\}$ is still an independent set. An IS is called \textit{maximal} if there are no free vertices. 
    The \textit{maximum independent set problem} (MIS) is that of finding an IS with maximum cardinality.
     The \textit{maximum weight independent set problem} (MWIS) is that of finding an IS with maximum weight. The weight of an independent set $\I$ is defined as $\omega(\I) = \sum_{v \in \I}\omega(v)$ and $\alpha_\omega(G)$ describes the weight of a MWIS of the corresponding graph. The complement of an independent set is a vertex cover, \ie a subset ${C \subseteq V}$ such that every edge $e \in E$ is covered by at least one vertex $v \in C$.
     An edge is \textit{covered} if it is incident to one vertex in the set~$C$. The minimum vertex cover problem, defined as looking for a vertex cover with minimum cardinality, is thereby complementary to the MIS problem. Another closely related concept are cliques. A \textit{clique} is a set $Q \subseteq V$ such that all vertices are pairwise adjacent. A clique in the complement graph $\overline{G}$ corresponds to an independent set in the original graph~$G$. A vertex is called \textit{isolated} or \textit{simplicial}, when its neighborhood forms a clique.

        The subdivision of the set of vertices $V$ into disjoint blocks $V_1$, $\ldots, V_k$ such that $V_1 \cup ...\cup V_k =~V$ is called a $k$-\textit{way partition} (see \cite{DBLP:reference/bdt/0003S19,DBLP:journals/corr/abs-2205-13202}). To ensure the blocks to be roughly of the same size, the balancing constraint $|V_i| \leq L_{max}\coloneqq \left(1+\varepsilon\right) \left\lceil \frac{|V|}{k}\right\rceil$ with the imbalance parameter $\varepsilon > 0$ is introduced. While satisfying this balance constraint, the \textit{edge separator} problem asks for minimizing the total cut, $\sum_{i<j}\omega(E_{ij})$, where $E_{ij}$ is defined by $E_{ij} \coloneqq \left\lbrace \{ u,v\}\in E : u \in V_i,\right. $ $\left. v \in V_j \right\rbrace$. The edge separator is the set of all edges in the cut. For the $k$-vertex separator problem on the other hand we look for a division of $V$ into ${k+1}$ blocks. In addition to the blocks $V_1,...,V_k$ a separator $S$ exists. This separator has to be chosen such that no edges between the blocks $V_1,...,V_k$ exist, but there is no balancing constraint on the separator $S$. However, as for the edge separator problem the balancing constraint on the blocks $|V_i| \leq L_{max}\coloneqq \left(1+\varepsilon\right) \left\lceil \frac{|V|}{k}\right\rceil$ has to hold. To solve the problem, the size of the separator $|S|$ has to be minimized. By removing the separator $S$ from the graph it results in at least $k$ connected components, since the different blocks $V_i$ are not connected.

        \section{Related Work}\label{sec:related_work}

        Here we give only a short overview of existing exact and heuristic techniques; for more details, see the expanded discussion in Appendix~\ref{sec:expanded_related_work} or the recent survey on practical data reduction~\cite{Abu-Khzam2022}.

        \subsection{Exact Methods}
        Exact algorithms usually compute optimal solutions by systematically exploring the space of solutions via variations of \emph{branch-and-} \emph{bound}~\cite{ostergaard2002fast,warren2006combinatorial}. 
        Branching schemes and better pruning methods use upper and lower bounds to exclude specific subtrees~\cite{balas1986finding, babel1994fast,li2017minimization}.  Of note, Warren and Hicks~\cite{warren2006combinatorial} proposed three branch-and-bound algorithms that use weighted clique covers as an upper bound, using a branching scheme first introduced by Balas and Yu~\cite{balas1986finding}.

        Data reduction rules are frequently intermixed with branching and bounding, yielding so-called \emph{branch-and-reduce} algorithms~\cite{akiba-tcs-2016}, which can improve their worst-case (and practical) running time. For the unweighted case, many branch-and-reduce
        algorithms have been developed. 
        In 2019 reduction rules for the MWIS problem were introduced, resulting in the first branch-and-reduce algorithm for MWIS by Lamm~\etal~\cite{DBLP:conf/alenex/Lamm0SWZ19}.
        They first introduce data reductions, together with a branch-and-reduce algorithm using pruning with weighted clique covers~\cite{warren2006combinatorial} for upper bounds and an adapted version of the \textsf{ARW} local search~\cite{andrade-2012} for lower bounds.

        Since this result, many new data reductions and branch-and-reduce algorithms have been introduced. Gellner~\etal~\cite{DBLP:conf/alenex/GellnerLSSZ21} integrated variants of the struction reduction~\cite{ebenegger1984pseudo,alexe2003struction} that may increase the graph size. 
        Recently, Xiao~\etal~\cite{DBLP:conf/www/XiaoHZD21} and Zheng~\etal~\cite{DBLP:conf/icde/ZhengGPY20} presented further data reductions and simple exact algorithms based on these data reduction rules and
        Huang~\etal~\cite{DBLP:conf/cocoon/HuangXC21} proposed a branch-and-reduce algorithm for maximum weight independent set with running time~$\mathcal{O}^*(\npthousandthpartsep{}\numprint{1.1443}^n)$.

        Finally, there are exact procedures which are either based on other extension of the branch-and-bound paradigm~\cite{rebennack2011branch,warrier2005branch,warrier2007branch}, or on the reformulation into other $\mathcal{NP}$-complete problems, such as SAT, for which a variety of solvers already exist~\cite{xu2016new}. 

        We additionally note that there are several recent works on the complementary maximum weighted clique problem that are able to handle large real-world networks~\cite{fang2016exact,jiang2017exact,held2012maximum}.
        However, using these solvers for the MWIS problem require computing complement graphs.
        Since large real-world networks are often very sparse, processing their complements quickly becomes infeasible due to their memory~requirement.
        \subsection{Heuristic Methods}
        A widely used heuristic approach is local search, which usually computes an initial solution and then tries to improve it by simple insertion, removal or swap operations. 
        Although local search generally offers no guarantees for solution quality, in practice local search algorithms find high-quality solutions significantly faster than exact procedures.

        For unweighted graphs, the iterated local search (\textsf{ARW}) by Andrade~\etal~\cite{andrade-2012}, is a very successful heuristic.
        It is based on so-called $(1,2)$-swaps which remove one vertex from the solution and add two new vertices to it, thus improving the current solution by one.
        Their algorithm is able to find (near-)optimal solutions for small to medium-size instances in milliseconds, but struggles on massive instances with millions of vertices and edges.

        The hybrid iterated local search (\hils) by Nogueira~\etal~\cite{hybrid-ils-2018} adapts the \textsf{ARW} algorithm for weighted graphs.
        In addition to weighted $(1,2)$-swaps, it also uses $(\omega,1)$-swaps that add one vertex $v$ into the current solution and exclude its $\omega$ neighbors. 

        Two other local searches, \textsf{DynWVC1} and \textsf{DynWVC2}, for the equivalent minimum weight vertex cover problem are presented by Cai~\etal~\cite{cai-dynwvc}.
        In practice, \textsf{DynWVC1} outperforms previous MWVC heuristics on map labeling instances and large scale networks, and \textsf{DynWVC2} provides further improvements on large scale networks but performs worse on map labeling instances.

        Li~\etal~\cite{li2019numwvc} presented a local search algorithm for the MWVC problem that  applies reduction rules during the construction phase of the initial solution.
        Experiments show that their algorithm outperforms state-of-the-art approaches on graphs of up to millions of vertices and on real-world~instances.

        Recently, a new hybrid method for MWVC was introduced by Langedal~\etal~\cite{langedal2022efficient},  combining elements from exact methods with local search, data reductions and graph neural networks. In their experiments they achieve improvements compared to \textsf{DynWVC2} and {\hils} in both solution quality and running~time.

        Lamm~\etal~\cite{lamm2015graph}~presented an evolutionary approach, \textsf{EvoMIS}, to tackle the MIS problem. %
        \textsf{ReduMIS} by Lamm~\etal~\cite{redumis-2017} combines branch-and-reduce approach with \textsf{EvoMIS}. In their experiments, \textsf{ReduMIS} outperformed the local search \textsf{ARW} as well as the pure evolutionary approach \textsf{EvoMIS}.

        Another reduction based heuristic called {\htwis} was presented recently by Gu~\etal~\cite{gu2021towards}. They repeatedly apply the following steps until reaching an empty graph. First they reduce exhaustively and then choose one vertex by a tie-breaking policy to add to the solution. Now this vertex as well as its neighbors can be removed from the graph and the reductions can be applied again.
        Their experiments prove a significant improvement in running time as well as solution quality compared to state-of-the-art~solvers. 

        Recently, a new metaheuristic was introduced by Dong~\etal~\cite{DBLP:conf/esa/DongGNPRS22} in particular for vehicle routing instances. With their algorithm \textsc{METAMIS} they developed a new local search algorithm combining a wide range of simple local search operations with a new variant of path-relinking to escape local optima. In their experiments they outperform {\hils} algorithm on a wide range of instances both in time and solution~quality.

        \section{Algorithm}

        We now present our memetic algorithm for the MWIS problem, which we call \textit{memetic maximum weight independent set} {\wmmis}. This algorithm is inspired by \textsf{ReduMIS} \cite{redumis-2017} and works in rounds, where each round can be split up into three parts. In the beginning of each round the exact reduction step takes place. Here the graph is reduced as far as possible using a wide range of data reduction rules. On the resulting reduced graph, we apply the memetic part of the algorithm as the second step. We represent a solution, also referred to as an individual, by using bitvectors. Meaning the independent set $\I$ is represented as an array $s \in \{0,1\}^n$. For each array entry it holds $s[v] = 1$ iff $v \in \I$. The memetic component itself works in rounds as well.
        Starting with an initial population $\mathcal{P}$, consisting of a set of individuals, this population is evolved over several rounds until a stopping criterion is fulfilled. 
        In the third part, we select a subset of vertices to be included in the independent set by considering the resulting population. Here, we implement different strategies to select vertices for inclusion. Including these vertices in the independent set enables us to remove them and their neighbors from the instance. This breaks up the reduction space, \ie further reductions might be applicable after the removal process.
         The steps of exact reduction, memetic search, and heuristic reduction are repeated until the remaining graph is empty or another stopping criterion is fulfilled.

        \begin{algorithm}[t]
                \begin{algorithmic}
                \small
                        \STATE   \textbf{input} graph $G=(V,E)$		
                        \STATE   \textbf{procedure} {\wmmis}($G$)
                        \STATE   \quad $\mathcal{W} = \emptyset$ 	\quad		 // best solution
                        \STATE   \quad \textbf{while} $G$ not empty and time limit not reached
                        \STATE   \qquad $(G,\mathcal{W}) \leftarrow$ \textsc{ExactReduce}$(G, \mathcal{W})$
                        \STATE   \qquad \textbf{if} $G$ is empty \textbf{then} \textbf{return} $\mathcal{W}$
                        \STATE   \qquad create initial population $\mathcal{P}$
                        \STATE   \qquad $\mathcal{P} \leftarrow$ \textsc{Evolve}$(G,\mathcal{P})$
                        \STATE   \qquad $(G,\mathcal{W}) \leftarrow$ \textsc{HeuristicReduce}$(G,\mathcal{P},\mathcal{W})$
                        \STATE   \textbf{return} $\mathcal{W}$
                \end{algorithmic}
                \caption{High Level Structure of {\wmmis}}\label{algo: overview}
                \vspace*{-.1cm}
        \end{algorithm}

        Following the order of the Algorithm \ref{algo: overview}, we first describe the \textsc{ExactReduce} routine in Section \ref{sec: reduction}.
        Section \ref{sec: memetic} is devoted to the memetic part, followed by the description of different vertex selection strategies used to heuristically reduce the instance and open up the reduction space in Section \ref{sec: vertex_selection}.

        \subsection{Exact Reductions}\label{sec: reduction}

        Especially for large instances, applying exact data reductions is a very important technique to reduce the problem size. In general, reductions allow the identification of vertices (1) as part of a solution to the MWIS problem, (2) as non-solution vertices or (3) as deferred, meaning the decision for this vertex is depending on additional information about neighboring vertices that will be obtained later. The resulting reduced graph, after no reduction rule can be applied anymore we denote by~$\mathcal{K}$. Once a solution on $\mathcal{K}$ is found, reductions can be undone to reconstruct a MWIS on the original graph.
        For the reduction process we apply a large set of reductions which we list here.
        \begin{figure*}
                \begin{minipage}{0.329\textwidth}
                        \centering
                        Case 1:\\
                        \includegraphics[width=0.75\textwidth]{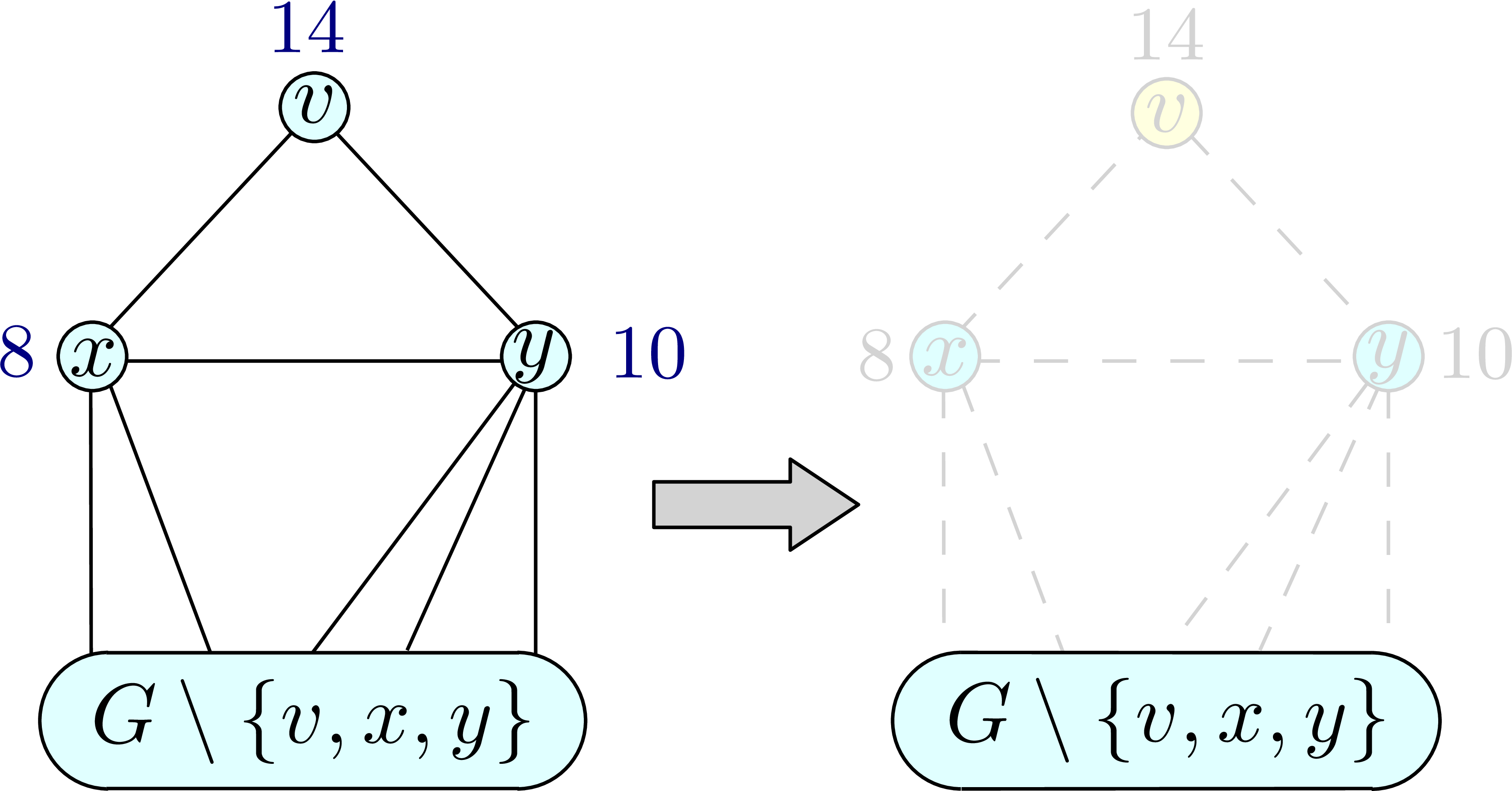}
                \end{minipage}
                \begin{minipage}{0.329\textwidth}
                        \centering
                        Case 2:\\
                        \includegraphics[width=0.75\textwidth]{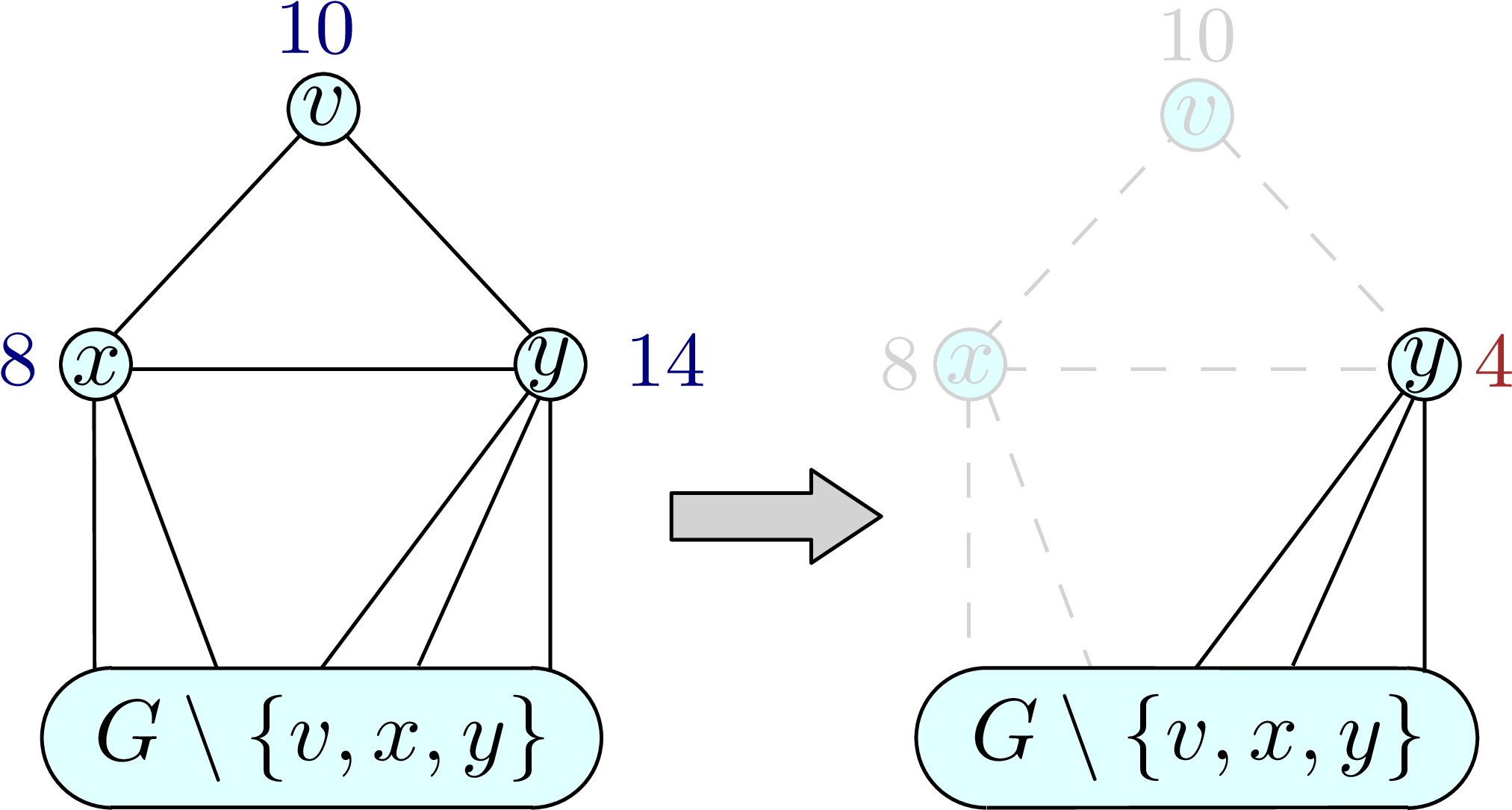}
                \end{minipage}
                \begin{minipage}{0.329\textwidth}
                        \centering
                        Case 3:\\
                        \includegraphics[width=0.75\textwidth]{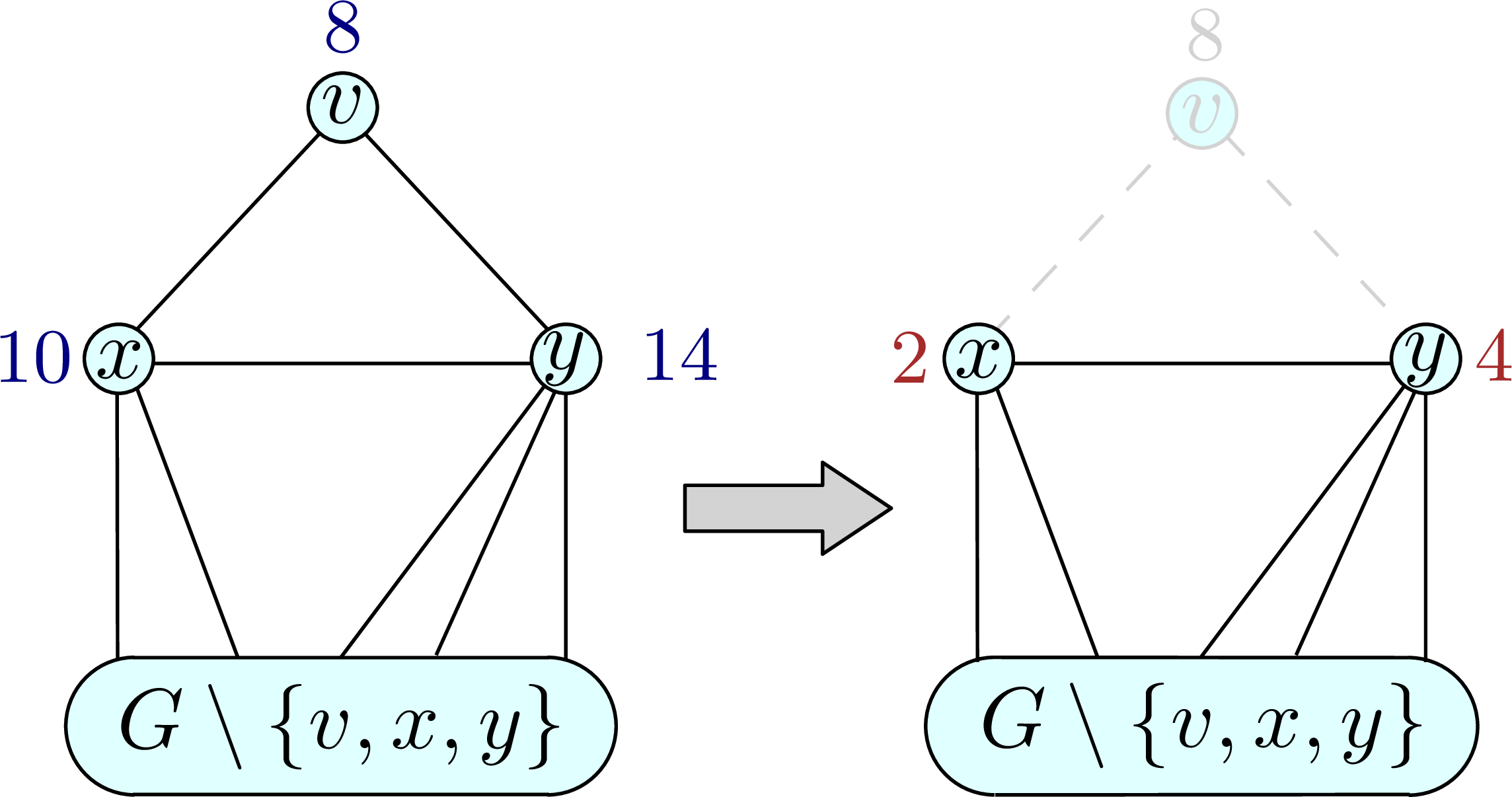}
                \end{minipage}
                \caption{Illustration of the three cases of the Triangle Reduction} \label{fig: triangle_reduction}
        \end{figure*}
        In the following list of reductions $\I$ refers to a MWIS of $G$, $\I'$ refers to a MWIS of the modified graph $G'$.

\begin{reduction}[Neighborhood Removal \cite{DBLP:conf/alenex/Lamm0SWZ19}] \label{red: neighborhood removal}
	For any $v\in V$, if $\omega(v)\geq \omega(N(v))$ then $v$ is in some MWIS of $G$. Let $G' = G[V\setminus N[v]]$ and $\alpha_\omega(G) = \alpha_\omega(G') + \omega(v)$.
\end{reduction}

\begin{reduction}[Degree-One \cite{gu2021towards}]\label{red: deg1}
	Let $v$ be a degree-one vertex with the neighbor $u$ in $G$.
	\begin{itemize}
		\item Case 1: if $\omega(v) \geq \omega(u)$, $v$ must be contained in some MWIS of $G$; thus $v$ can be removed from $G$, \ie $\alpha_\omega(G) = \alpha_\omega(G') + \omega(v)$, where $G'$ is the graph obtained by removing both $v$ and $u$.
		\item Case 2: if $\omega(v) < \omega(u)$, $\alpha_\omega(G) = \alpha_\omega(G') + \omega(v)$, where $G'$ is the graph obtained by removing $v$ and updating the weight of $u$ to be $\omega(u) = \omega(u) - \omega(v)$. It holds that $u \in \I $ iff $u \in \I'$.
	\end{itemize}

We note that Case 1 is a special case of Reduction~\ref{red: neighborhood removal}.
\end{reduction}

\begin{reduction}[Triangle \cite{gu2021towards}]\label{red: triangle}
	This reduction is illustrated in Figure~\ref{fig: triangle_reduction}. 
	Let $v$ be a degree-two vertex with two neighbors $x$ and $y$ in $G$, where edge ${\{x,y\} \in E}$. Without loss of generality, assume ${\omega(x) \leq \omega(y)}$.
	\begin{itemize}
            \item Case 1: if $\omega(v) \geq \omega(y)$, $v$ must be contained in some MWIS of $G$. This leads to $\alpha_\omega(G) = \alpha_\omega(G') + \omega(v)$, where $G'$ is obtained by removing $v$, $x$, $y$.
		\item Case 2: if $\omega(x) \leq \omega(v) < \omega(y)$, $\alpha_\omega(G) = \alpha_\omega(G')+ \omega(v)$, where $G'$ is the graph obtained by removing nodes $v$ and $x$, and updating $\omega(y)  = \omega(y) - \omega(v)$. It holds that $y \in \I$ iff $y \in \I'$.
		\item Case 3: if $\omega(v) < \omega(x)$, $\alpha_\omega(G) = \alpha_\omega(G')+ \omega(v)$, where $G'$ is the graph obtained by removing $v$, and updating $\omega(x)  = \omega(x) - \omega(v)$ as well as $\omega(y)  = \omega(y) - \omega(v)$. It holds for $z\in\{x,y\}$ that $z \in \I$ iff $z \in \I'$.
	\end{itemize}

\end{reduction}

\begin{reduction} [Extended V-Shape \cite{DBLP:conf/alenex/Lamm0SWZ19,gu2021towards}]\label{red: extended_v_shape}
	Let $v$ be a degree-two vertex with the neighbors $x$ and $y$ in $G$, where edge $\{x,y\} \notin E$. Without loss of generality, assume ${\omega(x) \leq \omega(y)}$.
	\begin{itemize}
		\item Case 1: \cite{DBLP:conf/alenex/Lamm0SWZ19} if $\omega(v) \geq \omega(y)$ 
		\begin{itemize} 
			\item if $\omega(v) \geq \omega(x) + \omega(y)$, $v$ must be contained in some MWIS of $G$ and this leads to $\alpha_\omega(G) = \alpha_\omega(G') + \omega(v)$, where $G'$ is obtained by removing $v, x, y$. 
			\item else we fold $v,x,y$ into a vertex $v'$ with weight $\omega(v') = \omega(x) + \omega(y) - \omega(v)$ forming a new graph $G'$. Then $\alpha_\omega(G) = \alpha_\omega(G') + \omega(v)$. If $v'\in \I'$ then $\{x,y\}\subset \I$, otherwise $v \in \I$.
			It holds that $\{x,y\} \subset \I$ iff $v' \in \I'$;\\
			
		\end{itemize}
		\item Case 2: \cite{gu2021towards} if $\omega(x) \leq \omega(v) < \omega(y)$, $\alpha_\omega(G) = \alpha_\omega(G') + \omega(v)$, where $G'$ is the graph obtained by removing $v$, updating $N(x) = N(x) \cup N(y)$ and $\omega(y) = \omega(y) - \omega(v)$. It holds that $\forall w \in \{x,y\}$, $w \in \I$ iff $w \in \I'$;\\
		\item Case 3: \cite{gu2021towards} if $\omega(x) > \omega(v)$, $\alpha_\omega(G) = \alpha_\omega(G') + \omega(v)$, where $G'$ is the graph obtained by updating $\omega(x) = \omega(x) - \omega(v)$, $\omega(y) = \omega(y) - \omega(v)$, and $N(v) = N(x) \cup N(y)$. It holds that $\{x,y\}\subseteq \I$ iff $\{x,y\}\subseteq \I'$.
	\end{itemize}
\end{reduction}

\begin{reduction}[Isolated Vertex Removal \cite{DBLP:conf/alenex/Lamm0SWZ19}] \label{red: clique} %
	Let $v\in V$ be isolated and $\omega(v)\geq \max_{u\in N(v)}\omega(u)$. Then $v$ is in some MWIS of $G$. Let $G' = G[V\setminus N[v]]$ and $\alpha_\omega(G) = \alpha_\omega(G') + \omega(v)$.
\end{reduction}

\begin{reduction}[Basic Single-Edge \cite{gu2021towards}]\label{red: bse}
	Given an edge $\{u,v\}$ $\in E_G$, if $\omega(v) + \omega(N(u) \setminus N(v)) \leq\omega(u)$, it holds that $\alpha_\omega(G) = \alpha_\omega(G')$, where $G'$ is obtained by removing $v$ from $G$.
\end{reduction}

\begin{reduction}[Extended Single-Edge \cite{gu2021towards}]\label{red: ese}
	For an edge $\{u,v\}$ $\in$ $E_G$ with $\omega(v) \geq \omega(N(v))-\omega(u)$, it holds that $\alpha_\omega(G) = \alpha_\omega(G')$, where $G'$ is obtained by removing all vertices in $N(u) \cap N(v)$.
\end{reduction}

\begin{reduction}[Domination \cite{DBLP:conf/alenex/Lamm0SWZ19}]\label{red: dom}
	Let $u,v\in V$ be vertices such that $N[u]\supseteq N[v]$ (i.e., $u$ dominates $v$). If $\omega(u)\leq \omega(v)$, there is an MWIS in $G$ that excludes $u$ and $\alpha_\omega(G) = \alpha_\omega(G[V\setminus \{u\}])$. Therefore, $u$ can be removed from the graph.
\end{reduction}

\begin{reduction}[Twin \cite{DBLP:conf/alenex/Lamm0SWZ19}]\label{red: twin}
	Let vertices $u$ and $v$ have equal neighborhoods $N(u) = N(v) = \{p,q,r\}$, forming an independent set.
	We have two cases:
	\begin{enumerate}
		\item If $\omega(\{u,v\}) \geq \omega(\{p,q,r\})$, then $u$ and $v$ are in some MWIS of $G$. Let $G' = G[V\setminus N[\{u,v\}]]$.
		\item If $\omega(\{u,v\}) < \omega(\{p,q,r\})$, but $\omega(\{u,v\}) > \omega(\{p,q,r\}) - \min_{x\in\{p,q,r\}} \omega(x)$, then we can fold $u, v, p, q, r$ into a new vertex $v'$ with weight $\omega(v') = \omega(\{p,q,r\}) - \omega(\{u,v\})$ and call this graph $G'$. Then we construct an MWIS $\I$ of $G$ as follows: if $v'\in \I'$ then $\I = (\I'\setminus \{v'\})\cup \{p,q,r\}$, if $v'\notin \I'$ then $\I = \I' \cup \{u,v\}$.
	\end{enumerate}
	Furthermore, $\alpha_\omega(G) = \alpha_\omega(G') + \omega(\{u,v\})$.
\end{reduction}

\begin{reduction}[Simplicial Weight Transfer \cite{DBLP:conf/alenex/Lamm0SWZ19}] \label{red: clique_neigh}%
	Let $v\in V$ be isolated, and suppose that the set of simplicial vertices $S(v)\subseteq N(v)$ is such that $\forall u\in S(v)$, $\omega(v) \geq \omega(u)$. We
	\begin{enumerate}
		\item remove all $u\in N(v)$ such that $\omega(u)\leq \omega(v)$, and let the remaining neighbors be denoted by $N'(v)$,
		\item remove $v$ and $\forall x\in N'(v)$ set its new weight to $\omega'(x) = \omega(x) - \omega(v)$, and
	\end{enumerate}
	let the resulting graph be denoted by $G'$. Then $\alpha_\omega(G) = \omega(v) + \alpha_\omega(G')$ and an MWIS $\I$ of $G$ can be constructed from an MWIS $\I'$ of $G'$ as follows: if $\I' \cap N'(v) = \emptyset$ then $\I = \I'\cup\{v\}$, otherwise $\I = \I'$.
\end{reduction}

\begin{reduction}[CWIS \cite{butenko-trukhanov}]\label{red: CWIS}
	Let $I_c\subseteq V$ be a critical weighted IS of $G$, \ie $\omega(\I_c) - \omega(N(\I_c)) = \max\{\omega(\I)-$ $\omega(N(\I)) : \I$ is an IS of $G\}$. Then $\I_c$ is in some MWIS of $G$. We set $G' = G[V\setminus N[\I_c]]$ and $\alpha_\omega(G) = \alpha_\omega(G') + \omega(\I_c)$.
\end{reduction}

\begin{reduction}[Neighborhood Folding \cite{DBLP:conf/alenex/Lamm0SWZ19}] \label{red: neighborhood folding}
Let $v \in V$, and suppose that $N(v)$ is independent. If $\omega(N(v)) > \omega(v)$, but $\omega(N(v)) - \min_{u\in N(v)}\{\omega(u)\} < \omega(v)$, then fold $v$ and $N(v)$ into a new vertex $v'$ with weight $\omega(v') = \omega(N(v)) - \omega(v)$. If $v'\in \I'$ then $\I = (\I'\setminus\{v'\}) \cup N(v)$, otherwise if $v\in \I'$ then $\I = \I' \cup \{v\}$. Furthermore, $\alpha_\omega(G) = \alpha_\omega(G') + \omega(v)$.
\end{reduction}

        In the \textsc{ExactReduce} routine, we test for each of these reductions whether they are applicable. This takes place in a predefined order. If one reduction is successfully applied, then the process of testing possible reductions starts from the beginning (according~to~this~order).
        If no more reductions can be applied, we obtained the reduced graph and continue with the next part of Algorithm~\ref{algo: overview}.

        The order in which reductions are applied has an effect on the weight offset and the size of the resulting reduced graph, as well as on the time needed for the computation. We give a detailed analysis in Appendix~\ref{subsec:ordering}.

        \subsection{Memetic Algorithm} \label{sec: memetic}

        \begin{algorithm}[t]
                \begin{algorithmic}
                \small
                        \STATE \textbf{input} graph $G=(V,E)$, current population $\mathcal{P}$ 
                        
                        \STATE \textbf{procedure} \textsc{Evolve}$(G,\mathcal{P})$
                        \STATE \quad \textbf{while} stopping criterion not fulfilled
                        \STATE \qquad randomly chose a combine operation \textsc{combine}
                        \STATE \qquad $k=$ number of individuals needed for \textsc{combine}
                        \STATE \qquad $\mathcal{IS} \leftarrow \emptyset$  // set of individuals
                        \STATE \qquad $\mathcal{IS} \leftarrow$ \textsc{tournamentSelect}($\mathcal{P}$) 
                        \STATE \qquad $\mathcal{OS} \leftarrow \emptyset$  // set of offspring
                        \STATE \qquad $\mathcal{OS} \leftarrow$ \textsc{combine}($\mathcal{IS}$)  
                        \STATE \qquad \textbf{if} mutate with probability 10\%
                        \STATE \qquad \quad $\mathcal{OS} \leftarrow$ \textsc{mutate}($\{\mathcal{OS}\}$)
                        \STATE \qquad \textbf{if} suitable replacement (different criteria)
                        \STATE \qquad \quad $\mathcal{P} \leftarrow$ \textsc{replace}($\mathcal{P}, \mathcal{O}$) 
                        \STATE \textbf{return} $\mathcal{P}$
                \end{algorithmic}
                
                \caption{High Level Structure of \textsc{Evolve}$(G,\mathcal{P})$}\label{algo: evolve}
        \end{algorithm}
        After \textsc{ExactReduce}, we apply the \textsc{Evolve} routine which is described in Algorithm \ref{algo: evolve} on the reduced graph $\mathcal{K}$. This starts by generating an initial \textit{population} of size $|\mathcal{P}|$ which we then evolve over several generational cycles~(rounds). For the evolution of the population two \textit{individuals} from the population are selected and combined to create an \textit{offspring}. 
        We also apply a mutation operation to this new solution by forcing new vertices into the solution and removing neighboring solution vertices.
        To keep the population size constant and still add a new offspring to the solution, we look for fit replacements. In this process, we search for individuals in the population, which have smaller weights than the new offspring. Among those, we look for the most similar solution by computing the intersection size of the new and existing individuals. We also added the possibility of forcing individuals into the population if it has not changed over a certain number of iterations, as well as rejecting the offspring if the solution with the smallest weight is still better than the new offspring. Note that the size $|\mathcal{P}|$ of the population does not change during this process.
        Additionally, at any time the individuals of our population are forming an independent set. 
        In the last step of the memetic algorithm we improve the solution by the {\hils} algorithm.
        The stopping criterion for the memetic procedure is either a specified number of unsuccessful combine operations or a time limit.
        In the following we discuss each of these steps in detail. We start with introducing the computation of the initial solution in Section~\ref{sec: initial solutions} and then explain the combine operations for the evolutionary process in Section~\ref{sec: combine} as well as the mutation operation in Section~\ref{sec: mutation}.

        \subsubsection{Initial Solutions}\label{sec: initial solutions}

        At the start of our memetic algorithm, we create an initial population of size $|\mathcal{P}|$. 
        To diversify as much as possible, this population contains solutions computed in six different ways, which we choose uniformly at random to create an individual. 
        Before applying the strategies we permute the order of the nodes such that different solutions are obtained for the same strategy by different tie breaking.

        \textsf{\emph{RandomMWIS}.} The first approach works by starting with an empty solution and adding free vertices uniformly at random until the solution is maximal. %

        \textsf{\emph{GreedyWeightMWIS}}.  For the \textsf{\emph{GreedyWeightMWIS}} strategy, we start with an empty solution. This is extended to a maximal independent set by adding free vertices ordered by their weight. Starting with the largest weight, we include this vertex and exclude all its neighbors until all vertices are labeled either included~or~excluded.

        \textsf{\emph{GreedyDegreeMWIS}}.  Via this greedy approach, we create initial solutions by successively choosing the next free vertex with the smallest residual degree. Each time a vertex is included, we label the neighboring vertices to be excluded.  

        \textsf{\emph{GreedyWeightVC}}.  In contrast to the previous approaches, here the vertex cover problem, the complementary problem to the independent set problem, is utilized. Therefore, an empty solution is extended by vertices of the smallest weight until a vertex cover is computed. As soon as the algorithm terminated, we compute the complement and have an initial solution to the MWIS problem.

        \textsf{\emph{GreedyDegreeVC}}. As in \textsf{\emph{GreedyWeightVC}} the complementary vertex cover problem is solved. However, for this approach, we choose those vertices to include in the solution, which cover the maximum number of currently uncovered edges.

        \textsf{\emph{CyclicFast}}.  We also add the possibility to compute an initial solution via the \textsc{CyclicFast} algorithm by Gellner~\etal~\cite{DBLP:conf/alenex/GellnerLSSZ21}. We set a time limit of \numprint{60}~seconds.

        \subsubsection{Combine Operations} \label{sec: combine}
        The common idea of our combine operations is to combine whole blocks of independent set vertices. To construct those blocks, we use the graph partitioning framework KaHIP \cite{kaHIPHomePage} which computes partitions of the graph $V=V_1\cup ... \cup V_n$. For $j=1,...,n$ the solution blocks $\I_j$ are defined by $\I_j = \I\cap V_j$.
        We created different offspring by using the following combine operations on those solution~blocks.
        \begin{center}
        \begin{figure}
                \includegraphics[width=0.45\textwidth]{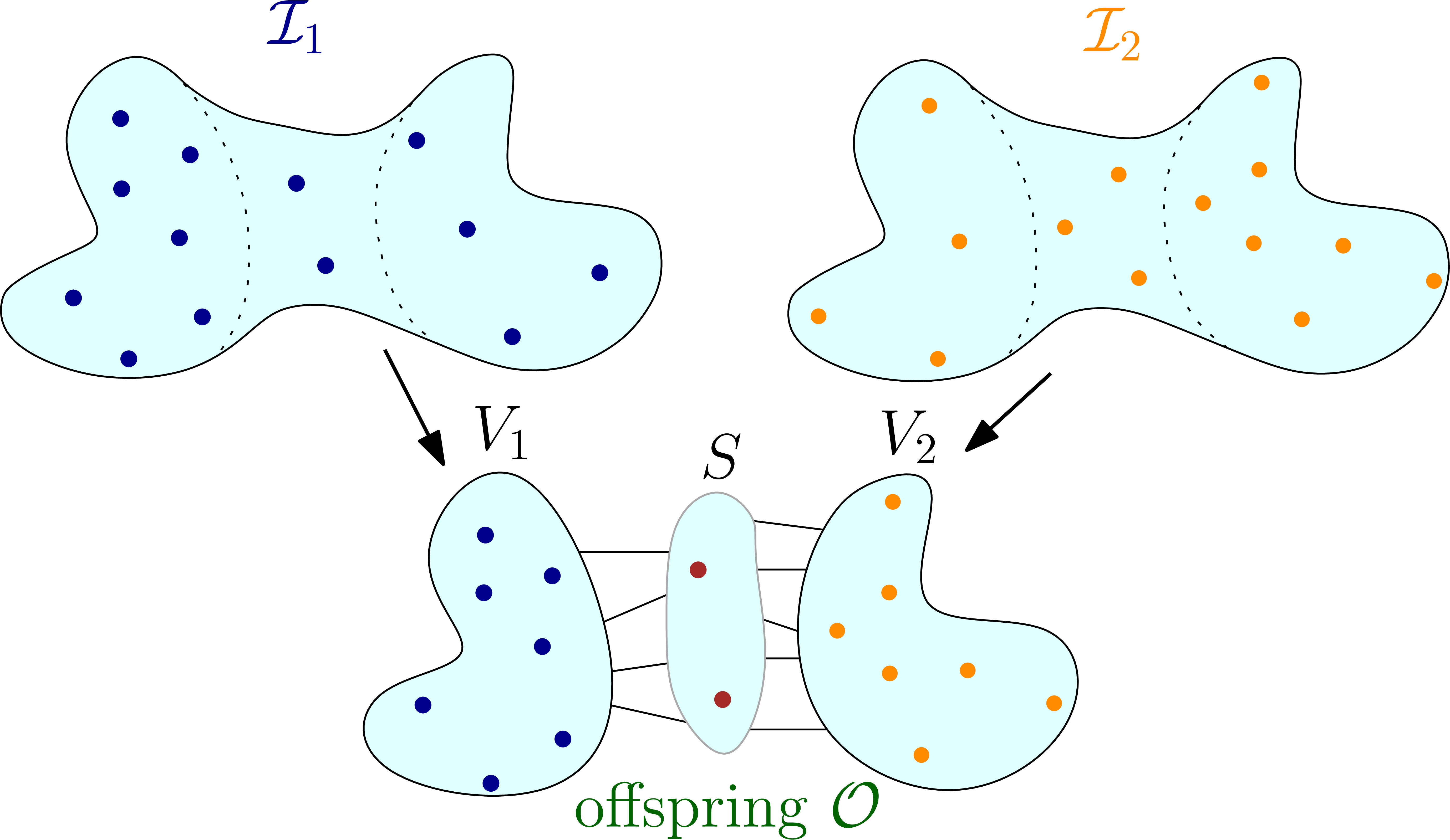}
                \caption{The vertex separator combine operation to create an offspring $\mathcal{O}$ out of two individuals $\I_1$ and $\I_2$. }\label{fig: node_sep}
        \end{figure}
        \end{center}
        The parents for the first two combine operations are chosen by two runs of the tournament selection~\cite{Miller95geneticalgorithms}, where the fittest individual \ie the solution with highest weight gets selected out of two random individuals from the population. Then we perform one of the combine operations outlined below and finally, after the combine operation, we use the {\hils} algorithm \cite{hybrid-ils-2018} to improve the computed offspring.

        \paragraph{Vertex Separator Combination.}
		
        The first operator works with a vertex separator $V = V_1 \cup V_2 \cup S$.
        We use a vertex separator to be able to exchange whole blocks of solutions without violating the independent set property. This can be done because no vertices belonging to different blocks are adjacent to one another. Neighboring vertices would either be part of the same block or one of them has to belong to the separator $S$. By this property the combination of those blocks will always result in a valid solution to the independent set problem. The two individuals selected by the tournament $\I_1$ and $\I_2$ are split up according to these partitions and then combined to generate two offspring $O_1 = \left( V_1 \cap \I_1 \right) \cup \left(V_2 \cap \I_2\right)$ and $O_2 = \left( V_1 \cap \I_2 \right) \cup \left(V_2 \cap \I_1\right)$. After that we add as many free vertices greedily by weight until the solution is maximal, we get a local optimum via one iteration of the weighted local search. See Figure~\ref{fig: node_sep} for an illustration.

        \paragraph{Multi-way Vertex Separator Combination.}

        We extended the previous described operator to the multi-way vertex separator, where multiple solutions can be used and combined. Therefore, we compute a $k$-vertex separator $V = V_1 \cup ... \cup V_k \cup S$ and select $k$ individuals. 
        Then for every pair of partition $V_i$ and individual $\I_j$ for $i,j \in \{1,...,k\}$ a score is computed. 
        This score is defined by $\sum_{v\in V_i\cap I_j} \omega(v)$.
        We start with the pair resulting in the highest score pair and then select pairs decreasingly. Once an individual or partition block is selected, we do not use it again. In contrast to the previous operator, this combination only results in one offspring.We then maximize this offspring and compute a local~maximum.

        \paragraph{Edge Separator Combination.}

        For this operator we exploit the duality to the weighted vertex cover problem. Starting with a partition $V= V_1 \cup V_2$ the operator computes temporary offspring for the weighted vertex cover problem. Let $\I_1$ and $\I_2$ be the individuals selected by the tournament rule. Let $C_i = V\setminus \I_i$ be the solution to the weighted vertex cover problem for $i \in \{1,2\}$. The new offspring are $O_1 = \left( V_1 \cap C_1 \right) \cup \left(V_2 \cap C_2\right)$ and $O_2 = \left( V_1 \cap C_2 \right) \cup \left(V_2 \cap C_1\right)$. 
        However, these offspring can contain some non-covered edges, which are a subset from the cut edges between the two partitions. The graph induced by the non-covered cut edges is bipartite. In this graph we compute a weighted vertex cover using maximum flows.

        \paragraph{Multi-way Edge Separator Combination.}

        Similar to the vertex separator also the edge separator can be extended to use multiple solutions. Therefore, a $k$-way-partition $V=V_1 \cup ...\cup V_k$ is computed. Equivalent to the multi-way vertex separator, we also select $k$ individuals and compute a score for each pair $V_j$ and~$\I_i$. For the scoring function, the complement of an independent set inside the given block is used to sum up the weights of the vertices of the vertex cover in this block. For the offspring computation, each block is combined with the individual with the lowest score. As in the basic edge separator combine operator there can be edges in the cut that are not covered. Since the induced graph here is not bipartite we handle this problem using a simple greedy strategy. Afterwards the solution is transformed to get the offspring for the independent set~individuals.

        \subsubsection{Mutation Operation}\label{sec: mutation}

        After each combine operation, a mutation operator can perturb the created offspring. This is done by forcing new vertices into the solution and removing the adjacent vertices to satisfy the independent set property. Those vertices are selected at random among all non solution nodes in the graph. Afterwards we improve the solution using the {\hils} algorithm.

        \subsection{Heuristic Reductions and Recursion}\label{sec: vertex_selection}

        After the memetic algorithm stops, we use a heuristic data reduction to open up the reduction space (and afterwards the next round of exact data reductions begins). We implemented different strategies to select vertices that we put into the solution. In each strategy, vertices are ordered by a rating function. Depending on the configuration, the algorithm either inserts the best vertex or a set of vertices into the solution. We now explain the different selection~strategies.

        \paragraph{Vertex Selection by Weight.}

        The first rating function is based on the weight $\omega(v)$ of a vertex $v$ (higher is better). The intuition here is that by adding a vertex, we want to increase the weight of our solution as much as possible. More precisely, the fittest individual from the population evolved by the memetic algorithm is selected. The fitness of an individual is defined as the solution weight. From this individual, we select the $x$ vertices from the independent set that have the highest weight and add them to our solution.
        Since we only consider vertices from one individual, $x$ can be freely chosen without violating the independent set property of our solution. 
        For example, we can choose to only add the highest weight vertex or select a fraction of those solution vertices.

        \paragraph{Vertex Selection by Degree.}

        Similar to the previous vertex selection strategy, we choose the fittest individual from which we add vertices to our solution. 
        Here, the vertices are rated by their degree $\mathrm{deg}(v)$ (smaller is better). 
        The intuition here is that adding vertices with a small degree to our solution will not remove too many other vertices from the graph that could be considered later. 

        \paragraph{Vertex Selection by Weight/Degree.}

        For this selection strategy, we rate the vertices $v$ of the fittest individual by the fraction $\frac{\omega(v)}{\mathrm{deg}(v)}$ (higher is better). This way we combine the two previous ratings. 

        \paragraph{Hybrid Vertex Selection.} 

        In the hybrid case, the solution vertices $v \in V$ are rated by the weight difference between a vertex and its neighbors $\omega(v) - \sum_{u \in N(v)} \omega(u)$ (higher is better). This value describes the minimum gain in solution weight we can achieve by adding the vertex $v$ to the solution. Note that Gu~\etal~\cite{gu2021towards} proposed this rule for their algorithm. The key difference here is that Gu~\etal~\cite{gu2021towards} use this function on all vertices, while our algorithm only considers solution vertices of the fittest solution of the memetic algorithm.

        \paragraph{Vertex Selection by Solution Participation.}

        In contrast to the previous strategies, this strategy considers the whole population. Moreover, here we consider \textit{each vertex} in the graph. We check the population and assign each vertex a value according to the number of times it is part of a solution. The maximum number a vertex can achieve is therefore bounded by the population size $|\mathcal{P}|$. Since many vertices achieved the same score, we subtracted $\omega(v)^{-1}$ for each vertex $v$ from the solution participation value for tie-breaking. Note that in this strategy we only add one vertex to the solution.

        \section{Experimental Evaluation}

        \paragraph{Methodology.}
        We implemented our algorithm using C++11. The code is compiled using g++ version 12.2 and full optimizations turned on (-O3).
        We compare our algorithm against the struction algorithm by Gellner~\etal~\cite{DBLP:conf/alenex/GellnerLSSZ21} and the (more recent) algorithm {\htwis} by Gu~\etal~\cite{gu2021towards}. We also compare the results with the branch-and-reduce algorithm by Lamm~\etal~\cite{DBLP:conf/alenex/Lamm0SWZ19}, as well as the {\hils} algorithm by Nogueira~\etal~\cite{hybrid-ils-2018}. 
        In most cases {\hils} outperforms \textsf{DynWVC1} and \textsf{DynWVC2} \cite{DBLP:conf/alenex/Lamm0SWZ19}. Hence, we omit comparisons to \textsf{DynWVC1} and \textsf{DynWVC2}.
        We generally run each configuration with four different seeds and a time limit of ten hours and report the mean results.
If a solver exceeded a memory threshold of \numprint{100} GB during a time limit of 10h for an instance we note this with a dash.
        In general, our algorithm does not test the time limit in the \textsc{ExactReduce} routine of {\wmmis} or during the calculation of the separator and partition pool. Hence, if the 10h mark is reached during these steps, the time limit can be exceeded.
        We used a machine equipped with a AMD EPYC 7702P (64 cores) processor  and 1 TB RAM running Ubuntu 20.04.1. %
        We used the fast configuration of the KaHIP graph partitioning package \cite{kaffpa,DBLP:conf/wea/Sanders016} for the computation of the graph partitions and vertex separators. 
        We also present extensive experiments regarding the impact of reduction ordering. Due to space constraints we give the results in the Appendix~\ref{subsec:ordering}. 
We conclude that the order in which we introduce the reductions from Reduction~\ref{red: neighborhood removal} to Reduction~\ref{red: neighborhood folding} is already robust and hence use it for the remaining experiments.

        \paragraph{Parameter Configuration.}
        We set up experiments to find good parameters for our algorithm. Here, we only show detailed analysis regarding the vertex selection strategy.
        In these previous experiments we found that increasing the amount of vertices added during \textsc{HeuristicReduce} reduces the total running time. 
    However, since the solution quality decreases as well, we focus on only adding one vertex during \textsc{HeuristicReduce} in all following experiments for {\wmmis}. 
        Similar to Lamm~\etal~\cite{redumis-2017}, we set the population size $|\mathcal{P}|$ to 250, the size of the partition and separator pool to 10 and the mutation rate to 10\%. 
        Local search is limited to \numprint{15000} iterations. 
        Finally, for the multi-way combine operations, we bound the number of blocks used by 64.

        \paragraph{Data Sets.}
        The set of instances for the experiments is built with graphs from different sources. We use all the instances used by Gellner~\etal~\cite{DBLP:conf/alenex/GellnerLSSZ21} and Gu~\etal~\cite{gu2021towards}. Our set consists of large social networks
        from the Stanford Large Network Dataset Repository (snap) \cite{snapnets}. Additionally, we added real-world graphs from OpenStreetMaps (osm) \cite{OSMWEB,barth-2016,cai-dynwvc}.
        Furthermore, as in Gu~\etal~\cite{gu2021towards} we took the same~6 graphs from the SuiteSparse Matrix Collection (ssmc) \cite{ssmcWEB,davis2011university} where weights correspond to population data. Each weight was increased by one, to avoid a large number of nodes assigned with zero weight.
        Additionally, we used instances from dual graphs of well-known triangle meshes (mesh) \cite{sander2008efficient}, as well as 3d meshes derived from simulations using the finite element method (fe) \cite{soper2004combined}. For unweighted graphs, we assigned each vertex a random weight that is uniformly distributed in the interval [1, 200].
        We also tested our algorithms on the kernels of osm instances as used by Dong \etal~\cite{DBLP:conf/esa/DongGNPRS22}. We do not compare our algorithm on the VR instances contained therein, as data reductions do not work on those instances~\cite{DBLP:conf/esa/DongGNPRS22} and the kernels are too large to be sufficiently explored by our memetic algorithm. We list all graphs in Table~\ref{tab: graphs}.

\begin{table}
	\caption{Vertex selection strategies comparing the number of best solutions computed and the geometric mean time.} \label{tab: compact_vertex_selection}
	\centering
	\small 	\begin{tabular}{r|r|r} 
\hline
{selection strategy} & {\# best} & {mean time}\\
\hline
{hybrid} & {26} & \numprint{10148.31} \\
 {weight}  & {26} & \numprint{8644.56} \\
 {degree}  &  {24} & \numprint{9852.82} \\
 {weight/degree}  &{20} & \numprint{6900.7} \\
 {sol. participation} &   {19}  & \numprint{5315.29} \\ 
\hline
\end{tabular} 

	\vspace*{-.5cm}
\end{table}
        \begin{table*}[t]
            \caption{Average solution weight $\omega$ and time $t$ (in seconds) required to compute it for a representative sample of our instances. The best solutions among all algorithms are marked \textbf{Bold}. Rows are colored \noindent\colorbox{lightergray} {\parbox{\widthof{gray}}{gray}} if branch reduce or struction are optimal.}\label{tab: compact_algos}
	\centering
	\setlength{\tabcolsep}{0.9ex}
		\small \begin{tabular}{lcccccccccccc} 
	\hline 
	\mc{1}{l|}{graphs} & \mc{1}{c}{$t$} & \mc{1}{c|}{$w$} & \mc{1}{c}{$t$} & \mc{1}{c|}{$w$} & \mc{1}{c}{$t$} & \mc{1}{c|}{$w$} & \mc{1}{c}{$t$} & \mc{1}{c|}{$w$} & \mc{1}{c}{$t$} & \mc{1}{c|}{$w$} & \mc{1}{c}{$t$} & \mc{1}{c}{$w$}\\ 
	\hline 
	\mc{1}{l|}{fe} & \mc{2}{c|}{branch reduce}  & \mc{2}{c|}{{\hils}}  & \mc{2}{c|}{{\htwis}}  & \mc{2}{c|}{{\wmmis}}  & \mc{2}{c|}{{\wmmiss}}  & \mc{2}{c}{struction} \\ 
	\hline 
	\mc{1}{l|}{\textit{body}} & \mc{2}{c|}{-} & \mc{1}{r}{\numprint{1259.67}} & \mc{1}{r|}{\numprint{1678510}} & \mc{1}{r}{\numprint{0.04}} & \mc{1}{r|}{\numprint{1645650}} & \mc{1}{r}{\numprint{29242.39}} & \mc{1}{r|}{\numprint{1679807}} & \mc{1}{r}{\numprint{96.95}} & \mc{1}{r|}{\textbf{\numprint{1680166}}} & \mc{2}{c}{-} \\ 
	\rowcolor{lightergray} \mc{1}{l|}{\textit{ocean}} & \mc{1}{r}{\numprint{4.88}} & \mc{1}{r|}{\textbf{\numprint{7248581}}} & \mc{1}{r}{\numprint{11142.43}} & \mc{1}{r|}{\numprint{7075329}} & \mc{1}{r}{\numprint{0.07}} & \mc{1}{r|}{\numprint{6803672}} & \mc{1}{r}{\numprint{44.58}} & \mc{1}{r|}{\textbf{\numprint{7248581}}} & \mc{1}{r}{\numprint{50.49}} & \mc{1}{r|}{\textbf{\numprint{7248581}}} & \mc{2}{c}{-} \\ 
	\mc{1}{l|}{\textit{pwt}} & \mc{2}{c|}{-} & \mc{1}{r}{\numprint{761.52}} & \mc{1}{r|}{\numprint{1175437}} & \mc{1}{r}{\numprint{0.03}} & \mc{1}{r|}{\numprint{1153600}} & \mc{1}{r}{\numprint{36050.99}} & \mc{1}{r|}{\numprint{1175149}} & \mc{1}{r}{\numprint{7590.48}} & \mc{1}{r|}{\textbf{\numprint{1178434}}} & \mc{2}{c}{-} \\ 
	\hline 
	\hline 
	\mc{1}{l|}{mesh} & \mc{2}{c|}{branch reduce}  & \mc{2}{c|}{{\hils}}  & \mc{2}{c|}{{\htwis}}  & \mc{2}{c|}{{\wmmis}}  & \mc{2}{c|}{{\wmmiss}}  & \mc{2}{c}{struction} \\ 
	\hline 
	\rowcolor{lightergray} \mc{1}{l|}{\textit{buddha}} & \mc{1}{r}{\numprint{51.87}} & \mc{1}{r|}{\textbf{\numprint{57555880}}} & \mc{1}{r}{\numprint{36000.07}} & \mc{1}{r|}{\numprint{57258790}} & \mc{1}{r}{\numprint{0.47}} & \mc{1}{r|}{\numprint{57508556}} & \mc{1}{r}{\numprint{30351.04}} & \mc{1}{r|}{\numprint{57555105}} & \mc{1}{r}{\numprint{9.14}} & \mc{1}{r|}{\textbf{\numprint{57555880}}} & \mc{1}{r}{\numprint{1.57}} & \mc{1}{r}{\textbf{\numprint{57555880}}} \\ 
	\rowcolor{lightergray} \mc{1}{l|}{\textit{dragon}} & \mc{1}{r}{\numprint{2.90}} & \mc{1}{r|}{\textbf{\numprint{7956530}}} & \mc{1}{r}{\numprint{9026.60}} & \mc{1}{r|}{\numprint{7947535}} & \mc{1}{r}{\numprint{0.04}} & \mc{1}{r|}{\numprint{7950526}} & \mc{1}{r}{\numprint{674.94}} & \mc{1}{r|}{\numprint{7956523}} & \mc{1}{r}{\numprint{0.82}} & \mc{1}{r|}{\textbf{\numprint{7956530}}} & \mc{1}{r}{\numprint{0.17}} & \mc{1}{r}{\textbf{\numprint{7956530}}} \\ 
	\rowcolor{lightergray} \mc{1}{l|}{\textit{ecat}} & \mc{1}{r}{\numprint{9.18}} & \mc{1}{r|}{\textbf{\numprint{36650298}}} & \mc{1}{r}{\numprint{36000.05}} & \mc{1}{r|}{\numprint{36562652}} & \mc{1}{r}{\numprint{0.50}} & \mc{1}{r|}{\numprint{36606394}} & \mc{1}{r}{\numprint{11626.77}} & \mc{1}{r|}{\numprint{36650108}} & \mc{1}{r}{\numprint{5.24}} & \mc{1}{r|}{\textbf{\numprint{36650298}}} & \mc{1}{r}{\numprint{1.92}} & \mc{1}{r}{\textbf{\numprint{36650298}}} \\ 
	\hline 
	\hline 
	\mc{1}{l|}{osm} & \mc{2}{c|}{branch reduce}  & \mc{2}{c|}{{\hils}}  & \mc{2}{c|}{{\htwis}}  & \mc{2}{c|}{{\wmmis}}  & \mc{2}{c|}{{\wmmiss}}  & \mc{2}{c}{struction} \\ 
	\hline 
	\rowcolor{lightergray} \mc{1}{l|}{\textit{florida-3}} & \mc{1}{r}{\numprint{1724.45}} & \mc{1}{r|}{\textbf{\numprint{237333}}} & \mc{1}{r}{\numprint{216.24}} & \mc{1}{r|}{\textbf{\numprint{237333}}} & \mc{1}{r}{\numprint{0.13}} & \mc{1}{r|}{\numprint{234218}} & \mc{1}{r}{\numprint{22.91}} & \mc{1}{r|}{\textbf{\numprint{237333}}} & \mc{1}{r}{\numprint{23.96}} & \mc{1}{r|}{\textbf{\numprint{237333}}} & \mc{1}{r}{\numprint{1.33}} & \mc{1}{r}{\textbf{\numprint{237333}}} \\ 
	\mc{1}{l|}{\textit{greenland-3}} & \mc{1}{r}{\numprint{36000.00}} & \mc{1}{r|}{\numprint{13894}} & \mc{1}{r}{\numprint{1226.78}} & \mc{1}{r|}{\numprint{14011}} & \mc{1}{r}{\numprint{32.75}} & \mc{1}{r|}{\numprint{12505}} & \mc{1}{r}{\numprint{1374.27}} & \mc{1}{r|}{\textbf{\numprint{14012}}} & \mc{1}{r}{\numprint{3595.51}} & \mc{1}{r|}{\textbf{\numprint{14012}}} & \mc{2}{c}{-} \\ 
	\rowcolor{lightergray} \mc{1}{l|}{\textit{utah-3}} & \mc{1}{r}{\numprint{239.50}} & \mc{1}{r|}{\textbf{\numprint{98847}}} & \mc{1}{r}{\numprint{72.21}} & \mc{1}{r|}{\textbf{\numprint{98847}}} & \mc{1}{r}{\numprint{0.04}} & \mc{1}{r|}{\numprint{97754}} & \mc{1}{r}{\numprint{6.81}} & \mc{1}{r|}{\textbf{\numprint{98847}}} & \mc{1}{r}{\numprint{5.48}} & \mc{1}{r|}{\textbf{\numprint{98847}}} & \mc{1}{r}{\numprint{0.08}} & \mc{1}{r}{\textbf{\numprint{98847}}} \\ 
	\hline 
	\hline 
	\mc{1}{l|}{snap} & \mc{2}{c|}{branch reduce}  & \mc{2}{c|}{{\hils}}  & \mc{2}{c|}{{\htwis}}  & \mc{2}{c|}{{\wmmis}}  & \mc{2}{c|}{{\wmmiss}}  & \mc{2}{c}{struction} \\ 
	\hline 
	\mc{1}{l|}{\textit{as-skitter}} & \mc{2}{c|}{-} & \mc{1}{r}{\numprint{36000.25}} & \mc{1}{r|}{\numprint{123994141}} & \mc{1}{r}{\numprint{1.04}} & \mc{1}{r|}{\numprint{124141373}} & \mc{1}{r}{\numprint{2564.30}} & \mc{1}{r|}{\textbf{\numprint{124157714}}} & \mc{1}{r}{\numprint{10231.98}} & \mc{1}{r|}{\numprint{124157712}} & \mc{2}{c}{-} \\ 
	\rowcolor{lightergray} \mc{1}{l|}{\textit{ca-GrQc}} & \mc{1}{r}{<\numprint{0.01}} & \mc{1}{r|}{\textbf{\numprint{286489}}} & \mc{1}{r}{\numprint{148.82}} & \mc{1}{r|}{\textbf{\numprint{286489}}} & \mc{1}{r}{<\numprint{0.01}} & \mc{1}{r|}{\numprint{286352}} & \mc{1}{r}{\numprint{3.48}} & \mc{1}{r|}{\textbf{\numprint{286489}}} & \mc{1}{r}{\numprint{3.93}} & \mc{1}{r|}{\textbf{\numprint{286489}}} & \mc{1}{r}{<\numprint{0.01}} & \mc{1}{r}{\textbf{\numprint{286489}}} \\ 
	\rowcolor{lightergray} \mc{1}{l|}{\textit{web-BS.}} & \mc{1}{r}{\numprint{36000.12}} & \mc{1}{r|}{\numprint{43891206}} & \mc{1}{r}{\numprint{36000.10}} & \mc{1}{r|}{\numprint{43888267}} & \mc{1}{r}{\numprint{9.94}} & \mc{1}{r|}{\numprint{43889843}} & \mc{1}{r}{\numprint{18698.57}} & \mc{1}{r|}{\numprint{43907225}} & \mc{1}{r}{\numprint{13.78}} & \mc{1}{r|}{\textbf{\numprint{43907482}}} & \mc{1}{r}{\numprint{6.52}} & \mc{1}{r}{\textbf{\numprint{43907482}}} \\ 
	\hline 	\hline 
	\mc{1}{l|}{ssmc} & \mc{2}{c|}{branch reduce}  & \mc{2}{c|}{{\hils}}  & \mc{2}{c|}{{\htwis}}  & \mc{2}{c|}{{\wmmis}}  & \mc{2}{c|}{{\wmmiss}}  & \mc{2}{c}{struction} \\ 
	\hline 
	\rowcolor{lightergray} \mc{1}{l|}{\textit{ga2010}} & \mc{1}{r}{\numprint{36000.10}} & \mc{1}{r|}{\numprint{4644324}} & \mc{1}{r}{\numprint{29522.41}} & \mc{1}{r|}{\numprint{4642807}} & \mc{1}{r}{\numprint{0.16}} & \mc{1}{r|}{\numprint{4639891}} & \mc{1}{r}{\numprint{30401.36}} & \mc{1}{r|}{\numprint{4644293}} & \mc{1}{r}{\numprint{3.83}} & \mc{1}{r|}{\textbf{\numprint{4644417}}} & \mc{1}{r}{\numprint{0.62}} & \mc{1}{r}{\textbf{\numprint{4644417}}} \\ 
	\rowcolor{lightergray} \mc{1}{l|}{\textit{nh2010}} & \mc{1}{r}{\numprint{36000.00}} & \mc{1}{r|}{\numprint{581637}} & \mc{1}{r}{\numprint{2163.80}} & \mc{1}{r|}{\numprint{588797}} & \mc{1}{r}{\numprint{0.03}} & \mc{1}{r|}{\numprint{587059}} & \mc{1}{r}{\numprint{3379.36}} & \mc{1}{r|}{\textbf{\numprint{588996}}} & \mc{1}{r}{\numprint{1.22}} & \mc{1}{r|}{\textbf{\numprint{588996}}} & \mc{1}{r}{\numprint{0.11}} & \mc{1}{r}{\textbf{\numprint{588996}}} \\ 
	\rowcolor{lightergray} \mc{1}{l|}{\textit{ri2010}} & \mc{1}{r}{\numprint{36000.00}} & \mc{1}{r|}{\numprint{447427}} & \mc{1}{r}{\numprint{782.49}} & \mc{1}{r|}{\numprint{458489}} & \mc{1}{r}{\numprint{0.02}} & \mc{1}{r|}{\numprint{457108}} & \mc{1}{r}{\numprint{25340.86}} & \mc{1}{r|}{\numprint{459227}} & \mc{1}{r}{\numprint{1.45}} & \mc{1}{r|}{\textbf{\numprint{459275}}} & \mc{1}{r}{\numprint{0.09}} & \mc{1}{r}{\textbf{\numprint{459275}}} \\ 
\hline 
\hline
\mc{1}{l|}{overall} & \mc{2}{c|}{branch reduce} & \mc{2}{c|}{{\hils}}  & \mc{2}{c|}{{\htwis}}   & \mc{2}{c|}{{\wmmis}}  & \mc{2}{c|}{{\wmmiss}} & \mc{2}{c}{struction}    \\ 
\hline 
\mc{1}{l|}{\# best} & \mc{2}{r|}{176/207} & \mc{2}{r|}{155/207} & \mc{2}{r|}{99/207} & \mc{2}{r|}{186/207} & \mc{2}{r|}{202/207} & \mc{2}{r}{189/207} \\ 
\mc{1}{l|}{mean time} & \mc{2}{r|}{-} & \mc{2}{r|}{\numprint{16,37}} & \mc{2}{r|}{$<$ \numprint{0.01}}  &  \mc{2}{r|}{\numprint{1.84}} & \mc{2}{r|}{\numprint{0.95}} & \mc{2}{r}{-} \\ 

\end{tabular} 

\end{table*}

\subsection{Heuristic Data Reduction Rules}

\label{sec: ExperimentsHeuristicReduce}
We now compare different vertex selection strategies presented in Section \ref{sec: vertex_selection}. Table~\ref{tab: compact_vertex_selection} summarizes our results. Detailed per-instance results are in the appendix, Tables~\ref{tab: combined_vertex_selection_best} and \ref{tab: combined_vertex_selection}. 
First, we note that each selection strategy is able to find the best solution for at least 19 instances. Overall, \textit{hybrid} and \textit{weight} are able to obtain the largest number of best solutions. 
However, Table~\ref{tab: combined_vertex_selection} in the appendix also shows that \textit{degree} can beat both \textit{hybrid} and \textit{weight} on some instances,~\eg \textit{ecat}. The other two strategies, \textit{weight/degree} and \textit{solution participation}, did not result in any improvement compared to \textit{hybrid} and \textit{weight}. 
Finally, a direct comparison of \textit{hybrid} and \textit{weight} (see appendix Table~\ref{tab: combined_vertex_selection_best}) shows that \textit{hybrid} is able to achieve a better solution than \textit{weight} on four instances at the cost of an 18\% higher geometric mean time. Since our primary focus is solution quality, we thus use \textit{hybrid} for the following experiments.

\subsection{Comparison against the State of the Art}\label{sec: ExperimentsSoA}

We now compare our algorithm {\wmmis} against a range of algorithms: {\htwis} by Gu~\etal~\cite{gu2021towards}, 
both struction-based algorithms by Gellner~\etal~\cite{DBLP:conf/alenex/GellnerLSSZ21} where we always report the better of the two results in the column named struction, the branch-and-reduce solver by Lamm~\etal~\cite{DBLP:conf/alenex/Lamm0SWZ19}, and {\hils} by Nogueira~\etal~\cite{hybrid-ils-2018}. 
We also include a variant of our algorithm, called {\wmmiss}, using \textsf{CyclicFast} from Gellner~\etal~\cite{DBLP:conf/alenex/GellnerLSSZ21} 
with a time limit of $60$ seconds to compute individuals for the initial population.
We present a representative sample of our full experiments in Table~\ref{tab: compact_algos}. 
In the last part we give a summary of all instances. This consists of the number of instances solved best and the geometric mean time respectively.
Detailed per-instance results are presented in the appendix, Tables~\ref{tab: soa_fe}--\ref{tab: soa_ssmc}.

Overall, we see that {\wmmiss} has the largest number of best solutions for our full set of $207$ instances.
In particular, it is able to compute the best solution for all but five instances,~\ie \textit{fe\_rotor}, \textit{hawaii-AM3}, \textit{kentucky-AM3}, \textit{as-skitter} and \textit{soc-pokec-relationships}. Only in two of these cases our algorithm was outperformed by a competitor, which is on \textit{soc-pokec-relationships} by {\htwis} and on \textit{fe\_rotor} by {\hils}. In the other cases {\wmmis} achieved a better result.
Additionally, {\wmmiss} is able to compute the best solutions for all graphs in the mesh and ssmc graph classes.
Finally, except for $18$ out of $207$ instances, {\wmmis} finds the best solution in less than $100$ seconds.
Without the use of the struction, our algorithm {\wmmis} is still able to compute the best solution for $186$ instances, including \textit{hawaii-AM3}, \textit{kentucky-AM3} and \textit{as-skitter}.
Except for osm instances, the geometric mean running time increases compared to {\wmmiss}.

When looking at the running times, we see that {\htwis} achieves the smallest geometric mean running time.
However, {\htwis} also has the least number of best solutions overall.
Furthermore, the quality is lower than {\wmmis} and {\wmmiss} on \emph{all but one} of the tested instances, with multiple instances having a significant difference in weight--larger than $\numprint{10000}$.
The running time achieved by the struction-based algorithms is also to be noted. These are for example the second fastest on ssmc instances, see the appendix, Table~\ref{tab: soa_ssmc}, however, these only find $188$ best solutions overall.
	
In terms of memory requirement, when the struction variants are able to solve the instance very fast, memory usage is usually below \numprint{1} GB and also a bit smaller than the the memory required by our algorithm. 
For more difficult instances, as for example \textit{fe\_body} where {\wmmiss} has a memory usage of \numprint{0,60} GB, the struction algorithm requires more than \numprint{100} GB. %

\subsubsection{Comparison to METAMIS}
In recent times, Dong~\etal~\cite{DBLP:conf/esa/DongGNPRS22} presented a novel heuristic for MWIS called \textsf{METAMIS}.
Since their code is not publicly available, we compare the solution quality of our algorithm against the results presented in their work. Detailed per-instance results can be found in the appendix, Table~\ref{tab: METAMIS}. 
We use the same pre-reduced osm instances as Dong~\etal~\cite{DBLP:conf/esa/DongGNPRS22}. In their experiments the time limit for \textsf{METAMIS} is \numprint{1500} seconds. However, as our algorithm unfolds its full potential over a long period of time, and our algorithm focused on higher-quality solutions and not fast running times, we stayed with a 10-hour time limit.
Moreover, note that the results have been computed on different machines.
Summarizing the results, we are able to compute the same or better solutions for all graphs. In total we were able to improve three solutions compared to the \textsf{METAMIS} results with both of our configurations. Especially for large instances, our algorithm outperforms the results stated in~\cite{DBLP:conf/esa/DongGNPRS22}.
 However, it is not clear whether \textsf{METAMIS} would compute equally good solutions for the instances where {\wmmis} performed~better.

\section{Conclusion and Future Work}
In this work, we developed a novel memetic algorithm for the maximum independent set problem. It repeatedly reduces the graph until an high-quality solution to the MWIS problem is found. After applying exact reductions, we use the best solution computed by the evolutionary algorithm on the reduced graph to identify vertices likely to be in a MWIS. These are removed from the graph which further opens the reduction space and creates the possibility to apply this process repeatedly. %

For future work, we are interested in an island-based approach to obtain a parallelization of our evolutionary approach, as well as parallelization of the reductions. Both the \textsc{ExactReduce} and the \textsc{HeuristicReduce} routine can result in a disconnected reduced graph. We are interested in solving the problem on each of the resulting connected components separately, which also enables new parallelization possibilities. We will release the code of our work in \url{https://github.com/KarlsruheMIS}.

\begin{acks}
We acknowledge support by DFG grant SCHU 2567/3-1. 
\end{acks}

\bibliographystyle{ACM-Reference-Format}
\bibliography{phdthesiscs,mwis}

\clearpage
\appendix

\section{Expanded Related Work}\label{sec:expanded_related_work}

We give a short overview of existing work on both exact and heuristic procedures. For more details, we refer the reader to the recent survey on data reduction techniques~\cite{Abu-Khzam2022}.%

\subsection{Exact Methods}
Exact algorithms usually compute optimal solutions by systematically exploring the solution space.
A frequently used paradigm in exact algorithms for combinatorial optimization problems is called \emph{branch-and-bound}~\cite{ostergaard2002fast,warren2006combinatorial}.
In case of the MWIS problem, these types of algorithms compute optimal solutions by case distinctions in which vertices are either included into the current solution or excluded from it, branching into two or more subproblems and resulting in a search tree.
Over the years, branch-and-bound methods have been improved by new branching schemes or better pruning methods using upper and lower bounds to exclude specific subtrees~\cite{balas1986finding, babel1994fast,li2017minimization}.
In particular, Warren and Hicks~\cite{warren2006combinatorial} proposed three branch-and-bound algorithms that combine the use of weighted clique covers and a branching scheme first introduced by Balas and Yu~\cite{balas1986finding}.
Their first approach extends the algorithm by Babel~\cite{babel1994fast} by using a more intricate data structures to improve its performance.
The second one is an adaptation of the algorithm of Balas and Yu, which uses a weighted clique heuristic that yields structurally similar results to the heuristic of Balas and Yu.
The last algorithm is a hybrid version that combines both algorithms and is able to compute optimal solutions on graphs with hundreds of vertices.

In recent years, reduction rules have frequently been added to branch-and-bound methods yielding so-called \emph{branch-and-reduce} algorithms~\cite{akiba-tcs-2016}. 
These algorithms are able to improve the worst-case runtime of branch-and-bound algorithms by applications of reduction rules to the current graph before each branching step.
For the unweighted case, a large number of branch-and-reduce
algorithms have been developed in the past. The currently best exact
solver~\cite{DBLP:conf/siamcsc/HespeL0S20}, which won the PACE challenge
2019~\cite{DBLP:conf/siamcsc/HespeL0S20, bogdan-pace, peaty-pace}, uses a portfolio of branch-and-reduce/bound solvers for the complementary  problems. Recently, novel branching strategies have been presented in~\cite{DBLP:conf/wea/HespeLS21} to further improve both branch-and-bound as well as branch-and-reduce approaches.

However, for a long time, virtually no weighted reduction rules were known, which is why hardly any branch-and-reduce algorithms exist for the MWIS problem. The first branch-and-reduce algorithm for the weighted case was presented by Lamm~\etal~\cite{DBLP:conf/alenex/Lamm0SWZ19}.
The authors first introduce two meta-reductions called neighborhood removal and neighborhood folding, from which they derive a new set of weighted reduction rules.
On this foundation a branch-and-reduce algorithm is developed using pruning with weighted clique covers similar to the approach by Warren and Hicks~\cite{warren2006combinatorial} for upper bounds and an adapted version of the ARW local search~\cite{andrade-2012} for lower bounds.

This algorithm was then extended by Gellner~\etal~\cite{DBLP:conf/alenex/GellnerLSSZ21} to utilize different variants of the struction, originally introduced by Ebenegger~\etal~\cite{ebenegger1984pseudo} and later improved by Alexe~\etal~\cite{alexe2003struction}. In contrast to previous reduction rules, these were not necessarily decreasing the graph size, but rather transforming the graph which later can lead to even further reduction possibilities. Those variants were integrated into the framework of Lamm~\etal~\cite{DBLP:conf/alenex/Lamm0SWZ19} in the preprocessing as well as in the reduce step.
The experimental evaluation shows that this algorithm can solve a large set of real-world instances and outperforms the branch-and-reduce algorithm by Lamm~\etal~\cite{DBLP:conf/alenex/Lamm0SWZ19}, as well as different state-of-the-art heuristic approaches such as the algorithm {\hils} presented by Nogueira~\cite{hybrid-ils-2018} as well as two other local search algorithms \textsf{DynWVC1} and \textsf{DynWVC2} by Cai~\etal~\cite{cai-dynwvc}. Recently, Xiao~\etal \cite{DBLP:conf/www/XiaoHZD21} present further data reductions for the weighted case as well as a simple exact algorithm based on these data reduction rules. Furthermore, in~\cite{DBLP:conf/icde/ZhengGPY20} a new reduction-and-branching algorithm was introduced using two new reduction rules.

Not long ago Huang~\etal~\cite{DBLP:conf/cocoon/HuangXC21} also presented a branch-and-bound algorithm using reduction rules working especially well on sparse graphs. In their work they additionally undertake a detailed analysis for the running time bound on special graphs. With the measure-and-conquer technique they show that for cubic graphs the running time of their algorithm is~$\mathcal{O}^*(\numprint{1.1443}^n)$ which is improving previous time bounds for this problem using polynomial space complexity and for graphs of average degree three.

Figiel~\etal~\cite{DBLP:conf/esa/FigielFNN22} introduced a new idea added to the state-of-the-art way of applying reductions. They propose to not only performing reductions, but also the possibility of undoing them during the reduction process. As they showed in their paper for the unweighted independent set problem, this can lead to new possibilities to apply further reductions and finally to smaller reduced graphs.

Finally, there are exact procedures which are either based on other extension of the branch-and-bound paradigm, e.g.~\cite{rebennack2011branch,warrier2005branch,warrier2007branch}, or on the reformulation into other $\mathcal{NP}$-complete problems, for which a variety of solvers already exist.
For instance, Xu~\etal~\cite{xu2016new} developed an algorithm called \textsf{SBMS}, which calculates an optimal solution for a given MWVC instance by solving a series of SAT instances. Also for the MWVC problem a new exact algorithm using the branch-and-bound idea combined with data reduction rules were recently presented \cite{DBLP:journals/corr/abs-1903-05948}.
We additionally note that there are several recent works on the complementary maximum weighted clique problem that are able to handle large real-world networks~\cite{fang2016exact,jiang2017exact,held2012maximum}.
However, using these solvers for the MWIS problem requires computing complement graphs.
Since large real-world networks are often very sparse, processing their complements quickly becomes infeasible due to their memory~requirement.

\subsection{Heuristic Methods}
A widely used heuristic approach is local search, which usually computes an initial solution and then tries to improve it by simple insertion, removal or swap operations. 
Although in theory local search generally offers no guarantees for the solution's quality, in practice they find high-quality solutions significantly faster than exact procedures.

For unweighted graphs, the iterated local search (ARW) by Andrade~\etal~\cite{andrade-2012}, is a very successful heuristic.
It is based on so-called $(1,2)$-swaps which remove one vertex from the solution and add two new vertices to it, thus improving the current solution by one.
Their algorithm uses special data structures which find such a $(1,2)$-swap in linear time in the number of edges or prove that none exists.
Their algorithm is able to find (near-)optimal solutions for small to medium-size instances in milliseconds, but struggles on massive instances with millions of vertices and edges.

The hybrid iterated local search (\hils) by Nogueira~\etal~\cite{hybrid-ils-2018} adapts the ARW algorithm for weighted graphs.
In addition to weighted $(1,2)$-swaps, it also uses $(\omega,1)$-swaps that add one vertex $v$ into the current solution and exclude its $\omega$ neighbors. 
These two types of neighborhoods are explored separately using variable neighborhood descent (VND).
Two other local searches, DynWVC1 and DynWVC2, for the equivalent minimum weight vertex cover problem are presented by Cai~\etal~\cite{cai-dynwvc}.
Their algorithms extend the existing FastWVC heuristic~\cite{li2017efficient} by dynamic selection strategies for vertices to be removed from the current solution.
In practice, DynWVC1 outperforms previous MWVC heuristics on map labeling instances and large scale networks, and DynWVC2 provides further improvements on large scale networks but performs worse on map~labeling~instances.

Li~\etal~\cite{li2019numwvc} presented a local search algorithm for the minimum weight vertex cover (MWVC) problem, which is complementary to the MWIS problem. Their algorithm applies reduction rules during the construction phase of the initial solution.
Furthermore, they adapt the configuration checking approach~\cite{cai2011local} to the MWVC problem which is used to reduce cycling, \ie returning to a solution that has been visited recently.
Finally, they develop a technique called self-adaptive-vertex-removing, which dynamically adjusts the number of removed vertices per iteration.
Experiments show that their algorithm outperforms state-of-the-art approaches on both graphs of up to millions of vertices and real-world~instances.

Recently, a hybrid method was introduced by Langedal~\etal~\cite{langedal2022efficient} to also solve the MWVC problem. For this approach they combined elements from exact methods with local search, data reductions and graph neural networks. In their experiments they achieve definite improvements compared to \textsf{DynWVC2} and the {\hils} algorithm in both solution quality and running~time.

With \textsf{EvoMIS}, Lamm~\etal~\cite{lamm2015graph} presented an evolutionary approach to tackle the maximum independent set problem. The key feature of their algorithm is to use graph partitioning to come up with natural combine operations, where whole blocks of solutions to the MIS problem can be exchanged easily. To these combine operations also local search algorithms were added to improve the solutions further.
Combining the branch-and-reduce approach with the evolutionary algorithm \textsf{EvoMIS}, a reduction evolution algorithm \textsf{ReduMIS} was presented by Lamm~\etal~\cite{redumis-2017}. In their experiments, \textsf{ReduMIS} outperformed the local search \textsf{ARW} as well as the pure evolutionary approach \textsf{EvoMIS}.
Another reduction based heuristic called {\htwis} was presented recently by Gu~\etal~\cite{gu2021towards}. %
The repeatedly apply their reductions exhaustively and then choose one vertex by a tie-breaking policy to add to the solution. Now this vertex as well as its neighbors can be removed from the graph and the reductions can be applied again.
Their experiments prove a significant improvement in running time as well as computed weights of the MWIS compared to the state-of-the-art~solvers. 

Recently, a new metaheuristic was introduced by Dong~\etal~\cite{DBLP:conf/esa/DongGNPRS22} in particular for vehicle routing instances. With their algorithm \textsc{METAMIS} they developed a new local search algorithm using a new variant of path-relinking to escape local optima. In their experiments they outperform {\hils} algorithm on a wide range of instances both in time and solution~quality.

\section{Experiments on Reduction Ordering}
\label{subsec:ordering}

Different orderings of applying data reductions yield different sizes of the reduced graph. Additionally, the ordering effects the running time of the \textsc{ExactReduce} routine from Algorithm \ref{algo: overview}. This effect has been described for example by Figiel~\etal~\cite{DBLP:conf/esa/FigielFNN22}. We now perform experiments to evaluate the impact different orderings may have. %
Here, we run the \textsc{ExactReduce}~routine, i.e. we apply the reductions in a given ordering exhaustively, and report the results.

\textit{Baseline:} Our starting point is an intuitive ordering, which we constructed by simplicity of the reductions, from simplest to most complex. Hereby we orientate us towards the ordering chosen by Akiba and Iwata~\cite{akiba-tcs-2016}. This initial ordering is precisely the order in which we introduce the reductions from Reduction~\ref{red: neighborhood removal} to Reduction~\ref{red: neighborhood folding}. In the following experiments, we use this ordering as a \emph{baseline} to compare against.

\subsection{Orderings Based on Impact of Single Reductions} The space of possible orderings of data reductions is very large. We start our evaluation by examining the impact of disabling single reductions in our baseline, \ie we run our baseline reductions and then build a set of reductions where exactly one data reduction of the baseline is disabled. The ordering of the remaining data reductions remains the same.
The time to apply all reductions exhaustively using our baseline ordering is denoted as $t_{all}$ and the weight of vertices added by those reductions to the solution is denoted~as~$\omega_{all}$. 
Then, we create different data reduction orderings from the baseline in which a single data reduction is disabled. 
For each of the available reductions $r$, we get a new time $t_{all \backslash r}$ and solution weight $\omega_{all \backslash r}$ which corresponds to running all reductions except $r$ of the baseline (and in its order). 
Based on these values, we derive three orderings, a time based ordering, a solution weight based ordering and a combination~of~both.

\textbf{Time-based Ordering.} For the \textit{time-based ordering} we rearranged the reductions such that the mean $\overline{t}_{all \backslash r}$ is decreasing. The intuition here is if removing a reduction from the baseline yields an ordering that has an excessive running time, then this reduction is important for running time and should be applied before a reduction that has a smaller impact. This results in the ordering (\ref{red: bse}, \ref{red: clique}, \ref{red: extended_v_shape}, \ref{red: twin}, \ref{red: deg1}, \ref{red: neighborhood removal}, \ref{red: ese}, \ref{red: extended_v_shape} (min), \ref{red: triangle}, \ref{red: dom}, \ref{red: clique_neigh}, \ref{red: CWIS}, \ref{red: neighborhood folding}). 

\textbf{Weight-based Ordering.} For the \textit{solution weight-based ordering} the reductions are reordered in increasing order according to the mean value $\overline{\omega}_{all \backslash r}$ over all graphs. The intuition here is that if  $\overline{\omega}_{all \backslash r}$ is small, then not using $r$ has a large impact on the weight of solution added and hence should be applied before a reduction that has a smaller impact. The resulting ordering is  (\ref{red: clique}, \ref{red: CWIS}, \ref{red: neighborhood folding}, \ref{red: bse}, \ref{red: extended_v_shape}, \ref{red: extended_v_shape} (min), \ref{red: twin}, \ref{red: dom}, \ref{red: neighborhood removal}, \ref{red: deg1}, \ref{red: triangle}, \ref{red: clique_neigh}, \ref{red: ese}).

\begin{table}[t]
	\begin{ThreePartTable}
		\caption{Comparing geometric mean of times and weights of orderings, relative to the initial ordering, as well as the average reduction in graph size $|\mathcal{K}|/|V|$, where $|\mathcal{K}|$ is the number of vertices in the reduced graph.}\label{tab: time_and_weight_rel}
		\vspace*{-.5cm}
		\centering
		\small \begin{center}     
\begin{tabular}{cccc} 
\hline
\mc{1}{l|}{Ordering} & \mc{1}{r}{$t/t_{initial}$} & \mc{1}{|r|}{$w/w_{initial}$} & \mc{1}{r}{$|\mathcal{K}|/|V|$}\\ 
\hline
\mc{1}{l|}{initial} & \mc{1}{r}{\numprint{1.0000}} & \mc{1}{|r|}{\numprint{1.0000}} & \mc{1}{r}{\numprint{0.3014}} \\ 
\mc{1}{l|}{time} & \mc{1}{r}{\numprint{0.9159}} & \mc{1}{|r|}{\numprint{1.0003}} & \mc{1}{r}{\numprint{0.3010}} \\ 
\mc{1}{l|}{weight} & \mc{1}{r}{\numprint{4.7734}} & \mc{1}{|r|}{\numprint{1.0029}} & \mc{1}{r}{\numprint{0.2997}} \\ 
\mc{1}{l|}{time/weight} & \mc{1}{r}{\numprint{2.0464}} & \mc{1}{|r|}{\numprint{1.0017}} & \mc{1}{r}{\numprint{0.3002}}\\
\hline
\end{tabular} 
\end{center}

		\vspace*{-.5cm}
	\end{ThreePartTable}
\end{table}

\textbf{Time and Weight-based Ordering.} Here we use a combination with $x_{all \backslash r} = \overline{t}_{all \backslash r} - 10\overline{\omega}_{all \backslash r}$ decreasingly to order reductions. We use a factor of 10 here, since solution quality is typically more important for applications than running time. This results in the following ordering of reductions
(\ref{red: clique}, \ref{red: bse}, \ref{red: CWIS}, \ref{red: extended_v_shape}, \ref{red: twin}, \ref{red: deg1}, \ref{red: extended_v_shape}(min), \ref{red: neighborhood removal}, \ref{red: dom}, \ref{red: ese}, \ref{red: triangle}, \ref{red: clique_neigh}, \ref{red: neighborhood folding}).

\emph{Discussion.} 
The results for the previous orderings are presented in Table \ref{tab: time_and_weight_rel}. We note that different orderings do not yield significant improvements compared to the initial ordering. We can observe, that the \textit{time ordering} can reduce the running time. The \textit{weight ordering} can improve the reduction offset in the thousandths range by a large expense of additional running time. 
Moreover, there are some reductions that remain at approximately the same position meaning they are either very important for solution size and quality (or the opposite). For example Reductions~\ref{red: clique} and~\ref{red: bse} are applied towards the beginning, whereas for example Reduction~\ref{red: clique_neigh} is applied towards the end. On the other hand there also are reductions that are on completely different positions,~\eg Reduction~\ref{red: CWIS}.
We conclude that the initial ordering (baseline) is already robust w.r.t. the orderings considered in this section.

\subsection{Orderings Based on Impact of Groups of Reductions}
We divide  the reductions into three groups of roughly similar complexity. The first group contains Reductions~\ref{red: neighborhood removal} and~\ref{red: deg1}, the second group consists of reductions for vertices of degree two, which are Reductions~\ref{red: triangle} and~\ref{red: extended_v_shape}. The third group contains all remaining reductions listed in Section~\ref{sec: reduction}.

\paragraph{Permutation of Ordering of Reductions in First and Second Group.} We now examine all permutations in the first and second group. The permutation in the ordering only takes place inside the groups. The groups themselves always stay in a fixed order,~\ie Reductions~\ref{red: neighborhood removal} and \ref{red: deg1} will always be the first two reductions applied. Reductions of the third group are applied as in the initial ordering. Overall, our experiments show that the order of the reductions in the first two groups has a negligible effect on running time and quality of a solution. 
Thus, for the remaining experiments we use the initial~ordering. %

\textbf{Permutation of Ordering of Reductions in Third Group.} 
We now examine permutations in the third group of reductions. We apply additional restrictions to reduce the number of permutations. We apply Reduction~\ref{red: neighborhood folding} always last, and Reduction~\ref{red: bse} is always followed by Reduction~\ref{red: ese}. The best performing permutation is (\ref{red: neighborhood removal}, \ref{red: deg1}, \ref{red: triangle}, \ref{red: extended_v_shape}, \ref{red: extended_v_shape} (min), \ref{red: clique}, \ref{red: twin}, \ref{red: CWIS}, \ref{red: clique_neigh}, \ref{red: dom}, \ref{red: bse}, \ref{red: ese}, \ref{red: neighborhood folding}). The geometric mean weight improvement is $w/w_{initial}=\numprint{1.0023}$ and the geometric mean time compared to the initial ordering is $t/t_{initial}=\numprint{2.9}$. 	

\textit{Conclusion.} Overall, there are some orderings that perform better than the initial ordering by complexity, however, these improvements are only on a few instances (and result in significantly higher running time). In most cases all orderings yielded the similar results. Among those, the initial ordering remains one of the fastest. 
We conclude that the initial ordering presents a very stable reduction ordering. Hence, we use it for the remaining experiments. 
For some graphs, it might be worth trying multiple runs of algorithms using one of the other orderings we presented in this section as~well.
\begin{table}[h]
	\centering
	\begin{ThreePartTable}
		\caption{Average solution weight $\omega$ and time $t$ in seconds required to compute it for the two best vertex selection strategies from Table \ref{tab: combined_vertex_selection}. \textbf{Bold} numbers indicate the best solution among all algorithms. We also report the number of best solutions and the geometric mean running time over all instances.}
		\label{tab: combined_vertex_selection_best}
		\small 	\begin{tabular}{lcccc} 
\hline
\mc{1}{l|}{graphs} & \mc{1}{c}{$t$} & \mc{1}{c|}{$w$} & \mc{1}{c}{$t$} & \mc{1}{c}{$w$}\\ 
\hline
fe & \mc{2}{|c|}{hybrid}  & \mc{2}{c}{weight} \\ 
\hline
\mc{1}{l|}{\textit{ocean}} & \mc{1}{r}{\numprint{38.67}} & \mc{1}{r|}{\textbf{\numprint{7248581}}} & \mc{1}{r}{\numprint{40.17}} & \mc{1}{r}{\textbf{\numprint{7248581}}} \\ 
\mc{1}{l|}{\textit{sphere}} & \mc{1}{r}{\numprint{36007.65}} & \mc{1}{r|}{\textbf{\numprint{616552}}} & \mc{1}{r}{\numprint{36006.32}} & \mc{1}{r}{\numprint{616400}} \\ 
\mc{1}{l|}{\textit{rotor}} & \mc{1}{r}{\numprint{36051.25}} & \mc{1}{r|}{\textbf{\numprint{2644428}}} & \mc{1}{r}{\numprint{36048.97}} & \mc{1}{r}{\numprint{2643498}} \\ 
\mc{1}{l|}{\textit{pwt}} & \mc{1}{r}{\numprint{36020.62}} & \mc{1}{r|}{\numprint{1174368}} & \mc{1}{r}{\numprint{36011.25}} & \mc{1}{r}{\textbf{\numprint{1174644}}} \\ 
\mc{1}{l|}{\textit{body}} & \mc{1}{r}{\numprint{36008.02}} & \mc{1}{r|}{\numprint{1679750}} & \mc{1}{r}{\numprint{18947.76}} & \mc{1}{r}{\textbf{\numprint{1679854}}} \\ 

\hline
\hline
mesh & \mc{2}{|c|}{hybrid}  & \mc{2}{c}{weight} \\ 
\hline
\mc{1}{l|}{\textit{buddha}} & \mc{1}{r}{\numprint{19363.72}} & \mc{1}{r|}{\numprint{57555044}} & \mc{1}{r}{\numprint{26143.17}} & \mc{1}{r}{\textbf{\numprint{57555070}}} \\ 
\mc{1}{l|}{\textit{dragon}} & \mc{1}{r}{\numprint{292.29}} & \mc{1}{r|}{\textbf{\numprint{7956524}}} & \mc{1}{r}{\numprint{938.47}} & \mc{1}{r}{\textbf{\numprint{7956524}}} \\ 
\mc{1}{l|}{\textit{ecat}} & \mc{1}{r}{\numprint{14174.83}} & \mc{1}{r|}{\textbf{\numprint{36650097}}} & \mc{1}{r}{\numprint{3096.31}} & \mc{1}{r}{\numprint{36650048}} \\ 

\hline
\hline
osm & \mc{2}{|c|}{hybrid}  & \mc{2}{c}{weight} \\ 
\hline
\mc{1}{l|}{\textit{alabama-2}} & \mc{1}{r}{\numprint{0.31}} & \mc{1}{r|}{\textbf{\numprint{174309}}} & \mc{1}{r}{\numprint{0.33}} & \mc{1}{r}{\textbf{\numprint{174309}}} \\ 
\mc{1}{l|}{\textit{florida-3}} & \mc{1}{r}{\numprint{18.55}} & \mc{1}{r|}{\textbf{\numprint{237333}}} & \mc{1}{r}{\numprint{19.36}} & \mc{1}{r}{\textbf{\numprint{237333}}} \\ 
\mc{1}{l|}{\textit{georgia-3}} & \mc{1}{r}{\numprint{7.24}} & \mc{1}{r|}{\textbf{\numprint{222652}}} & \mc{1}{r}{\numprint{7.88}} & \mc{1}{r}{\textbf{\numprint{222652}}} \\ 
\mc{1}{l|}{\textit{greenland-3}} & \mc{1}{r}{\numprint{9307.76}} & \mc{1}{r|}{\textbf{\numprint{14012}}} & \mc{1}{r}{\numprint{7073.15}} & \mc{1}{r}{\textbf{\numprint{14012}}} \\ 
\mc{1}{l|}{\textit{new-hampshire-3}} & \mc{1}{r}{\numprint{2.98}} & \mc{1}{r|}{\textbf{\numprint{116060}}} & \mc{1}{r}{\numprint{2.91}} & \mc{1}{r}{\textbf{\numprint{116060}}} \\ 
\mc{1}{l|}{\textit{rhode-island-2}} & \mc{1}{r}{\numprint{9.06}} & \mc{1}{r|}{\textbf{\numprint{184596}}} & \mc{1}{r}{\numprint{9.77}} & \mc{1}{r}{\textbf{\numprint{184596}}} \\ 
\mc{1}{l|}{\textit{utah-3}} & \mc{1}{r}{\numprint{5.67}} & \mc{1}{r|}{\textbf{\numprint{98847}}} & \mc{1}{r}{\numprint{5.65}} & \mc{1}{r}{\textbf{\numprint{98847}}} \\ 

\hline
\hline
snap & \mc{2}{|c|}{hybrid}  & \mc{2}{c}{weight} \\ 
\hline
\mc{1}{l|}{\textit{as-skitter}} & \mc{1}{r}{\numprint{663.07}} & \mc{1}{r|}{\numprint{123993350}} & \mc{1}{r}{\numprint{10510.85}} & \mc{1}{r}{\textbf{\numprint{123993368}}} \\ 
\mc{1}{l|}{\textit{ca-AstroPh}} & \mc{1}{r}{\numprint{19.41}} & \mc{1}{r|}{\textbf{\numprint{797510}}} & \mc{1}{r}{\numprint{19.33}} & \mc{1}{r}{\textbf{\numprint{797510}}} \\ 
\mc{1}{l|}{\textit{ca-CondMat}} & \mc{1}{r}{\numprint{18.93}} & \mc{1}{r|}{\textbf{\numprint{1145292}}} & \mc{1}{r}{\numprint{18.69}} & \mc{1}{r}{\textbf{\numprint{1145292}}} \\ 
\mc{1}{l|}{\textit{ca-GrQc}} & \mc{1}{r}{\numprint{3.61}} & \mc{1}{r|}{\textbf{\numprint{289356}}} & \mc{1}{r}{\numprint{3.51}} & \mc{1}{r}{\textbf{\numprint{289356}}} \\ 
\mc{1}{l|}{\textit{com-amazon}} & \mc{1}{r}{\numprint{1.09}} & \mc{1}{r|}{\textbf{\numprint{19271031}}} & \mc{1}{r}{\numprint{1.08}} & \mc{1}{r}{\textbf{\numprint{19271031}}} \\ 
\mc{1}{l|}{\textit{com-youtube}} & \mc{1}{r}{\numprint{1.06}} & \mc{1}{r|}{\textbf{\numprint{90295294}}} & \mc{1}{r}{\numprint{1.17}} & \mc{1}{r}{\textbf{\numprint{90295294}}} \\ 
\mc{1}{l|}{\textit{email-Enron}} & \mc{1}{r}{\numprint{21.46}} & \mc{1}{r|}{\textbf{\numprint{2457578}}} & \mc{1}{r}{\numprint{23.24}} & \mc{1}{r}{\textbf{\numprint{2457578}}} \\ 
\mc{1}{l|}{\textit{email-EuAll}} & \mc{1}{r}{\numprint{70.80}} & \mc{1}{r|}{\textbf{\numprint{25286322}}} & \mc{1}{r}{\numprint{66.26}} & \mc{1}{r}{\textbf{\numprint{25286322}}} \\ 
\mc{1}{l|}{\textit{loc-gowalla\_edges}} & \mc{1}{r}{\numprint{3.71}} & \mc{1}{r|}{\textbf{\numprint{12276929}}} & \mc{1}{r}{\numprint{4.13}} & \mc{1}{r}{\textbf{\numprint{12276929}}} \\ 
\mc{1}{l|}{\textit{p2p-Gnutella06}} & \mc{1}{r}{\numprint{3.65}} & \mc{1}{r|}{\textbf{\numprint{548612}}} & \mc{1}{r}{\numprint{3.72}} & \mc{1}{r}{\textbf{\numprint{548612}}} \\ 
\mc{1}{l|}{\textit{roadNet-PA}} & \mc{1}{r}{\numprint{34376.97}} & \mc{1}{r|}{\textbf{\numprint{61707635}}} & \mc{1}{r}{\numprint{7348.31}} & \mc{1}{r}{\numprint{61707560}} \\ 
\mc{1}{l|}{\textit{web-BerkStan}} & \mc{1}{r}{\numprint{17832.98}} & \mc{1}{r|}{\textbf{\numprint{43916421}}} & \mc{1}{r}{\numprint{9630.48}} & \mc{1}{r}{\numprint{43916363}} \\ 
\mc{1}{l|}{\textit{web-Google}} & \mc{1}{r}{\numprint{3.18}} & \mc{1}{r|}{\textbf{\numprint{56326504}}} & \mc{1}{r}{\numprint{3.71}} & \mc{1}{r}{\textbf{\numprint{56326504}}} \\ 
\mc{1}{l|}{\textit{web-NotreDame}} & \mc{1}{r}{\numprint{4.69}} & \mc{1}{r|}{\textbf{\numprint{25965069}}} & \mc{1}{r}{\numprint{4.66}} & \mc{1}{r}{\textbf{\numprint{25965069}}} \\ 
\mc{1}{l|}{\textit{wiki-Vote}} & \mc{1}{r}{\numprint{4.66}} & \mc{1}{r|}{\textbf{\numprint{500079}}} & \mc{1}{r}{\numprint{4.92}} & \mc{1}{r}{\textbf{\numprint{500079}}} \\ 

\hline
\hline
ssmc & \mc{2}{|c|}{hybrid}  & \mc{2}{c}{weight} \\ 
\hline
\mc{1}{l|}{\textit{ca2010}} & \mc{1}{r}{\numprint{36509.75}} & \mc{1}{r|}{\textbf{\numprint{16843380}}} & \mc{1}{r}{\numprint{36498.48}} & \mc{1}{r}{\numprint{16842694}} \\ 
\mc{1}{l|}{\textit{ga2010}} & \mc{1}{r}{\numprint{31547.16}} & \mc{1}{r|}{\textbf{\numprint{4644286}}} & \mc{1}{r}{\numprint{28515.69}} & \mc{1}{r}{\numprint{4644284}} \\ 
\mc{1}{l|}{\textit{il2010}} & \mc{1}{r}{\numprint{36278.16}} & \mc{1}{r|}{\textbf{\numprint{5985214}}} & \mc{1}{r}{\numprint{36290.48}} & \mc{1}{r}{\numprint{5984378}} \\ 
\mc{1}{l|}{\textit{nh2010}} & \mc{1}{r}{\numprint{6050.72}} & \mc{1}{r|}{\textbf{\numprint{588996}}} & \mc{1}{r}{\numprint{178.27}} & \mc{1}{r}{\numprint{588991}} \\ 
\mc{1}{l|}{\textit{ri2010}} & \mc{1}{r}{\numprint{4467.22}} & \mc{1}{r|}{\numprint{459234}} & \mc{1}{r}{\numprint{9081.21}} & \mc{1}{r}{\textbf{\numprint{459251}}} \\ 

\hline
\hline
overall & \mc{2}{|c|}{hybrid}  & \mc{2}{c}{weight} \\ 
\hline
\# best     & \mc{2}{|r|}{30/35} & \mc{2}{r}{26/35}  \\ 
mean time   & \mc{2}{|r|}{\numprint{10148.31}} & \mc{2}{r}{\numprint{8644.56}} \\ 
\hline
 
\end{tabular} 

	\end{ThreePartTable}
\end{table}

\begin{table*}[h]
	\centering
	\begin{ThreePartTable}
		\caption{Average solution weight $\omega$ and time $t$ in seconds required to compute it for different vertex selection strategies. \textbf{Bold} numbers indicate the best solution among all algorithms. We also report the number of best solutions and the geometric mean running time over all instances.}
		\label{tab: combined_vertex_selection}
		\setlength{\tabcolsep}{1.1ex}
		\small 	\begin{tabular}{lcccccccccc} 
\hline
\mc{1}{l|}{graphs} & \mc{1}{c}{$t$} & \mc{1}{c|}{$w$} & \mc{1}{c}{$t$} & \mc{1}{c|}{$w$} & \mc{1}{c}{$t$} & \mc{1}{c|}{$w$} & \mc{1}{c}{$t$} & \mc{1}{c|}{$w$} & \mc{1}{c}{$t$} & \mc{1}{c}{$w$}\\ 
\hline
fe & \mc{2}{|c|}{hybrid}  & \mc{2}{c|}{weight}  & \mc{2}{c|}{degree}  & \mc{2}{c|}{weight/degree}  & \mc{2}{c}{solution participation} \\ 
\hline
\mc{1}{l|}{\textit{ocean}} & \mc{1}{r}{\numprint{38.67}} & \mc{1}{r|}{\textbf{\numprint{7248581}}} & \mc{1}{r}{\numprint{40.17}} & \mc{1}{r|}{\textbf{\numprint{7248581}}} & \mc{1}{r}{\numprint{37.22}} & \mc{1}{r|}{\textbf{\numprint{7248581}}} & \mc{1}{r}{\numprint{43.19}} & \mc{1}{r|}{\textbf{\numprint{7248581}}} & \mc{1}{r}{\numprint{40.54}} & \mc{1}{r}{\textbf{\numprint{7248581}}} \\ 
\mc{1}{l|}{\textit{sphere}} & \mc{1}{r}{\numprint{36007.65}} & \mc{1}{r|}{\textbf{\numprint{616552}}} & \mc{1}{r}{\numprint{36006.32}} & \mc{1}{r|}{\numprint{616400}} & \mc{1}{r}{\numprint{36007.27}} & \mc{1}{r|}{\numprint{616469}} & \mc{1}{r}{\numprint{18305.97}} & \mc{1}{r|}{\numprint{616400}} & \mc{1}{r}{\numprint{36008.05}} & \mc{1}{r}{\numprint{616358}} \\ 
\mc{1}{l|}{\textit{rotor}} & \mc{1}{r}{\numprint{36051.25}} & \mc{1}{r|}{\textbf{\numprint{2644428}}} & \mc{1}{r}{\numprint{36048.97}} & \mc{1}{r|}{\numprint{2643498}} & \mc{1}{r}{\numprint{36062.19}} & \mc{1}{r|}{\numprint{2643867}} & \mc{1}{r}{\numprint{36048.55}} & \mc{1}{r|}{\numprint{2643685}} & \mc{1}{r}{\numprint{36061.17}} & \mc{1}{r}{\numprint{2643981}} \\ 
\mc{1}{l|}{\textit{pwt}} & \mc{1}{r}{\numprint{36020.62}} & \mc{1}{r|}{\numprint{1174368}} & \mc{1}{r}{\numprint{36011.25}} & \mc{1}{r|}{\textbf{\numprint{1174644}}} & \mc{1}{r}{\numprint{36017.32}} & \mc{1}{r|}{\numprint{1174550}} & \mc{1}{r}{\numprint{36011.27}} & \mc{1}{r|}{\numprint{1174410}} & \mc{1}{r}{\numprint{36014.77}} & \mc{1}{r}{\numprint{1174583}} \\ 
\mc{1}{l|}{\textit{body}} & \mc{1}{r}{\numprint{36008.02}} & \mc{1}{r|}{\numprint{1679750}} & \mc{1}{r}{\numprint{18947.76}} & \mc{1}{r|}{\textbf{\numprint{1679854}}} & \mc{1}{r}{\numprint{34529.38}} & \mc{1}{r|}{\numprint{1679774}} & \mc{1}{r}{\numprint{17445.31}} & \mc{1}{r|}{\numprint{1679784}} & \mc{1}{r}{\numprint{3151.43}} & \mc{1}{r}{\numprint{1679695}} \\ 

\hline
\hline
mesh & \mc{2}{|c|}{hybrid}  & \mc{2}{c|}{weight}  & \mc{2}{c|}{degree}  & \mc{2}{c|}{weight/degree}  & \mc{2}{c}{solution participation} \\ 
\hline
\mc{1}{l|}{\textit{buddha}} & \mc{1}{r}{\numprint{19363.72}} & \mc{1}{r|}{\numprint{57555044}} & \mc{1}{r}{\numprint{26143.17}} & \mc{1}{r|}{\textbf{\numprint{57555070}}} & \mc{1}{r}{\numprint{6134.64}} & \mc{1}{r|}{\numprint{57554994}} & \mc{1}{r}{\numprint{7844.39}} & \mc{1}{r|}{\numprint{57555049}} & \mc{1}{r}{\numprint{422.34}} & \mc{1}{r}{\numprint{57554909}} \\ 
\mc{1}{l|}{\textit{dragon}} & \mc{1}{r}{\numprint{292.29}} & \mc{1}{r|}{\textbf{\numprint{7956524}}} & \mc{1}{r}{\numprint{938.47}} & \mc{1}{r|}{\textbf{\numprint{7956524}}} & \mc{1}{r}{\numprint{250.63}} & \mc{1}{r|}{\numprint{7956523}} & \mc{1}{r}{\numprint{27.25}} & \mc{1}{r|}{\numprint{7956518}} & \mc{1}{r}{\numprint{26.38}} & \mc{1}{r}{\numprint{7956518}} \\ 
\mc{1}{l|}{\textit{ecat}} & \mc{1}{r}{\numprint{14174.83}} & \mc{1}{r|}{\numprint{36650097}} & \mc{1}{r}{\numprint{3096.31}} & \mc{1}{r|}{\numprint{36650048}} & \mc{1}{r}{\numprint{7969.31}} & \mc{1}{r|}{\textbf{\numprint{36650120}}} & \mc{1}{r}{\numprint{23.54}} & \mc{1}{r|}{\numprint{36649991}} & \mc{1}{r}{\numprint{65.50}} & \mc{1}{r}{\numprint{36650009}} \\ 

\hline
\hline
osm & \mc{2}{|c|}{hybrid}  & \mc{2}{c|}{weight}  & \mc{2}{c|}{degree}  & \mc{2}{c|}{weight/degree}  & \mc{2}{c}{solution participation} \\ 
\hline
\mc{1}{l|}{\textit{alabama-2}} & \mc{1}{r}{\numprint{0.31}} & \mc{1}{r|}{\textbf{\numprint{174309}}} & \mc{1}{r}{\numprint{0.33}} & \mc{1}{r|}{\textbf{\numprint{174309}}} & \mc{1}{r}{\numprint{0.29}} & \mc{1}{r|}{\textbf{\numprint{174309}}} & \mc{1}{r}{\numprint{0.32}} & \mc{1}{r|}{\textbf{\numprint{174309}}} & \mc{1}{r}{\numprint{0.32}} & \mc{1}{r}{\textbf{\numprint{174309}}} \\ 
\mc{1}{l|}{\textit{florida-3}} & \mc{1}{r}{\numprint{18.55}} & \mc{1}{r|}{\textbf{\numprint{237333}}} & \mc{1}{r}{\numprint{19.36}} & \mc{1}{r|}{\textbf{\numprint{237333}}} & \mc{1}{r}{\numprint{19.33}} & \mc{1}{r|}{\textbf{\numprint{237333}}} & \mc{1}{r}{\numprint{20.76}} & \mc{1}{r|}{\textbf{\numprint{237333}}} & \mc{1}{r}{\numprint{20.50}} & \mc{1}{r}{\textbf{\numprint{237333}}} \\ 
\mc{1}{l|}{\textit{georgia-3}} & \mc{1}{r}{\numprint{7.24}} & \mc{1}{r|}{\textbf{\numprint{222652}}} & \mc{1}{r}{\numprint{7.88}} & \mc{1}{r|}{\textbf{\numprint{222652}}} & \mc{1}{r}{\numprint{7.96}} & \mc{1}{r|}{\textbf{\numprint{222652}}} & \mc{1}{r}{\numprint{7.74}} & \mc{1}{r|}{\textbf{\numprint{222652}}} & \mc{1}{r}{\numprint{8.12}} & \mc{1}{r}{\textbf{\numprint{222652}}} \\ 
\mc{1}{l|}{\textit{greenland-3}} & \mc{1}{r}{\numprint{9307.76}} & \mc{1}{r|}{\textbf{\numprint{14012}}} & \mc{1}{r}{\numprint{7073.15}} & \mc{1}{r|}{\textbf{\numprint{14012}}} & \mc{1}{r}{\numprint{11670.21}} & \mc{1}{r|}{\textbf{\numprint{14012}}} & \mc{1}{r}{\numprint{6623.16}} & \mc{1}{r|}{\textbf{\numprint{14012}}} & \mc{1}{r}{\numprint{436.13}} & \mc{1}{r}{\numprint{14011}} \\ 
\mc{1}{l|}{\textit{new-hampshire-3}} & \mc{1}{r}{\numprint{2.98}} & \mc{1}{r|}{\textbf{\numprint{116060}}} & \mc{1}{r}{\numprint{2.91}} & \mc{1}{r|}{\textbf{\numprint{116060}}} & \mc{1}{r}{\numprint{3.05}} & \mc{1}{r|}{\textbf{\numprint{116060}}} & \mc{1}{r}{\numprint{3.04}} & \mc{1}{r|}{\textbf{\numprint{116060}}} & \mc{1}{r}{\numprint{3.02}} & \mc{1}{r}{\textbf{\numprint{116060}}} \\ 
\mc{1}{l|}{\textit{rhode-island-2}} & \mc{1}{r}{\numprint{9.06}} & \mc{1}{r|}{\textbf{\numprint{184596}}} & \mc{1}{r}{\numprint{9.77}} & \mc{1}{r|}{\textbf{\numprint{184596}}} & \mc{1}{r}{\numprint{10.42}} & \mc{1}{r|}{\textbf{\numprint{184596}}} & \mc{1}{r}{\numprint{10.47}} & \mc{1}{r|}{\textbf{\numprint{184596}}} & \mc{1}{r}{\numprint{9.74}} & \mc{1}{r}{\textbf{\numprint{184596}}} \\ 
\mc{1}{l|}{\textit{utah-3}} & \mc{1}{r}{\numprint{5.67}} & \mc{1}{r|}{\textbf{\numprint{98847}}} & \mc{1}{r}{\numprint{5.65}} & \mc{1}{r|}{\textbf{\numprint{98847}}} & \mc{1}{r}{\numprint{4.97}} & \mc{1}{r|}{\textbf{\numprint{98847}}} & \mc{1}{r}{\numprint{5.70}} & \mc{1}{r|}{\textbf{\numprint{98847}}} & \mc{1}{r}{\numprint{4.81}} & \mc{1}{r}{\textbf{\numprint{98847}}} \\ 

\hline
\hline
snap & \mc{2}{|c|}{hybrid}  & \mc{2}{c|}{weight}  & \mc{2}{c|}{degree}  & \mc{2}{c|}{weight/degree}  & \mc{2}{c}{solution participation} \\ 
\hline
\mc{1}{l|}{\textit{as-skitter}} & \mc{1}{r}{\numprint{663.07}} & \mc{1}{r|}{\numprint{123993350}} & \mc{1}{r}{\numprint{10510.85}} & \mc{1}{r|}{\textbf{\numprint{123993368}}} & \mc{1}{r}{\numprint{21898.14}} & \mc{1}{r|}{\numprint{123993360}} & \mc{1}{r}{\numprint{19196.14}} & \mc{1}{r|}{\numprint{123993367}} & \mc{1}{r}{\numprint{92.74}} & \mc{1}{r}{\numprint{123993310}} \\ 
\mc{1}{l|}{\textit{ca-AstroPh}} & \mc{1}{r}{\numprint{19.41}} & \mc{1}{r|}{\textbf{\numprint{797510}}} & \mc{1}{r}{\numprint{19.33}} & \mc{1}{r|}{\textbf{\numprint{797510}}} & \mc{1}{r}{\numprint{19.21}} & \mc{1}{r|}{\textbf{\numprint{797510}}} & \mc{1}{r}{\numprint{18.38}} & \mc{1}{r|}{\textbf{\numprint{797510}}} & \mc{1}{r}{\numprint{18.63}} & \mc{1}{r}{\textbf{\numprint{797510}}} \\ 
\mc{1}{l|}{\textit{ca-CondMat}} & \mc{1}{r}{\numprint{18.93}} & \mc{1}{r|}{\textbf{\numprint{1145292}}} & \mc{1}{r}{\numprint{18.69}} & \mc{1}{r|}{\textbf{\numprint{1145292}}} & \mc{1}{r}{\numprint{18.48}} & \mc{1}{r|}{\textbf{\numprint{1145292}}} & \mc{1}{r}{\numprint{17.96}} & \mc{1}{r|}{\textbf{\numprint{1145292}}} & \mc{1}{r}{\numprint{16.85}} & \mc{1}{r}{\textbf{\numprint{1145292}}} \\ 
\mc{1}{l|}{\textit{ca-GrQc}} & \mc{1}{r}{\numprint{3.61}} & \mc{1}{r|}{\textbf{\numprint{289356}}} & \mc{1}{r}{\numprint{3.51}} & \mc{1}{r|}{\textbf{\numprint{289356}}} & \mc{1}{r}{\numprint{3.57}} & \mc{1}{r|}{\textbf{\numprint{289356}}} & \mc{1}{r}{\numprint{3.21}} & \mc{1}{r|}{\textbf{\numprint{289356}}} & \mc{1}{r}{\numprint{3.27}} & \mc{1}{r}{\textbf{\numprint{289356}}} \\ 
\mc{1}{l|}{\textit{com-amazon}} & \mc{1}{r}{\numprint{1.09}} & \mc{1}{r|}{\textbf{\numprint{19271031}}} & \mc{1}{r}{\numprint{1.08}} & \mc{1}{r|}{\textbf{\numprint{19271031}}} & \mc{1}{r}{\numprint{1.01}} & \mc{1}{r|}{\textbf{\numprint{19271031}}} & \mc{1}{r}{\numprint{1.09}} & \mc{1}{r|}{\textbf{\numprint{19271031}}} & \mc{1}{r}{\numprint{0.96}} & \mc{1}{r}{\textbf{\numprint{19271031}}} \\ 
\mc{1}{l|}{\textit{com-youtube}} & \mc{1}{r}{\numprint{1.06}} & \mc{1}{r|}{\textbf{\numprint{90295294}}} & \mc{1}{r}{\numprint{1.17}} & \mc{1}{r|}{\textbf{\numprint{90295294}}} & \mc{1}{r}{\numprint{1.12}} & \mc{1}{r|}{\textbf{\numprint{90295294}}} & \mc{1}{r}{\numprint{1.14}} & \mc{1}{r|}{\textbf{\numprint{90295294}}} & \mc{1}{r}{\numprint{1.15}} & \mc{1}{r}{\textbf{\numprint{90295294}}} \\ 
\mc{1}{l|}{\textit{email-Enron}} & \mc{1}{r}{\numprint{21.46}} & \mc{1}{r|}{\textbf{\numprint{2457578}}} & \mc{1}{r}{\numprint{23.24}} & \mc{1}{r|}{\textbf{\numprint{2457578}}} & \mc{1}{r}{\numprint{22.28}} & \mc{1}{r|}{\textbf{\numprint{2457578}}} & \mc{1}{r}{\numprint{21.66}} & \mc{1}{r|}{\textbf{\numprint{2457578}}} & \mc{1}{r}{\numprint{21.57}} & \mc{1}{r}{\textbf{\numprint{2457578}}} \\ 
\mc{1}{l|}{\textit{email-EuAll}} & \mc{1}{r}{\numprint{70.80}} & \mc{1}{r|}{\textbf{\numprint{25286322}}} & \mc{1}{r}{\numprint{66.26}} & \mc{1}{r|}{\textbf{\numprint{25286322}}} & \mc{1}{r}{\numprint{74.46}} & \mc{1}{r|}{\textbf{\numprint{25286322}}} & \mc{1}{r}{\numprint{63.01}} & \mc{1}{r|}{\textbf{\numprint{25286322}}} & \mc{1}{r}{\numprint{69.86}} & \mc{1}{r}{\textbf{\numprint{25286322}}} \\ 
\mc{1}{l|}{\textit{loc-gowalla\_edges}} & \mc{1}{r}{\numprint{3.71}} & \mc{1}{r|}{\textbf{\numprint{12276929}}} & \mc{1}{r}{\numprint{4.13}} & \mc{1}{r|}{\textbf{\numprint{12276929}}} & \mc{1}{r}{\numprint{4.28}} & \mc{1}{r|}{\textbf{\numprint{12276929}}} & \mc{1}{r}{\numprint{4.14}} & \mc{1}{r|}{\textbf{\numprint{12276929}}} & \mc{1}{r}{\numprint{4.10}} & \mc{1}{r}{\textbf{\numprint{12276929}}} \\ 
\mc{1}{l|}{\textit{p2p-Gnutella06}} & \mc{1}{r}{\numprint{3.65}} & \mc{1}{r|}{\textbf{\numprint{548612}}} & \mc{1}{r}{\numprint{3.72}} & \mc{1}{r|}{\textbf{\numprint{548612}}} & \mc{1}{r}{\numprint{3.66}} & \mc{1}{r|}{\textbf{\numprint{548612}}} & \mc{1}{r}{\numprint{3.67}} & \mc{1}{r|}{\textbf{\numprint{548612}}} & \mc{1}{r}{\numprint{3.64}} & \mc{1}{r}{\textbf{\numprint{548612}}} \\ 
\mc{1}{l|}{\textit{roadNet-PA}} & \mc{1}{r}{\numprint{34376.97}} & \mc{1}{r|}{\textbf{\numprint{61707635}}} & \mc{1}{r}{\numprint{7348.31}} & \mc{1}{r|}{\numprint{61707560}} & \mc{1}{r}{\numprint{13063.29}} & \mc{1}{r|}{\numprint{61707608}} & \mc{1}{r}{\numprint{2805.07}} & \mc{1}{r|}{\numprint{61707491}} & \mc{1}{r}{\numprint{119.02}} & \mc{1}{r}{\numprint{61707493}} \\ 
\mc{1}{l|}{\textit{web-BerkStan}} & \mc{1}{r}{\numprint{17832.98}} & \mc{1}{r|}{\numprint{43916421}} & \mc{1}{r}{\numprint{9630.48}} & \mc{1}{r|}{\numprint{43916363}} & \mc{1}{r}{\numprint{21836.54}} & \mc{1}{r|}{\textbf{\numprint{43916464}}} & \mc{1}{r}{\numprint{3111.59}} & \mc{1}{r|}{\numprint{43916383}} & \mc{1}{r}{\numprint{185.84}} & \mc{1}{r}{\numprint{43916286}} \\ 
\mc{1}{l|}{\textit{web-Google}} & \mc{1}{r}{\numprint{3.18}} & \mc{1}{r|}{\textbf{\numprint{56326504}}} & \mc{1}{r}{\numprint{3.71}} & \mc{1}{r|}{\textbf{\numprint{56326504}}} & \mc{1}{r}{\numprint{3.78}} & \mc{1}{r|}{\textbf{\numprint{56326504}}} & \mc{1}{r}{\numprint{3.75}} & \mc{1}{r|}{\textbf{\numprint{56326504}}} & \mc{1}{r}{\numprint{3.75}} & \mc{1}{r}{\textbf{\numprint{56326504}}} \\ 
\mc{1}{l|}{\textit{web-NotreDame}} & \mc{1}{r}{\numprint{4.69}} & \mc{1}{r|}{\textbf{\numprint{25965069}}} & \mc{1}{r}{\numprint{4.66}} & \mc{1}{r|}{\textbf{\numprint{25965069}}} & \mc{1}{r}{\numprint{4.55}} & \mc{1}{r|}{\textbf{\numprint{25965069}}} & \mc{1}{r}{\numprint{4.74}} & \mc{1}{r|}{\textbf{\numprint{25965069}}} & \mc{1}{r}{\numprint{4.53}} & \mc{1}{r}{\textbf{\numprint{25965069}}} \\ 
\mc{1}{l|}{\textit{wiki-Vote}} & \mc{1}{r}{\numprint{4.66}} & \mc{1}{r|}{\textbf{\numprint{500079}}} & \mc{1}{r}{\numprint{4.92}} & \mc{1}{r|}{\textbf{\numprint{500079}}} & \mc{1}{r}{\numprint{4.91}} & \mc{1}{r|}{\textbf{\numprint{500079}}} & \mc{1}{r}{\numprint{4.92}} & \mc{1}{r|}{\textbf{\numprint{500079}}} & \mc{1}{r}{\numprint{4.90}} & \mc{1}{r}{\textbf{\numprint{500079}}} \\ 

\hline
\hline
ssmc & \mc{2}{|c|}{hybrid}  & \mc{2}{c|}{weight}  & \mc{2}{c|}{degree}  & \mc{2}{c|}{weight/degree}  & \mc{2}{c}{solution participation} \\ 
\hline
\mc{1}{l|}{\textit{ca2010}} & \mc{1}{r}{\numprint{36509.75}} & \mc{1}{r|}{\numprint{16843380}} & \mc{1}{r}{\numprint{36498.48}} & \mc{1}{r|}{\numprint{16842694}} & \mc{1}{r}{\numprint{36526.09}} & \mc{1}{r|}{\textbf{\numprint{16843911}}} & \mc{1}{r}{\numprint{36523.53}} & \mc{1}{r|}{\numprint{16843123}} & \mc{1}{r}{\numprint{36512.31}} & \mc{1}{r}{\numprint{16843769}} \\ 
\mc{1}{l|}{\textit{ga2010}} & \mc{1}{r}{\numprint{31547.16}} & \mc{1}{r|}{\numprint{4644286}} & \mc{1}{r}{\numprint{28515.69}} & \mc{1}{r|}{\numprint{4644284}} & \mc{1}{r}{\numprint{18639.77}} & \mc{1}{r|}{\textbf{\numprint{4644295}}} & \mc{1}{r}{\numprint{13474.85}} & \mc{1}{r|}{\numprint{4644294}} & \mc{1}{r}{\numprint{350.83}} & \mc{1}{r}{\numprint{4644235}} \\ 
\mc{1}{l|}{\textit{il2010}} & \mc{1}{r}{\numprint{36278.16}} & \mc{1}{r|}{\textbf{\numprint{5985214}}} & \mc{1}{r}{\numprint{36290.48}} & \mc{1}{r|}{\numprint{5984378}} & \mc{1}{r}{\numprint{36267.74}} & \mc{1}{r|}{\numprint{5985084}} & \mc{1}{r}{\numprint{36284.69}} & \mc{1}{r|}{\numprint{5984593}} & \mc{1}{r}{\numprint{36288.38}} & \mc{1}{r}{\numprint{5984524}} \\ 
\mc{1}{l|}{\textit{nh2010}} & \mc{1}{r}{\numprint{6050.72}} & \mc{1}{r|}{\textbf{\numprint{588996}}} & \mc{1}{r}{\numprint{178.27}} & \mc{1}{r|}{\numprint{588991}} & \mc{1}{r}{\numprint{6168.09}} & \mc{1}{r|}{\numprint{588995}} & \mc{1}{r}{\numprint{5910.69}} & \mc{1}{r|}{\numprint{588994}} & \mc{1}{r}{\numprint{32.06}} & \mc{1}{r}{\numprint{588989}} \\ 
\mc{1}{l|}{\textit{ri2010}} & \mc{1}{r}{\numprint{4467.22}} & \mc{1}{r|}{\numprint{459234}} & \mc{1}{r}{\numprint{9081.21}} & \mc{1}{r|}{\textbf{\numprint{459251}}} & \mc{1}{r}{\numprint{21563.71}} & \mc{1}{r|}{\numprint{459250}} & \mc{1}{r}{\numprint{1649.44}} & \mc{1}{r|}{\numprint{459222}} & \mc{1}{r}{\numprint{27.97}} & \mc{1}{r}{\numprint{459192}} \\ 

\hline
\hline
overall & \mc{2}{|c|}{hybrid}  & \mc{2}{c|}{weight}  & \mc{2}{c|}{degree}  & \mc{2}{c|}{weight/degree}  & \mc{2}{c}{solution participation} \\ 
\hline
\# best     & \mc{2}{|r|}{26/35}      & \mc{2}{r|}{26/35} & \mc{2}{r|}{24/35} & \mc{2}{r|}{20/35} & \mc{2}{r}{19/35} \\ 
mean time   & \mc{2}{|r|}{\numprint{10148.31}} & \mc{2}{r|}{\numprint{8644.56}} & \mc{2}{r|}{\numprint{9852.82}} & \mc{2}{r|}{\numprint{6900.7}} & \mc{2}{r}{\numprint{5315.29}} \\
\hline
\end{tabular} 

	\end{ThreePartTable}
\end{table*}

\clearpage
\onecolumn
\section{Additional Detailed Per-Instance Results}

\begin{table*}[h]
	\centering
	\begin{ThreePartTable}
		\caption{Average solution weight $\omega$ and time $t$ in seconds required to compute it for our set of fe instances. \textbf{Bold} numbers indicate the best solution among all algorithms. Rows have a \noindent\colorbox{lightergray} {\parbox{\widthof{gray}}{gray}} background color, if branch reduce or struction computed an exact solution. We also report the number of best solutions and the geometric mean running time over all instances. }\label{tab: soa_fe}
		\setlength{\tabcolsep}{1.1ex}
		\small 	\begin{tabular}{lcccccccccccc} 
 \hline 
\mc{1}{l|}{graphs} & \mc{1}{c}{$t$} & \mc{1}{c|}{$w$} & \mc{1}{c}{$t$} & \mc{1}{c|}{$w$} & \mc{1}{c}{$t$} & \mc{1}{c|}{$w$} & \mc{1}{c}{$t$} & \mc{1}{c|}{$w$} & \mc{1}{c}{$t$} & \mc{1}{c|}{$w$} & \mc{1}{c}{$t$} & \mc{1}{c}{$w$}\\ 
 \hline 
\mc{1}{l|}{fe} & \mc{2}{c|}{branch reduce}  & \mc{2}{c|}{{\hils}}  & \mc{2}{c|}{{\htwis}}  & \mc{2}{c|}{{\wmmis}}  & \mc{2}{c|}{{\wmmiss}}  & \mc{2}{c}{struction} \\ 
\hline 
\mc{1}{l|}{\textit{body}} & \mc{2}{c|}{-} & \mc{1}{r}{\numprint{1259.67}} & \mc{1}{r|}{\numprint{1678510}} & \mc{1}{r}{\numprint{0.04}} & \mc{1}{r|}{\numprint{1645650}} & \mc{1}{r}{\numprint{29242.39}} & \mc{1}{r|}{\numprint{1679807}} & \mc{1}{r}{\numprint{96.95}} & \mc{1}{r|}{\textbf{\numprint{1680166}}} & \mc{2}{c}{-} \\ 
 \rowcolor{lightergray} \mc{1}{l|}{\textit{ocean}} & \mc{1}{r}{\numprint{4.88}} & \mc{1}{r|}{\textbf{\numprint{7248581}}} & \mc{1}{r}{\numprint{11142.43}} & \mc{1}{r|}{\numprint{7075329}} & \mc{1}{r}{\numprint{0.07}} & \mc{1}{r|}{\numprint{6803672}} & \mc{1}{r}{\numprint{44.58}} & \mc{1}{r|}{\textbf{\numprint{7248581}}} & \mc{1}{r}{\numprint{50.49}} & \mc{1}{r|}{\textbf{\numprint{7248581}}} & \mc{2}{c}{-} \\ 
\mc{1}{l|}{\textit{pwt}} & \mc{2}{c|}{-} & \mc{1}{r}{\numprint{761.52}} & \mc{1}{r|}{\numprint{1175437}} & \mc{1}{r}{\numprint{0.03}} & \mc{1}{r|}{\numprint{1153600}} & \mc{1}{r}{\numprint{36050.99}} & \mc{1}{r|}{\numprint{1175149}} & \mc{1}{r}{\numprint{7590.48}} & \mc{1}{r|}{\textbf{\numprint{1178434}}} & \mc{2}{c}{-} \\ 
\mc{1}{l|}{\textit{rotor}} & \mc{2}{c|}{-} & \mc{1}{r}{\numprint{6503.27}} & \mc{1}{r|}{\textbf{\numprint{2650018}}} & \mc{1}{r}{\numprint{0.24}} & \mc{1}{r|}{\numprint{2591456}} & \mc{1}{r}{\numprint{36169.87}} & \mc{1}{r|}{\numprint{2643433}} & \mc{1}{r}{\numprint{36219.34}} & \mc{1}{r|}{\numprint{2642600}} & \mc{2}{c}{-} \\ 
 \rowcolor{lightergray} \mc{1}{l|}{\textit{sphere}} & \mc{2}{c|}{-} & \mc{1}{r}{\numprint{257.42}} & \mc{1}{r|}{\numprint{615958}} & \mc{1}{r}{\numprint{0.02}} & \mc{1}{r|}{\numprint{608401}} & \mc{1}{r}{\numprint{36007.30}} & \mc{1}{r|}{\numprint{616663}} & \mc{1}{r}{\numprint{6.08}} & \mc{1}{r|}{\textbf{\numprint{617816}}} & \mc{1}{r}{\numprint{0.57}} & \mc{1}{r}{\textbf{\numprint{617816}}} \\ 
\hline 
\mc{1}{l|}{overall} & \mc{2}{c|}{branch reduce}  & \mc{2}{c|}{{\hils}}  & \mc{2}{c|}{{\htwis}}  & \mc{2}{c|}{{\wmmis}}  & \mc{2}{c|}{{\wmmiss}}  & \mc{2}{c}{struction} \\ 
\hline 
\mc{1}{l|}{\# best} & \mc{2}{r|}{1/5} & \mc{2}{r|}{1/5} & \mc{2}{r|}{0/5} & \mc{2}{r|}{1/5} & \mc{2}{r|}{4/5} & \mc{2}{r}{1/5}\\ 
\mc{1}{l|}{mean time} & \mc{2}{r|}{-} & \mc{2}{r|}{\numprint{1780.48}} & \mc{2}{r|}{\numprint{0.05}} & \mc{2}{r|}{\numprint{9064.97}} & \mc{2}{r|}{\numprint{382.46}} & \mc{2}{r}{-}\\ 
\hline 

\end{tabular} 

	\end{ThreePartTable}
\end{table*}
\
\begin{table*}[h]
	\centering
	\begin{ThreePartTable}
		\caption{Average solution weight $\omega$ and time $t$ in seconds required to compute it for our set of mesh instances. \textbf{Bold} numbers indicate the best solution among all algorithms. Rows have a \noindent\colorbox{lightergray} {\parbox{\widthof{gray}}{gray}} background color, if branch reduce or struction computed an exact solution. We also report the number of best solutions and the geometric mean running time over all instances. }\label{tab: soa_mesh}
		\setlength{\tabcolsep}{1.1ex}
		\small 	\begin{tabular}{lcccccccccccc} 
 \hline 
\mc{1}{l|}{graphs} & \mc{1}{c}{$t$} & \mc{1}{c|}{$w$} & \mc{1}{c}{$t$} & \mc{1}{c|}{$w$} & \mc{1}{c}{$t$} & \mc{1}{c|}{$w$} & \mc{1}{c}{$t$} & \mc{1}{c|}{$w$} & \mc{1}{c}{$t$} & \mc{1}{c|}{$w$} & \mc{1}{c}{$t$} & \mc{1}{c}{$w$}\\ 
 \hline 
\mc{1}{l|}{mesh} & \mc{2}{c|}{branch reduce}  & \mc{2}{c|}{{\hils}}  & \mc{2}{c|}{{\htwis}}  & \mc{2}{c|}{{\wmmis}}  & \mc{2}{c|}{{\wmmiss}}  & \mc{2}{c}{struction} \\ 
\hline 
 \rowcolor{lightergray} \mc{1}{l|}{\textit{blob}} & \mc{1}{r}{\numprint{0.14}} & \mc{1}{r|}{\textbf{\numprint{855547}}} & \mc{1}{r}{\numprint{260.10}} & \mc{1}{r|}{\numprint{854803}} & \mc{1}{r}{\numprint{0.01}} & \mc{1}{r|}{\numprint{854484}} & \mc{1}{r}{\numprint{0.23}} & \mc{1}{r|}{\textbf{\numprint{855547}}} & \mc{1}{r}{\numprint{0.13}} & \mc{1}{r|}{\textbf{\numprint{855547}}} & \mc{1}{r}{\numprint{0.02}} & \mc{1}{r}{\textbf{\numprint{855547}}} \\ 
 \rowcolor{lightergray} \mc{1}{l|}{\textit{buddha}} & \mc{1}{r}{\numprint{51.87}} & \mc{1}{r|}{\textbf{\numprint{57555880}}} & \mc{1}{r}{\numprint{36000.07}} & \mc{1}{r|}{\numprint{57258790}} & \mc{1}{r}{\numprint{0.47}} & \mc{1}{r|}{\numprint{57508556}} & \mc{1}{r}{\numprint{30351.04}} & \mc{1}{r|}{\numprint{57555105}} & \mc{1}{r}{\numprint{9.14}} & \mc{1}{r|}{\textbf{\numprint{57555880}}} & \mc{1}{r}{\numprint{1.57}} & \mc{1}{r}{\textbf{\numprint{57555880}}} \\ 
 \rowcolor{lightergray} \mc{1}{l|}{\textit{bunny}} & \mc{1}{r}{\numprint{0.48}} & \mc{1}{r|}{\textbf{\numprint{3686960}}} & \mc{1}{r}{\numprint{1927.84}} & \mc{1}{r|}{\numprint{3680587}} & \mc{1}{r}{\numprint{0.03}} & \mc{1}{r|}{\numprint{3682356}} & \mc{1}{r}{\numprint{0.77}} & \mc{1}{r|}{\textbf{\numprint{3686960}}} & \mc{1}{r}{\numprint{0.43}} & \mc{1}{r|}{\textbf{\numprint{3686960}}} & \mc{1}{r}{\numprint{0.09}} & \mc{1}{r}{\textbf{\numprint{3686960}}} \\ 
 \rowcolor{lightergray} \mc{1}{l|}{\textit{cow}} & \mc{1}{r}{\numprint{0.04}} & \mc{1}{r|}{\textbf{\numprint{269543}}} & \mc{1}{r}{\numprint{52.05}} & \mc{1}{r|}{\numprint{269336}} & \mc{1}{r}{<\numprint{0.01}} & \mc{1}{r|}{\numprint{269304}} & \mc{1}{r}{\numprint{0.15}} & \mc{1}{r|}{\textbf{\numprint{269543}}} & \mc{1}{r}{\numprint{0.08}} & \mc{1}{r|}{\textbf{\numprint{269543}}} & \mc{1}{r}{\numprint{0.01}} & \mc{1}{r}{\textbf{\numprint{269543}}} \\ 
 \rowcolor{lightergray} \mc{1}{l|}{\textit{dragon}} & \mc{1}{r}{\numprint{2.90}} & \mc{1}{r|}{\textbf{\numprint{7956530}}} & \mc{1}{r}{\numprint{9026.60}} & \mc{1}{r|}{\numprint{7947535}} & \mc{1}{r}{\numprint{0.04}} & \mc{1}{r|}{\numprint{7950526}} & \mc{1}{r}{\numprint{674.94}} & \mc{1}{r|}{\numprint{7956523}} & \mc{1}{r}{\numprint{0.82}} & \mc{1}{r|}{\textbf{\numprint{7956530}}} & \mc{1}{r}{\numprint{0.17}} & \mc{1}{r}{\textbf{\numprint{7956530}}} \\ 
 \rowcolor{lightergray} \mc{1}{l|}{\textit{dragonsub}} & \mc{1}{r}{\numprint{4.76}} & \mc{1}{r|}{\textbf{\numprint{32213898}}} & \mc{1}{r}{\numprint{36000.07}} & \mc{1}{r|}{\numprint{32148544}} & \mc{1}{r}{\numprint{0.24}} & \mc{1}{r|}{\numprint{32163872}} & \mc{1}{r}{\numprint{30066.25}} & \mc{1}{r|}{\numprint{32213637}} & \mc{1}{r}{\numprint{3.73}} & \mc{1}{r|}{\textbf{\numprint{32213898}}} & \mc{1}{r}{\numprint{0.93}} & \mc{1}{r}{\textbf{\numprint{32213898}}} \\ 
 \rowcolor{lightergray} \mc{1}{l|}{\textit{ecat}} & \mc{1}{r}{\numprint{9.18}} & \mc{1}{r|}{\textbf{\numprint{36650298}}} & \mc{1}{r}{\numprint{36000.05}} & \mc{1}{r|}{\numprint{36562652}} & \mc{1}{r}{\numprint{0.50}} & \mc{1}{r|}{\numprint{36606394}} & \mc{1}{r}{\numprint{11626.77}} & \mc{1}{r|}{\numprint{36650108}} & \mc{1}{r}{\numprint{5.24}} & \mc{1}{r|}{\textbf{\numprint{36650298}}} & \mc{1}{r}{\numprint{1.92}} & \mc{1}{r}{\textbf{\numprint{36650298}}} \\ 
 \rowcolor{lightergray} \mc{1}{l|}{\textit{face}} & \mc{1}{r}{\numprint{0.16}} & \mc{1}{r|}{\textbf{\numprint{1219418}}} & \mc{1}{r}{\numprint{403.52}} & \mc{1}{r|}{\numprint{1218433}} & \mc{1}{r}{\numprint{0.01}} & \mc{1}{r|}{\numprint{1218515}} & \mc{1}{r}{\numprint{0.25}} & \mc{1}{r|}{\textbf{\numprint{1219418}}} & \mc{1}{r}{\numprint{0.14}} & \mc{1}{r|}{\textbf{\numprint{1219418}}} & \mc{1}{r}{\numprint{0.02}} & \mc{1}{r}{\textbf{\numprint{1219418}}} \\ 
 \rowcolor{lightergray} \mc{1}{l|}{\textit{fandisk}} & \mc{1}{r}{\numprint{0.04}} & \mc{1}{r|}{\textbf{\numprint{463288}}} & \mc{1}{r}{\numprint{114.01}} & \mc{1}{r|}{\numprint{462794}} & \mc{1}{r}{<\numprint{0.01}} & \mc{1}{r|}{\numprint{462765}} & \mc{1}{r}{\numprint{0.14}} & \mc{1}{r|}{\textbf{\numprint{463288}}} & \mc{1}{r}{\numprint{0.09}} & \mc{1}{r|}{\textbf{\numprint{463288}}} & \mc{1}{r}{\numprint{0.01}} & \mc{1}{r}{\textbf{\numprint{463288}}} \\ 
 \rowcolor{lightergray} \mc{1}{l|}{\textit{feline}} & \mc{1}{r}{\numprint{0.34}} & \mc{1}{r|}{\textbf{\numprint{2207219}}} & \mc{1}{r}{\numprint{794.54}} & \mc{1}{r|}{\numprint{2204454}} & \mc{1}{r}{\numprint{0.02}} & \mc{1}{r|}{\numprint{2204947}} & \mc{1}{r}{\numprint{0.86}} & \mc{1}{r|}{\textbf{\numprint{2207219}}} & \mc{1}{r}{\numprint{0.27}} & \mc{1}{r|}{\textbf{\numprint{2207219}}} & \mc{1}{r}{\numprint{0.05}} & \mc{1}{r}{\textbf{\numprint{2207219}}} \\ 
 \rowcolor{lightergray} \mc{1}{l|}{\textit{gameguy}} & \mc{1}{r}{\numprint{0.10}} & \mc{1}{r|}{\textbf{\numprint{2325878}}} & \mc{1}{r}{\numprint{789.78}} & \mc{1}{r|}{\numprint{2322814}} & \mc{1}{r}{\numprint{0.02}} & \mc{1}{r|}{\numprint{2324088}} & \mc{1}{r}{\numprint{0.20}} & \mc{1}{r|}{\textbf{\numprint{2325878}}} & \mc{1}{r}{\numprint{0.13}} & \mc{1}{r|}{\textbf{\numprint{2325878}}} & \mc{1}{r}{\numprint{0.04}} & \mc{1}{r}{\textbf{\numprint{2325878}}} \\ 
 \rowcolor{lightergray} \mc{1}{l|}{\textit{gargoyle}} & \mc{1}{r}{\numprint{0.22}} & \mc{1}{r|}{\textbf{\numprint{1059559}}} & \mc{1}{r}{\numprint{346.19}} & \mc{1}{r|}{\numprint{1058536}} & \mc{1}{r}{\numprint{0.01}} & \mc{1}{r|}{\numprint{1058656}} & \mc{1}{r}{\numprint{0.36}} & \mc{1}{r|}{\textbf{\numprint{1059559}}} & \mc{1}{r}{\numprint{0.17}} & \mc{1}{r|}{\textbf{\numprint{1059559}}} & \mc{1}{r}{\numprint{0.02}} & \mc{1}{r}{\textbf{\numprint{1059559}}} \\ 
 \rowcolor{lightergray} \mc{1}{l|}{\textit{turtle}} & \mc{1}{r}{\numprint{3.98}} & \mc{1}{r|}{\textbf{\numprint{14263005}}} & \mc{1}{r}{\numprint{20430.40}} & \mc{1}{r|}{\numprint{14245854}} & \mc{1}{r}{\numprint{0.09}} & \mc{1}{r|}{\numprint{14247883}} & \mc{1}{r}{\numprint{1631.40}} & \mc{1}{r|}{\numprint{14263001}} & \mc{1}{r}{\numprint{1.48}} & \mc{1}{r|}{\textbf{\numprint{14263005}}} & \mc{1}{r}{\numprint{0.35}} & \mc{1}{r}{\textbf{\numprint{14263005}}} \\ 
 \rowcolor{lightergray} \mc{1}{l|}{\textit{venus}} & \mc{1}{r}{\numprint{0.02}} & \mc{1}{r|}{\textbf{\numprint{305749}}} & \mc{1}{r}{\numprint{59.48}} & \mc{1}{r|}{\numprint{305556}} & \mc{1}{r}{<\numprint{0.01}} & \mc{1}{r|}{\numprint{305182}} & \mc{1}{r}{\numprint{0.13}} & \mc{1}{r|}{\textbf{\numprint{305749}}} & \mc{1}{r}{\numprint{0.07}} & \mc{1}{r|}{\textbf{\numprint{305749}}} & \mc{1}{r}{\numprint{0.01}} & \mc{1}{r}{\textbf{\numprint{305749}}} \\ 
\hline 
\mc{1}{l|}{overall} & \mc{2}{c|}{branch reduce}  & \mc{2}{c|}{{\hils}}  & \mc{2}{c|}{{\htwis}}  & \mc{2}{c|}{{\wmmis}}  & \mc{2}{c|}{{\wmmiss}}  & \mc{2}{c}{struction} \\ 
\hline 
\mc{1}{l|}{\# best} & \mc{2}{r|}{14/14} & \mc{2}{r|}{0/14} & \mc{2}{r|}{0/14} & \mc{2}{r|}{9/14} & \mc{2}{r|}{14/14} & \mc{2}{r}{14/14}\\ 
\mc{1}{l|}{mean time} & \mc{2}{r|}{\numprint{0.50}} & \mc{2}{r|}{\numprint{1418.48}} & \mc{2}{r|}{\numprint{0.03}} & \mc{2}{r|}{\numprint{9.92}} & \mc{2}{r|}{\numprint{0.42}} & \mc{2}{r}{\numprint{0.07}}\\ 
\hline 

\end{tabular} 

	\end{ThreePartTable}
\end{table*} 

\begin{table*}
	\centering
	\begin{ThreePartTable} 
		\caption{Average solution weight $\omega$ and time $t$ in seconds required to compute it for our set of osm instances. \textbf{Bold} numbers indicate the best solution among all algorithms. Rows have a \noindent\colorbox{lightergray} {\parbox{\widthof{gray}}{gray}} background color, if branch reduce or struction computed an exact solution. We also report the number of best solutions and the geometric mean running time over all instances. }\label{tab: soa_osm}
		\setlength{\tabcolsep}{1.1ex}
		\small 	
\begin{tabular}{lcccccccccccc} 
 \hline 
\mc{1}{l|}{graphs} & \mc{1}{c}{$t$} & \mc{1}{c|}{$w$} & \mc{1}{c}{$t$} & \mc{1}{c|}{$w$} & \mc{1}{c}{$t$} & \mc{1}{c|}{$w$} & \mc{1}{c}{$t$} & \mc{1}{c|}{$w$} & \mc{1}{c}{$t$} & \mc{1}{c|}{$w$} & \mc{1}{c}{$t$} & \mc{1}{c}{$w$}\\ 
 \hline 
\mc{1}{l|}{osm} & \mc{2}{c|}{branch reduce}  & \mc{2}{c|}{{\hils}}  & \mc{2}{c|}{{\htwis}}  & \mc{2}{c|}{{\wmmis}}  & \mc{2}{c|}{{\wmmiss}}  & \mc{2}{c}{struction} \\ 
\hline 
 \rowcolor{lightergray} \mc{1}{l|}{\textit{alabama-1}} & \mc{1}{r}{<\numprint{0.01}} & \mc{1}{r|}{\textbf{\numprint{167588}}} & \mc{1}{r}{\numprint{5.60}} & \mc{1}{r|}{\textbf{\numprint{167588}}} & \mc{1}{r}{<\numprint{0.01}} & \mc{1}{r|}{\textbf{\numprint{167588}}} & \mc{1}{r}{\numprint{0.14}} & \mc{1}{r|}{\textbf{\numprint{167588}}} & \mc{1}{r}{\numprint{0.14}} & \mc{1}{r|}{\textbf{\numprint{167588}}} & \mc{1}{r}{<\numprint{0.01}} & \mc{1}{r}{\textbf{\numprint{167588}}} \\ 
 \rowcolor{lightergray} \mc{1}{l|}{\textit{alabama-2}} & \mc{1}{r}{\numprint{0.35}} & \mc{1}{r|}{\textbf{\numprint{174309}}} & \mc{1}{r}{\numprint{41.71}} & \mc{1}{r|}{\textbf{\numprint{174309}}} & \mc{1}{r}{\numprint{0.01}} & \mc{1}{r|}{\numprint{172797}} & \mc{1}{r}{\numprint{0.44}} & \mc{1}{r|}{\textbf{\numprint{174309}}} & \mc{1}{r}{\numprint{0.28}} & \mc{1}{r|}{\textbf{\numprint{174309}}} & \mc{1}{r}{\numprint{0.01}} & \mc{1}{r}{\textbf{\numprint{174309}}} \\ 
 \rowcolor{lightergray} \mc{1}{l|}{\textit{alabama-3}} & \mc{1}{r}{\numprint{36000.00}} & \mc{1}{r|}{\numprint{185707}} & \mc{1}{r}{\numprint{321.23}} & \mc{1}{r|}{\textbf{\numprint{185744}}} & \mc{1}{r}{\numprint{0.40}} & \mc{1}{r|}{\numprint{182667}} & \mc{1}{r}{\numprint{21.49}} & \mc{1}{r|}{\textbf{\numprint{185744}}} & \mc{1}{r}{\numprint{56.06}} & \mc{1}{r|}{\textbf{\numprint{185744}}} & \mc{1}{r}{\numprint{1.53}} & \mc{1}{r}{\textbf{\numprint{185744}}} \\
\hline

\end{tabular} 

	\end{ThreePartTable}
\end{table*}

\begin{table*}
	\begin{ThreePartTable} 
		\caption*{Table \ref{tab: soa_osm} continued.}
		\setlength{\tabcolsep}{1.1ex}
		\small 	\begin{tabular}{lcccccccccccc} 
 \hline 
\mc{1}{l|}{graphs} & \mc{1}{c}{$t$} & \mc{1}{c|}{$w$} & \mc{1}{c}{$t$} & \mc{1}{c|}{$w$} & \mc{1}{c}{$t$} & \mc{1}{c|}{$w$} & \mc{1}{c}{$t$} & \mc{1}{c|}{$w$} & \mc{1}{c}{$t$} & \mc{1}{c|}{$w$} & \mc{1}{c}{$t$} & \mc{1}{c}{$w$}\\ 
 \hline 
\mc{1}{l|}{osm} & \mc{2}{c|}{branch reduce}  & \mc{2}{c|}{{\hils}}  & \mc{2}{c|}{{\htwis}}  & \mc{2}{c|}{{\wmmis}}  & \mc{2}{c|}{{\wmmiss}}  & \mc{2}{c}{struction} \\ 
\hline 
 \rowcolor{lightergray} \mc{1}{l|}{\textit{kansas-1}} & \mc{1}{r}{<\numprint{0.01}} & \mc{1}{r|}{\textbf{\numprint{84449}}} & \mc{1}{r}{\numprint{4.13}} & \mc{1}{r|}{\textbf{\numprint{84449}}} & \mc{1}{r}{<\numprint{0.01}} & \mc{1}{r|}{\textbf{\numprint{84449}}} & \mc{1}{r}{\numprint{0.01}} & \mc{1}{r|}{\textbf{\numprint{84449}}} & \mc{1}{r}{\numprint{0.01}} & \mc{1}{r|}{\textbf{\numprint{84449}}} & \mc{1}{r}{<\numprint{0.01}} & \mc{1}{r}{\textbf{\numprint{84449}}} \\ 
 \rowcolor{lightergray} \mc{1}{l|}{\textit{kansas-2}} & \mc{1}{r}{\numprint{0.15}} & \mc{1}{r|}{\textbf{\numprint{85942}}} & \mc{1}{r}{\numprint{31.26}} & \mc{1}{r|}{\textbf{\numprint{85942}}} & \mc{1}{r}{\numprint{0.01}} & \mc{1}{r|}{\numprint{85687}} & \mc{1}{r}{\numprint{0.88}} & \mc{1}{r|}{\textbf{\numprint{85942}}} & \mc{1}{r}{\numprint{0.88}} & \mc{1}{r|}{\textbf{\numprint{85942}}} & \mc{1}{r}{\numprint{0.01}} & \mc{1}{r}{\textbf{\numprint{85942}}} \\ 
 \rowcolor{lightergray} \mc{1}{l|}{\textit{kansas-3}} & \mc{1}{r}{\numprint{36001.02}} & \mc{1}{r|}{\numprint{87963}} & \mc{1}{r}{\numprint{759.69}} & \mc{1}{r|}{\textbf{\numprint{87976}}} & \mc{1}{r}{\numprint{2.43}} & \mc{1}{r|}{\numprint{87424}} & \mc{1}{r}{\numprint{57.95}} & \mc{1}{r|}{\textbf{\numprint{87976}}} & \mc{1}{r}{\numprint{96.49}} & \mc{1}{r|}{\textbf{\numprint{87976}}} & \mc{1}{r}{\numprint{16.79}} & \mc{1}{r}{\textbf{\numprint{87976}}} \\
\hline 

\end{tabular} 

	\end{ThreePartTable}
\end{table*}

\begin{table*}
	\begin{ThreePartTable} 
		\caption*{Table \ref{tab: soa_osm} continued.}
		\setlength{\tabcolsep}{1.1ex}
		\small 	\begin{tabular}{lcccccccccccc} 
 \hline 
\mc{1}{l|}{graphs} & \mc{1}{c}{$t$} & \mc{1}{c|}{$w$} & \mc{1}{c}{$t$} & \mc{1}{c|}{$w$} & \mc{1}{c}{$t$} & \mc{1}{c|}{$w$} & \mc{1}{c}{$t$} & \mc{1}{c|}{$w$} & \mc{1}{c}{$t$} & \mc{1}{c|}{$w$} & \mc{1}{c}{$t$} & \mc{1}{c}{$w$}\\ 
 \hline 
\mc{1}{l|}{osm} & \mc{2}{c|}{branch reduce}  & \mc{2}{c|}{{\hils}}  & \mc{2}{c|}{{\htwis}}  & \mc{2}{c|}{{\wmmis}}  & \mc{2}{c|}{{\wmmiss}}  & \mc{2}{c}{struction} \\ 
\hline 
 \rowcolor{lightergray} \mc{1}{l|}{\textit{new-york-1}} & \mc{1}{r}{<\numprint{0.01}} & \mc{1}{r|}{\textbf{\numprint{13187}}} & \mc{1}{r}{\numprint{1.74}} & \mc{1}{r|}{\textbf{\numprint{13187}}} & \mc{1}{r}{<\numprint{0.01}} & \mc{1}{r|}{\numprint{13166}} & \mc{1}{r}{\numprint{0.02}} & \mc{1}{r|}{\textbf{\numprint{13187}}} & \mc{1}{r}{\numprint{0.02}} & \mc{1}{r|}{\textbf{\numprint{13187}}} & \mc{1}{r}{<\numprint{0.01}} & \mc{1}{r}{\textbf{\numprint{13187}}} \\ 
 \rowcolor{lightergray} \mc{1}{l|}{\textit{new-york-2}} & \mc{1}{r}{\numprint{0.20}} & \mc{1}{r|}{\textbf{\numprint{14330}}} & \mc{1}{r}{\numprint{19.63}} & \mc{1}{r|}{\textbf{\numprint{14330}}} & \mc{1}{r}{<\numprint{0.01}} & \mc{1}{r|}{\textbf{\numprint{14330}}} & \mc{1}{r}{\numprint{0.31}} & \mc{1}{r|}{\textbf{\numprint{14330}}} & \mc{1}{r}{\numprint{0.30}} & \mc{1}{r|}{\textbf{\numprint{14330}}} & \mc{1}{r}{\numprint{0.02}} & \mc{1}{r}{\textbf{\numprint{14330}}} \\ 
 \rowcolor{lightergray} \mc{1}{l|}{\textit{new-york-3}} & \mc{1}{r}{\numprint{13823.10}} & \mc{1}{r|}{\textbf{\numprint{16268}}} & \mc{1}{r}{\numprint{105.03}} & \mc{1}{r|}{\textbf{\numprint{16268}}} & \mc{1}{r}{\numprint{0.14}} & \mc{1}{r|}{\numprint{15293}} & \mc{1}{r}{\numprint{10.83}} & \mc{1}{r|}{\textbf{\numprint{16268}}} & \mc{1}{r}{\numprint{67.99}} & \mc{1}{r|}{\textbf{\numprint{16268}}} & \mc{1}{r}{\numprint{0.34}} & \mc{1}{r}{\textbf{\numprint{16268}}} \\
\hline 

\end{tabular} 

	\end{ThreePartTable}
\end{table*}

\begin{table*}
	\centering
	\begin{ThreePartTable}
		\caption{Average solution weight $\omega$ and time $t$ in seconds required to compute it for our set of snap instances. \textbf{Bold} numbers indicate the best solution among all algorithms. Rows have a \noindent\colorbox{lightergray} {\parbox{\widthof{gray}}{gray}} background color, if branch reduce or struction computed an exact solution. We also report the number of best solutions and the geometric mean running time over all instances. }\label{tab: soa_snap}
		\setlength{\tabcolsep}{.7ex}
		 \small 	\begin{tabular}{lcccccccccccc} 
 \hline 
\mc{1}{l|}{graphs} & \mc{1}{c}{$t$} & \mc{1}{c|}{$w$} & \mc{1}{c}{$t$} & \mc{1}{c|}{$w$} & \mc{1}{c}{$t$} & \mc{1}{c|}{$w$} & \mc{1}{c}{$t$} & \mc{1}{c|}{$w$} & \mc{1}{c}{$t$} & \mc{1}{c|}{$w$} & \mc{1}{c}{$t$} & \mc{1}{c}{$w$}\\ 
 \hline 
\mc{1}{l|}{snap} & \mc{2}{c|}{branch reduce}  & \mc{2}{c|}{{\hils}}  & \mc{2}{c|}{{\htwis}}  & \mc{2}{c|}{{\wmmis}}  & \mc{2}{c|}{{\wmmiss}}  & \mc{2}{c}{struction} \\ 
\hline 
\mc{1}{l|}{\textit{as-skitter}} & \mc{2}{c|}{-} & \mc{1}{r}{\numprint{36000.25}} & \mc{1}{r|}{\numprint{123994141}} & \mc{1}{r}{\numprint{1.04}} & \mc{1}{r|}{\numprint{124141373}} & \mc{1}{r}{\numprint{2564.30}} & \mc{1}{r|}{\textbf{\numprint{124157714}}} & \mc{1}{r}{\numprint{10231.98}} & \mc{1}{r|}{\numprint{124157712}} & \mc{2}{c}{-} \\ 
 \rowcolor{lightergray} \mc{1}{l|}{\textit{ca-AstroPh}} & \mc{1}{r}{\numprint{0.02}} & \mc{1}{r|}{\textbf{\numprint{797510}}} & \mc{1}{r}{\numprint{924.29}} & \mc{1}{r|}{\numprint{797508}} & \mc{1}{r}{\numprint{0.02}} & \mc{1}{r|}{\numprint{797363}} & \mc{1}{r}{\numprint{20.37}} & \mc{1}{r|}{\textbf{\numprint{797510}}} & \mc{1}{r}{\numprint{26.50}} & \mc{1}{r|}{\textbf{\numprint{797510}}} & \mc{1}{r}{\numprint{0.02}} & \mc{1}{r}{\textbf{\numprint{797510}}} \\ 
 \rowcolor{lightergray} \mc{1}{l|}{\textit{ca-CondMat}} & \mc{1}{r}{\numprint{0.02}} & \mc{1}{r|}{\textbf{\numprint{1147950}}} & \mc{1}{r}{\numprint{1015.80}} & \mc{1}{r|}{\numprint{1147947}} & \mc{1}{r}{\numprint{0.01}} & \mc{1}{r|}{\textbf{\numprint{1147950}}} & \mc{1}{r}{\numprint{24.25}} & \mc{1}{r|}{\textbf{\numprint{1147950}}} & \mc{1}{r}{\numprint{21.09}} & \mc{1}{r|}{\textbf{\numprint{1147950}}} & \mc{1}{r}{\numprint{0.01}} & \mc{1}{r}{\textbf{\numprint{1147950}}} \\ 
 \rowcolor{lightergray} \mc{1}{l|}{\textit{ca-GrQc}} & \mc{1}{r}{<\numprint{0.01}} & \mc{1}{r|}{\textbf{\numprint{286489}}} & \mc{1}{r}{\numprint{148.82}} & \mc{1}{r|}{\textbf{\numprint{286489}}} & \mc{1}{r}{<\numprint{0.01}} & \mc{1}{r|}{\numprint{286352}} & \mc{1}{r}{\numprint{3.48}} & \mc{1}{r|}{\textbf{\numprint{286489}}} & \mc{1}{r}{\numprint{3.93}} & \mc{1}{r|}{\textbf{\numprint{286489}}} & \mc{1}{r}{<\numprint{0.01}} & \mc{1}{r}{\textbf{\numprint{286489}}} \\ 
 \rowcolor{lightergray} \mc{1}{l|}{\textit{ca-HepPh}} & \mc{1}{r}{\numprint{0.01}} & \mc{1}{r|}{\textbf{\numprint{581039}}} & \mc{1}{r}{\numprint{589.17}} & \mc{1}{r|}{\textbf{\numprint{581039}}} & \mc{1}{r}{\numprint{0.01}} & \mc{1}{r|}{\numprint{580864}} & \mc{1}{r}{\numprint{13.85}} & \mc{1}{r|}{\textbf{\numprint{581039}}} & \mc{1}{r}{\numprint{14.54}} & \mc{1}{r|}{\textbf{\numprint{581039}}} & \mc{1}{r}{\numprint{0.01}} & \mc{1}{r}{\textbf{\numprint{581039}}} \\ 
 \rowcolor{lightergray} \mc{1}{l|}{\textit{ca-HepTh}} & \mc{1}{r}{\numprint{0.01}} & \mc{1}{r|}{\textbf{\numprint{562004}}} & \mc{1}{r}{\numprint{279.70}} & \mc{1}{r|}{\textbf{\numprint{562004}}} & \mc{1}{r}{<\numprint{0.01}} & \mc{1}{r|}{\numprint{561736}} & \mc{1}{r}{\numprint{5.90}} & \mc{1}{r|}{\textbf{\numprint{562004}}} & \mc{1}{r}{\numprint{5.89}} & \mc{1}{r|}{\textbf{\numprint{562004}}} & \mc{1}{r}{<\numprint{0.01}} & \mc{1}{r}{\textbf{\numprint{562004}}} \\ 
 \rowcolor{lightergray} \mc{1}{l|}{\textit{com-amazon}} & \mc{1}{r}{\numprint{0.48}} & \mc{1}{r|}{\textbf{\numprint{19271031}}} & \mc{1}{r}{\numprint{33178.00}} & \mc{1}{r|}{\numprint{19270284}} & \mc{1}{r}{\numprint{0.14}} & \mc{1}{r|}{\numprint{19270078}} & \mc{1}{r}{\numprint{1.38}} & \mc{1}{r|}{\textbf{\numprint{19271031}}} & \mc{1}{r}{\numprint{1.07}} & \mc{1}{r|}{\textbf{\numprint{19271031}}} & \mc{1}{r}{\numprint{0.36}} & \mc{1}{r}{\textbf{\numprint{19271031}}} \\ 
 \rowcolor{lightergray} \mc{1}{l|}{\textit{com-youtube}} & \mc{1}{r}{\numprint{0.76}} & \mc{1}{r|}{\textbf{\numprint{90295294}}} & \mc{1}{r}{\numprint{36000.10}} & \mc{1}{r|}{\numprint{90289947}} & \mc{1}{r}{\numprint{0.34}} & \mc{1}{r|}{\numprint{90295285}} & \mc{1}{r}{\numprint{1.76}} & \mc{1}{r|}{\textbf{\numprint{90295294}}} & \mc{1}{r}{\numprint{1.87}} & \mc{1}{r|}{\textbf{\numprint{90295294}}} & \mc{1}{r}{\numprint{0.69}} & \mc{1}{r}{\textbf{\numprint{90295294}}} \\ 
 \rowcolor{lightergray} \mc{1}{l|}{\textit{email-Enron}} & \mc{1}{r}{\numprint{0.03}} & \mc{1}{r|}{\textbf{\numprint{2464935}}} & \mc{1}{r}{\numprint{1498.60}} & \mc{1}{r|}{\numprint{2464920}} & \mc{1}{r}{\numprint{0.01}} & \mc{1}{r|}{\textbf{\numprint{2464935}}} & \mc{1}{r}{\numprint{26.64}} & \mc{1}{r|}{\textbf{\numprint{2464935}}} & \mc{1}{r}{\numprint{26.79}} & \mc{1}{r|}{\textbf{\numprint{2464935}}} & \mc{1}{r}{\numprint{0.02}} & \mc{1}{r}{\textbf{\numprint{2464935}}} \\ 
 \rowcolor{lightergray} \mc{1}{l|}{\textit{email-EuAll}} & \mc{1}{r}{\numprint{0.07}} & \mc{1}{r|}{\textbf{\numprint{25286322}}} & \mc{1}{r}{\numprint{17439.93}} & \mc{1}{r|}{\textbf{\numprint{25286322}}} & \mc{1}{r}{\numprint{0.03}} & \mc{1}{r|}{\numprint{25265214}} & \mc{1}{r}{\numprint{87.55}} & \mc{1}{r|}{\textbf{\numprint{25286322}}} & \mc{1}{r}{\numprint{91.68}} & \mc{1}{r|}{\textbf{\numprint{25286322}}} & \mc{1}{r}{\numprint{0.04}} & \mc{1}{r}{\textbf{\numprint{25286322}}} \\ 
 \rowcolor{lightergray} \mc{1}{l|}{\textit{loc-gowalla}} & \mc{2}{c|}{-} & \mc{1}{r}{\numprint{17018.50}} & \mc{1}{r|}{\numprint{12275375}} & \mc{1}{r}{\numprint{0.08}} & \mc{1}{r|}{\numprint{12276781}} & \mc{1}{r}{\numprint{4.38}} & \mc{1}{r|}{\textbf{\numprint{12276929}}} & \mc{1}{r}{\numprint{7.62}} & \mc{1}{r|}{\textbf{\numprint{12276929}}} & \mc{1}{r}{\numprint{1.32}} & \mc{1}{r}{\textbf{\numprint{12276929}}} \\ 
 \rowcolor{lightergray} \mc{1}{l|}{\textit{p2p-G.04}} & \mc{1}{r}{\numprint{0.01}} & \mc{1}{r|}{\textbf{\numprint{679111}}} & \mc{1}{r}{\numprint{250.11}} & \mc{1}{r|}{\numprint{679110}} & \mc{1}{r}{<\numprint{0.01}} & \mc{1}{r|}{\numprint{679085}} & \mc{1}{r}{\numprint{4.57}} & \mc{1}{r|}{\textbf{\numprint{679111}}} & \mc{1}{r}{\numprint{5.03}} & \mc{1}{r|}{\textbf{\numprint{679111}}} & \mc{1}{r}{\numprint{0.01}} & \mc{1}{r}{\textbf{\numprint{679111}}} \\ 
 \rowcolor{lightergray} \mc{1}{l|}{\textit{p2p-G.05}} & \mc{1}{r}{\numprint{0.01}} & \mc{1}{r|}{\textbf{\numprint{554943}}} & \mc{1}{r}{\numprint{192.45}} & \mc{1}{r|}{\textbf{\numprint{554943}}} & \mc{1}{r}{<\numprint{0.01}} & \mc{1}{r|}{\textbf{\numprint{554943}}} & \mc{1}{r}{\numprint{3.96}} & \mc{1}{r|}{\textbf{\numprint{554943}}} & \mc{1}{r}{\numprint{3.85}} & \mc{1}{r|}{\textbf{\numprint{554943}}} & \mc{1}{r}{<\numprint{0.01}} & \mc{1}{r}{\textbf{\numprint{554943}}} \\ 
 \rowcolor{lightergray} \mc{1}{l|}{\textit{p2p-G.06}} & \mc{1}{r}{\numprint{0.01}} & \mc{1}{r|}{\textbf{\numprint{548612}}} & \mc{1}{r}{\numprint{183.91}} & \mc{1}{r|}{\textbf{\numprint{548612}}} & \mc{1}{r}{<\numprint{0.01}} & \mc{1}{r|}{\textbf{\numprint{548612}}} & \mc{1}{r}{\numprint{3.79}} & \mc{1}{r|}{\textbf{\numprint{548612}}} & \mc{1}{r}{\numprint{3.72}} & \mc{1}{r|}{\textbf{\numprint{548612}}} & \mc{1}{r}{<\numprint{0.01}} & \mc{1}{r}{\textbf{\numprint{548612}}} \\ 
 \rowcolor{lightergray} \mc{1}{l|}{\textit{p2p-G.08}} & \mc{1}{r}{<\numprint{0.01}} & \mc{1}{r|}{\textbf{\numprint{434577}}} & \mc{1}{r}{\numprint{109.72}} & \mc{1}{r|}{\textbf{\numprint{434577}}} & \mc{1}{r}{<\numprint{0.01}} & \mc{1}{r|}{\textbf{\numprint{434577}}} & \mc{1}{r}{\numprint{2.71}} & \mc{1}{r|}{\textbf{\numprint{434577}}} & \mc{1}{r}{\numprint{2.53}} & \mc{1}{r|}{\textbf{\numprint{434577}}} & \mc{1}{r}{<\numprint{0.01}} & \mc{1}{r}{\textbf{\numprint{434577}}} \\ 
 \rowcolor{lightergray} \mc{1}{l|}{\textit{p2p-G.09}} & \mc{1}{r}{<\numprint{0.01}} & \mc{1}{r|}{\textbf{\numprint{568439}}} & \mc{1}{r}{\numprint{152.25}} & \mc{1}{r|}{\textbf{\numprint{568439}}} & \mc{1}{r}{<\numprint{0.01}} & \mc{1}{r|}{\textbf{\numprint{568439}}} & \mc{1}{r}{\numprint{3.32}} & \mc{1}{r|}{\textbf{\numprint{568439}}} & \mc{1}{r}{\numprint{3.40}} & \mc{1}{r|}{\textbf{\numprint{568439}}} & \mc{1}{r}{<\numprint{0.01}} & \mc{1}{r}{\textbf{\numprint{568439}}} \\ 
 \rowcolor{lightergray} \mc{1}{l|}{\textit{p2p-G.24}} & \mc{1}{r}{\numprint{0.01}} & \mc{1}{r|}{\textbf{\numprint{1984567}}} & \mc{1}{r}{\numprint{569.39}} & \mc{1}{r|}{\textbf{\numprint{1984567}}} & \mc{1}{r}{\numprint{0.01}} & \mc{1}{r|}{\textbf{\numprint{1984567}}} & \mc{1}{r}{\numprint{9.99}} & \mc{1}{r|}{\textbf{\numprint{1984567}}} & \mc{1}{r}{\numprint{10.25}} & \mc{1}{r|}{\textbf{\numprint{1984567}}} & \mc{1}{r}{\numprint{0.01}} & \mc{1}{r}{\textbf{\numprint{1984567}}} \\ 
 \rowcolor{lightergray} \mc{1}{l|}{\textit{p2p-G.25}} & \mc{1}{r}{\numprint{0.01}} & \mc{1}{r|}{\textbf{\numprint{1701967}}} & \mc{1}{r}{\numprint{467.31}} & \mc{1}{r|}{\textbf{\numprint{1701967}}} & \mc{1}{r}{\numprint{0.01}} & \mc{1}{r|}{\textbf{\numprint{1701967}}} & \mc{1}{r}{\numprint{7.78}} & \mc{1}{r|}{\textbf{\numprint{1701967}}} & \mc{1}{r}{\numprint{7.93}} & \mc{1}{r|}{\textbf{\numprint{1701967}}} & \mc{1}{r}{\numprint{0.01}} & \mc{1}{r}{\textbf{\numprint{1701967}}} \\ 
 \rowcolor{lightergray} \mc{1}{l|}{\textit{p2p-G.30}} & \mc{1}{r}{\numprint{0.02}} & \mc{1}{r|}{\textbf{\numprint{2787907}}} & \mc{1}{r}{\numprint{810.63}} & \mc{1}{r|}{\textbf{\numprint{2787907}}} & \mc{1}{r}{\numprint{0.01}} & \mc{1}{r|}{\numprint{2787902}} & \mc{1}{r}{\numprint{13.35}} & \mc{1}{r|}{\textbf{\numprint{2787907}}} & \mc{1}{r}{\numprint{12.71}} & \mc{1}{r|}{\textbf{\numprint{2787907}}} & \mc{1}{r}{\numprint{0.01}} & \mc{1}{r}{\textbf{\numprint{2787907}}} \\ 
 \rowcolor{lightergray} \mc{1}{l|}{\textit{p2p-G.31}} & \mc{1}{r}{\numprint{0.03}} & \mc{1}{r|}{\textbf{\numprint{4776986}}} & \mc{1}{r}{\numprint{1795.27}} & \mc{1}{r|}{\numprint{4776969}} & \mc{1}{r}{\numprint{0.01}} & \mc{1}{r|}{\numprint{4776925}} & \mc{1}{r}{\numprint{26.13}} & \mc{1}{r|}{\textbf{\numprint{4776986}}} & \mc{1}{r}{\numprint{27.62}} & \mc{1}{r|}{\textbf{\numprint{4776986}}} & \mc{1}{r}{\numprint{0.02}} & \mc{1}{r}{\textbf{\numprint{4776986}}} \\ 
 \rowcolor{lightergray} \mc{1}{l|}{\textit{roadNet-CA}} & \mc{1}{r}{\numprint{279.50}} & \mc{1}{r|}{\textbf{\numprint{111360828}}} & \mc{1}{r}{\numprint{36000.15}} & \mc{1}{r|}{\numprint{109991788}} & \mc{1}{r}{\numprint{0.61}} & \mc{1}{r|}{\numprint{111325524}} & \mc{1}{r}{\numprint{28447.76}} & \mc{1}{r|}{\numprint{111360436}} & \mc{1}{r}{\numprint{8.35}} & \mc{1}{r|}{\textbf{\numprint{111360828}}} & \mc{1}{r}{\numprint{1.54}} & \mc{1}{r}{\textbf{\numprint{111360828}}} \\ 
 \rowcolor{lightergray} \mc{1}{l|}{\textit{roadNet-PA}} & \mc{1}{r}{\numprint{16.44}} & \mc{1}{r|}{\textbf{\numprint{61731589}}} & \mc{1}{r}{\numprint{36000.07}} & \mc{1}{r|}{\numprint{61549659}} & \mc{1}{r}{\numprint{0.33}} & \mc{1}{r|}{\numprint{61710606}} & \mc{1}{r}{\numprint{5584.85}} & \mc{1}{r|}{\numprint{61731489}} & \mc{1}{r}{\numprint{4.13}} & \mc{1}{r|}{\textbf{\numprint{61731589}}} & \mc{1}{r}{\numprint{0.85}} & \mc{1}{r}{\textbf{\numprint{61731589}}} \\ 
 \rowcolor{lightergray} \mc{1}{l|}{\textit{roadNet-TX}} & \mc{1}{r}{\numprint{15.76}} & \mc{1}{r|}{\textbf{\numprint{78599946}}} & \mc{1}{r}{\numprint{36000.10}} & \mc{1}{r|}{\numprint{78164327}} & \mc{1}{r}{\numprint{0.42}} & \mc{1}{r|}{\numprint{78575460}} & \mc{1}{r}{\numprint{24452.19}} & \mc{1}{r|}{\numprint{78599705}} & \mc{1}{r}{\numprint{5.78}} & \mc{1}{r|}{\textbf{\numprint{78599946}}} & \mc{1}{r}{\numprint{1.05}} & \mc{1}{r}{\textbf{\numprint{78599946}}} \\ 
 \rowcolor{lightergray} \mc{1}{l|}{\textit{soc-Ep.1}} & \mc{1}{r}{\numprint{0.05}} & \mc{1}{r|}{\textbf{\numprint{5690970}}} & \mc{1}{r}{\numprint{2813.40}} & \mc{1}{r|}{\numprint{5690859}} & \mc{1}{r}{\numprint{0.02}} & \mc{1}{r|}{\textbf{\numprint{5690970}}} & \mc{1}{r}{\numprint{40.04}} & \mc{1}{r|}{\textbf{\numprint{5690970}}} & \mc{1}{r}{\numprint{43.38}} & \mc{1}{r|}{\textbf{\numprint{5690970}}} & \mc{1}{r}{\numprint{0.05}} & \mc{1}{r}{\textbf{\numprint{5690970}}} \\ 
\mc{1}{l|}{\textit{soc-LiveJ.1}} & \mc{1}{r}{\numprint{36002.35}} & \mc{1}{r|}{\numprint{284008877}} & \mc{1}{r}{\numprint{36000.67}} & \mc{1}{r|}{\numprint{281688778}} & \mc{1}{r}{\numprint{12.20}} & \mc{1}{r|}{\numprint{283922214}} & \mc{1}{r}{\numprint{27065.63}} & \mc{1}{r|}{\numprint{284036182}} & \mc{1}{r}{\numprint{688.97}} & \mc{1}{r|}{\textbf{\numprint{284036236}}} & \mc{2}{c}{-} \\ 
 \rowcolor{lightergray} \mc{1}{l|}{\textit{soc-Sl.0811}} & \mc{1}{r}{\numprint{0.06}} & \mc{1}{r|}{\textbf{\numprint{5660899}}} & \mc{1}{r}{\numprint{4106.47}} & \mc{1}{r|}{\numprint{5660734}} & \mc{1}{r}{\numprint{0.02}} & \mc{1}{r|}{\textbf{\numprint{5660899}}} & \mc{1}{r}{\numprint{58.67}} & \mc{1}{r|}{\textbf{\numprint{5660899}}} & \mc{1}{r}{\numprint{56.73}} & \mc{1}{r|}{\textbf{\numprint{5660899}}} & \mc{1}{r}{\numprint{0.06}} & \mc{1}{r}{\textbf{\numprint{5660899}}} \\ 
 \rowcolor{lightergray} \mc{1}{l|}{\textit{soc-Sl.0902}} & \mc{1}{r}{\numprint{0.07}} & \mc{1}{r|}{\textbf{\numprint{5971849}}} & \mc{1}{r}{\numprint{4260.67}} & \mc{1}{r|}{\numprint{5971574}} & \mc{1}{r}{\numprint{0.02}} & \mc{1}{r|}{\numprint{5971821}} & \mc{1}{r}{\numprint{62.50}} & \mc{1}{r|}{\textbf{\numprint{5971849}}} & \mc{1}{r}{\numprint{63.13}} & \mc{1}{r|}{\textbf{\numprint{5971849}}} & \mc{1}{r}{\numprint{0.07}} & \mc{1}{r}{\textbf{\numprint{5971849}}} \\ 
\mc{1}{l|}{\textit{soc-p.-rel.}} & \mc{1}{r}{\numprint{36059.01}} & \mc{1}{r|}{\numprint{82778214}} & \mc{1}{r}{\numprint{36000.42}} & \mc{1}{r|}{\numprint{83696885}} & \mc{1}{r}{\numprint{55.41}} & \mc{1}{r|}{\textbf{\numprint{83920370}}} & \mc{1}{r}{\numprint{42132.17}} & \mc{1}{r|}{\numprint{83720129}} & \mc{1}{r}{\numprint{45807.65}} & \mc{1}{r|}{\numprint{83720972}} & \mc{1}{r}{\numprint{634.61}} & \mc{1}{r}{\numprint{79620979}} \\ 
 \rowcolor{lightergray} \mc{1}{l|}{\textit{web-BS.}} & \mc{1}{r}{\numprint{36000.12}} & \mc{1}{r|}{\numprint{43891206}} & \mc{1}{r}{\numprint{36000.10}} & \mc{1}{r|}{\numprint{43888267}} & \mc{1}{r}{\numprint{9.94}} & \mc{1}{r|}{\numprint{43889843}} & \mc{1}{r}{\numprint{18698.57}} & \mc{1}{r|}{\numprint{43907225}} & \mc{1}{r}{\numprint{13.78}} & \mc{1}{r|}{\textbf{\numprint{43907482}}} & \mc{1}{r}{\numprint{6.52}} & \mc{1}{r}{\textbf{\numprint{43907482}}} \\ 
 \rowcolor{lightergray} \mc{1}{l|}{\textit{web-Google}} & \mc{1}{r}{\numprint{2.33}} & \mc{1}{r|}{\textbf{\numprint{56326504}}} & \mc{1}{r}{\numprint{36000.15}} & \mc{1}{r|}{\numprint{56319614}} & \mc{1}{r}{\numprint{0.65}} & \mc{1}{r|}{\numprint{56323382}} & \mc{1}{r}{\numprint{5.15}} & \mc{1}{r|}{\textbf{\numprint{56326504}}} & \mc{1}{r}{\numprint{4.37}} & \mc{1}{r|}{\textbf{\numprint{56326504}}} & \mc{1}{r}{\numprint{1.52}} & \mc{1}{r}{\textbf{\numprint{56326504}}} \\ 
 \rowcolor{lightergray} \mc{1}{l|}{\textit{web-ND.}} & \mc{1}{r}{\numprint{496.64}} & \mc{1}{r|}{\textbf{\numprint{26016941}}} & \mc{1}{r}{\numprint{27389.91}} & \mc{1}{r|}{\numprint{26014810}} & \mc{1}{r}{\numprint{0.12}} & \mc{1}{r|}{\numprint{26013830}} & \mc{1}{r}{\numprint{6.55}} & \mc{1}{r|}{\textbf{\numprint{26016941}}} & \mc{1}{r}{\numprint{2.79}} & \mc{1}{r|}{\textbf{\numprint{26016941}}} & \mc{1}{r}{\numprint{1.36}} & \mc{1}{r}{\textbf{\numprint{26016941}}} \\ 
 \rowcolor{lightergray} \mc{1}{l|}{\textit{web-Stanford}} & \mc{2}{c|}{-} & \mc{1}{r}{\numprint{35324.07}} & \mc{1}{r|}{\numprint{17789989}} & \mc{1}{r}{\numprint{0.50}} & \mc{1}{r|}{\numprint{17789430}} & \mc{1}{r}{\numprint{3.92}} & \mc{1}{r|}{\textbf{\numprint{17792930}}} & \mc{1}{r}{\numprint{2.99}} & \mc{1}{r|}{\textbf{\numprint{17792930}}} & \mc{1}{r}{\numprint{1.36}} & \mc{1}{r}{\textbf{\numprint{17792930}}} \\ 
 \rowcolor{lightergray} \mc{1}{l|}{\textit{wiki-Talk}} & \mc{1}{r}{\numprint{1.08}} & \mc{1}{r|}{\textbf{\numprint{235837346}}} & \mc{1}{r}{\numprint{36000.12}} & \mc{1}{r|}{\numprint{235837287}} & \mc{1}{r}{\numprint{0.41}} & \mc{1}{r|}{\textbf{\numprint{235837346}}} & \mc{1}{r}{\numprint{881.14}} & \mc{1}{r|}{\textbf{\numprint{235837346}}} & \mc{1}{r}{\numprint{887.71}} & \mc{1}{r|}{\textbf{\numprint{235837346}}} & \mc{1}{r}{\numprint{0.95}} & \mc{1}{r}{\textbf{\numprint{235837346}}} \\ 
 \rowcolor{lightergray} \mc{1}{l|}{\textit{wiki-Vote}} & \mc{1}{r}{\numprint{0.01}} & \mc{1}{r|}{\textbf{\numprint{500079}}} & \mc{1}{r}{\numprint{201.84}} & \mc{1}{r|}{\textbf{\numprint{500079}}} & \mc{1}{r}{\numprint{0.01}} & \mc{1}{r|}{\numprint{499740}} & \mc{1}{r}{\numprint{5.14}} & \mc{1}{r|}{\textbf{\numprint{500079}}} & \mc{1}{r}{\numprint{5.21}} & \mc{1}{r|}{\textbf{\numprint{500079}}} & \mc{1}{r}{\numprint{0.01}} & \mc{1}{r}{\textbf{\numprint{500079}}} \\ 
\hline 
\mc{1}{l|}{overall} & \mc{2}{c|}{branch reduce}  & \mc{2}{c|}{{\hils}}  & \mc{2}{c|}{{\htwis}}  & \mc{2}{c|}{{\wmmis}}  & \mc{2}{c|}{{\wmmiss}}  & \mc{2}{c}{struction} \\ 
\hline 
\mc{1}{l|}{\# best} & \mc{2}{r|}{28/34} & \mc{2}{r|}{12/34} & \mc{2}{r|}{12/34} & \mc{2}{r|}{28/34} & \mc{2}{r|}{32/34} & \mc{2}{r}{31/34}\\ 
\mc{1}{l|}{mean time} & \mc{2}{r|}{-} & \mc{2}{r|}{\numprint{3346.24}} & \mc{2}{r|}{\numprint{0.06}} & \mc{2}{r|}{\numprint{47.59}} & \mc{2}{r|}{\numprint{17.74}} & \mc{2}{r}{-}\\ 
\hline 

\end{tabular} 

	\end{ThreePartTable}
\end{table*}

\begin{table*}
	\centering
	\begin{ThreePartTable}
		\caption{Average solution weight $\omega$ and time $t$ in seconds required to compute it for our set of ssmc instances. \textbf{Bold} numbers indicate the best solution among all algorithms. Rows have a \noindent\colorbox{lightergray} {\parbox{\widthof{gray}}{gray}} background color, if branch reduce or struction computed an exact solution. We also report the number of best solutions and the geometric mean running time over all instances. }\label{tab: soa_ssmc}
		\setlength{\tabcolsep}{1.1ex}
		 \small 	\begin{tabular}{lcccccccccccc} 
 \hline 
\mc{1}{l|}{graphs} & \mc{1}{c}{$t$} & \mc{1}{c|}{$w$} & \mc{1}{c}{$t$} & \mc{1}{c|}{$w$} & \mc{1}{c}{$t$} & \mc{1}{c|}{$w$} & \mc{1}{c}{$t$} & \mc{1}{c|}{$w$} & \mc{1}{c}{$t$} & \mc{1}{c|}{$w$} & \mc{1}{c}{$t$} & \mc{1}{c}{$w$}\\ 
 \hline 
\mc{1}{l|}{ssmc} & \mc{2}{c|}{branch reduce}  & \mc{2}{c|}{{\hils}}  & \mc{2}{c|}{{\htwis}}  & \mc{2}{c|}{{\wmmis}}  & \mc{2}{c|}{{\wmmiss}}  & \mc{2}{c}{struction} \\ 
\hline 
 \rowcolor{lightergray} \mc{1}{l|}{\textit{ca2010}} & \mc{2}{c|}{-} & \mc{1}{r}{\numprint{36000.07}} & \mc{1}{r|}{\numprint{16828547}} & \mc{1}{r}{\numprint{0.47}} & \mc{1}{r|}{\numprint{16792827}} & \mc{1}{r}{\numprint{36612.45}} & \mc{1}{r|}{\numprint{16843620}} & \mc{1}{r}{\numprint{60.95}} & \mc{1}{r|}{\textbf{\numprint{16869550}}} & \mc{1}{r}{\numprint{6.18}} & \mc{1}{r}{\textbf{\numprint{16869550}}} \\ 
 \rowcolor{lightergray} \mc{1}{l|}{\textit{fl2010}} & \mc{1}{r}{\numprint{36000.10}} & \mc{1}{r|}{\numprint{8638961}} & \mc{1}{r}{\numprint{36000.05}} & \mc{1}{r|}{\numprint{8732113}} & \mc{1}{r}{\numprint{0.44}} & \mc{1}{r|}{\numprint{8719272}} & \mc{1}{r}{\numprint{36248.97}} & \mc{1}{r|}{\numprint{8738319}} & \mc{1}{r}{\numprint{16.62}} & \mc{1}{r|}{\textbf{\numprint{8743506}}} & \mc{1}{r}{\numprint{2.02}} & \mc{1}{r}{\textbf{\numprint{8743506}}} \\ 
 \rowcolor{lightergray} \mc{1}{l|}{\textit{ga2010}} & \mc{1}{r}{\numprint{36000.10}} & \mc{1}{r|}{\numprint{4644324}} & \mc{1}{r}{\numprint{29522.41}} & \mc{1}{r|}{\numprint{4642807}} & \mc{1}{r}{\numprint{0.16}} & \mc{1}{r|}{\numprint{4639891}} & \mc{1}{r}{\numprint{30401.36}} & \mc{1}{r|}{\numprint{4644293}} & \mc{1}{r}{\numprint{3.83}} & \mc{1}{r|}{\textbf{\numprint{4644417}}} & \mc{1}{r}{\numprint{0.62}} & \mc{1}{r}{\textbf{\numprint{4644417}}} \\ 
 \rowcolor{lightergray} \mc{1}{l|}{\textit{il2010}} & \mc{1}{r}{\numprint{36000.10}} & \mc{1}{r|}{\numprint{5852296}} & \mc{1}{r}{\numprint{36000.00}} & \mc{1}{r|}{\numprint{5983871}} & \mc{1}{r}{\numprint{0.31}} & \mc{1}{r|}{\numprint{5963974}} & \mc{1}{r}{\numprint{36332.33}} & \mc{1}{r|}{\numprint{5984484}} & \mc{1}{r}{\numprint{48.42}} & \mc{1}{r|}{\textbf{\numprint{5998539}}} & \mc{1}{r}{\numprint{2.33}} & \mc{1}{r}{\textbf{\numprint{5998539}}} \\ 
 \rowcolor{lightergray} \mc{1}{l|}{\textit{nh2010}} & \mc{1}{r}{\numprint{36000.00}} & \mc{1}{r|}{\numprint{581637}} & \mc{1}{r}{\numprint{2163.80}} & \mc{1}{r|}{\numprint{588797}} & \mc{1}{r}{\numprint{0.03}} & \mc{1}{r|}{\numprint{587059}} & \mc{1}{r}{\numprint{3379.36}} & \mc{1}{r|}{\textbf{\numprint{588996}}} & \mc{1}{r}{\numprint{1.22}} & \mc{1}{r|}{\textbf{\numprint{588996}}} & \mc{1}{r}{\numprint{0.11}} & \mc{1}{r}{\textbf{\numprint{588996}}} \\ 
 \rowcolor{lightergray} \mc{1}{l|}{\textit{ri2010}} & \mc{1}{r}{\numprint{36000.00}} & \mc{1}{r|}{\numprint{447427}} & \mc{1}{r}{\numprint{782.49}} & \mc{1}{r|}{\numprint{458489}} & \mc{1}{r}{\numprint{0.02}} & \mc{1}{r|}{\numprint{457108}} & \mc{1}{r}{\numprint{25340.86}} & \mc{1}{r|}{\numprint{459227}} & \mc{1}{r}{\numprint{1.45}} & \mc{1}{r|}{\textbf{\numprint{459275}}} & \mc{1}{r}{\numprint{0.09}} & \mc{1}{r}{\textbf{\numprint{459275}}} \\ 
\hline 
\mc{1}{l|}{overall} & \mc{2}{c|}{branch reduce}  & \mc{2}{c|}{{\hils}}  & \mc{2}{c|}{{\htwis}}  & \mc{2}{c|}{{\wmmis}}  & \mc{2}{c|}{{\wmmiss}}  & \mc{2}{c}{struction} \\ 
\hline 
\mc{1}{l|}{\# best} & \mc{2}{r|}{0/6} & \mc{2}{r|}{0/6} & \mc{2}{r|}{0/6} & \mc{2}{r|}{1/6} & \mc{2}{r|}{6/6} & \mc{2}{r}{6/6}\\ 
\mc{1}{l|}{mean time} & \mc{2}{r|}{-} & \mc{2}{r|}{\numprint{11515.77}} & \mc{2}{r|}{\numprint{0.13}} & \mc{2}{r|}{\numprint{22376.62}} & \mc{2}{r|}{\numprint{8.32}} & \mc{2}{r}{\numprint{0.75}}\\ 
\hline 

\end{tabular} 

	\end{ThreePartTable}
\end{table*}
\clearpage 
\section{Detailed Data for Comparison with METAMIS}
\begin{table*}[h]
	\centering
	\caption{Comparison to quality of \textsf{METAMIS} from~\cite{DBLP:conf/esa/DongGNPRS22} for osm instances. \textbf{Bold} numbers indicate the best solution among the algorithms. As done by Dong~\etal~\cite{DBLP:conf/esa/DongGNPRS22} we reduced the osm instances in advance using \textsf{KaMIS}~\cite{DBLP:conf/alenex/Lamm0SWZ19}. For \textsf{METAMIS} the best result out of five runs is reported; we report the best solution out of four runs, each with a 10h time limit.}\label{tab: METAMIS}
	\small 	
\begin{tabular}{lcccccc} 
 \hline 
\mc{1}{l|}{graphs} & \mc{1}{c}{$t$} & \mc{1}{c|}{$w$} & \mc{1}{c}{$t$} & \mc{1}{c|}{$w$} & \mc{1}{c}{$t$} & \mc{1}{c}{$w$}\\ 
 \hline 
\mc{1}{l|}{reducedOsm} & \mc{2}{c|}{metamis}  & \mc{2}{c|}{{\wmmis}}  & \mc{2}{c}{{\wmmiss}} \\ 
\hline 
\mc{1}{l|}{\textit{alabama-3}} & \mc{1}{r}{\numprint{5.41}} & \mc{1}{r|}{\textbf{\numprint{45449}}} & \mc{1}{r}{\numprint{15.44}} & \mc{1}{r|}{\textbf{\numprint{45449}}} & \mc{1}{r}{\numprint{37.50}} & \mc{1}{r}{\textbf{\numprint{45449}}} \\ 
\mc{1}{l|}{\textit{d.o.c.-2}} & \mc{1}{r}{\numprint{58.38}} & \mc{1}{r|}{\textbf{\numprint{100302}}} & \mc{1}{r}{\numprint{4819.35}} & \mc{1}{r|}{\textbf{\numprint{100302}}} & \mc{1}{r}{\numprint{76.66}} & \mc{1}{r}{\textbf{\numprint{100302}}} \\ 
\mc{1}{l|}{\textit{d.o.c.-3}} & \mc{1}{r}{\numprint{1347.00}} & \mc{1}{r|}{\numprint{142910}} & \mc{1}{r}{\numprint{3238.98}} & \mc{1}{r|}{\textbf{\numprint{143052}}} & \mc{1}{r}{\numprint{4003.63}} & \mc{1}{r}{\numprint{143028}} \\ 
\mc{1}{l|}{\textit{florida-3}} & \mc{1}{r}{\numprint{3.08}} & \mc{1}{r|}{\textbf{\numprint{46132}}} & \mc{1}{r}{\numprint{14.19}} & \mc{1}{r|}{\textbf{\numprint{46132}}} & \mc{1}{r}{\numprint{14.92}} & \mc{1}{r}{\textbf{\numprint{46132}}} \\ 
\mc{1}{l|}{\textit{greenland-3}} & \mc{1}{r}{\numprint{28.02}} & \mc{1}{r|}{\textbf{\numprint{11960}}} & \mc{1}{r}{\numprint{5466.57}} & \mc{1}{r|}{\textbf{\numprint{11960}}} & \mc{1}{r}{\numprint{20355.60}} & \mc{1}{r}{\textbf{\numprint{11960}}} \\ 
\mc{1}{l|}{\textit{hawaii-3}} & \mc{1}{r}{\numprint{1207.00}} & \mc{1}{r|}{\numprint{58819}} & \mc{1}{r}{\numprint{19273.20}} & \mc{1}{r|}{\textbf{\numprint{58887}}} & \mc{1}{r}{\numprint{9534.49}} & \mc{1}{r}{\numprint{58858}} \\
\hline 

\end{tabular} 

\end{table*}

\clearpage 

\section{Graph Properties}
\begin{table*}[h]
	\caption{Graph properties.}\label{tab: graphs}
	\centering
	\setlength{\tabcolsep}{1.1ex}
	\footnotesize \begin{tabular}{lllclllclllclll}
    \cline{1-3} \cline{5-7} \cline{9-11} \cline{13-15}
\mc{1}{l|}{fe} & \mc{1}{c|}{$\vert V \vert$}  & \mc{1}{c}{$\vert E \vert$}  &\quad \qquad &\mc{1}{l|}{osm} & \mc{1}{c|}{$\vert V \vert$}  & \mc{1}{c}{$\vert E \vert$}  &\quad \qquad &\mc{1}{l|}{osm} & \mc{1}{c|}{$\vert V \vert$}  & \mc{1}{c}{$\vert E \vert$} &\quad \qquad &\mc{1}{l|}{ssmc} & \mc{1}{c|}{$\vert V \vert$}  & \mc{1}{c}{$\vert E \vert$}\\
    \cline{1-3} \cline{5-7} \cline{9-11}\cline{13-15}
\mc{1}{l|}{\textit{body}} & \mc{1}{r|}{\numprint{45087}} & \mc{1}{r}{\numprint{327468}} &\ &\mc{1}{l|}{\textit{iowa-1}} & \mc{1}{r|}{\numprint{90}} & \mc{1}{r}{\numprint{328}} &\ &\mc{1}{l|}{\textit{puerto-rico-1}} & \mc{1}{r|}{\numprint{60}} & \mc{1}{r}{\numprint{126}} &\ &\mc{1}{l|}{\textit{ca2010}} & \mc{1}{r|}{\numprint{710145}} & \mc{1}{r}{\numprint{3489366}}\\ 
\mc{1}{l|}{\textit{ocean}} & \mc{1}{r|}{\numprint{143437}} & \mc{1}{r}{\numprint{819186}} &\ &\mc{1}{l|}{\textit{iowa-2}} & \mc{1}{r|}{\numprint{155}} & \mc{1}{r}{\numprint{1908}} &\ &\mc{1}{l|}{\textit{puerto-rico-2}} & \mc{1}{r|}{\numprint{165}} & \mc{1}{r}{\numprint{2570}} &\ &\mc{1}{l|}{\textit{fl2010}} & \mc{1}{r|}{\numprint{484481}} & \mc{1}{r}{\numprint{2346294}}\\ 
\mc{1}{l|}{\textit{pwt}} & \mc{1}{r|}{\numprint{36519}} & \mc{1}{r}{\numprint{289588}} &\ &\mc{1}{l|}{\textit{kansas-1}} & \mc{1}{r|}{\numprint{190}} & \mc{1}{r}{\numprint{800}} &\ &\mc{1}{l|}{\textit{puerto-rico-3}} & \mc{1}{r|}{\numprint{494}} & \mc{1}{r}{\numprint{53852}} &\ &\mc{1}{l|}{\textit{ga2010}} & \mc{1}{r|}{\numprint{291086}} & \mc{1}{r}{\numprint{1418056}}\\ 
\mc{1}{l|}{\textit{rotor}} & \mc{1}{r|}{\numprint{99617}} & \mc{1}{r}{\numprint{1324862}} &\ &\mc{1}{l|}{\textit{kansas-2}} & \mc{1}{r|}{\numprint{602}} & \mc{1}{r}{\numprint{32948}} &\ &\mc{1}{l|}{\textit{rhode-is.-1}} & \mc{1}{r|}{\numprint{455}} & \mc{1}{r}{\numprint{3946}} &\ &\mc{1}{l|}{\textit{il2010}} & \mc{1}{r|}{\numprint{451554}} & \mc{1}{r}{\numprint{2164464}}\\
\end{tabular}

\end{table*}

\end{document}